\documentclass[prb,aps,twocolumn,floatfix,showpacs,superscriptaddress]{revtex4-1}
\usepackage{graphics}
\usepackage{epsfig}
\usepackage{times}
\usepackage{bm}
\usepackage{braket}
\usepackage{color}

\usepackage{amsmath}	

\begin{document}
\title{Variety of order-by-disorder phases in the asymmetric $J_1-J_2$ zigzag ladder:
From the delta chain to the $J_1-J_2$ chain}
\author{Tomoki Yamaguchi}
\affiliation{Department of physics, Chiba University, Japan}
\author{Stefan-Ludwig Drechsler}
\affiliation{Institute for Theoretical Solid State Physics, IFW Dresden, 01069 Dresden, Germany}
\author{Yukinori Ohta}
\affiliation{Department of physics, Chiba University, Japan}
\author{Satoshi Nishimoto}
\affiliation{Institute for Theoretical Solid State Physics, IFW Dresden, 01069 Dresden, Germany}
\affiliation{Department of Physics, Technical University Dresden, 01069 Dresden, Germany}

\date{\today}

\begin{abstract}
We study an asymmetric $J_1$-$J_2$ zigzag ladder consisting of two different spin-$\frac{1}{2}$ antiferromagnetic (AFM; $J_2$, $\gamma J_2>0$) Heisenberg legs coupled by zigzag-shaped ferromagnetic (FM; $J_1<0$) inter-leg interaction. On the basis of density-matrix renormalization group based calculations the ground-state phase diagram is obtained as functions of $\gamma$ and $J_2/|J_1|$. It contains four kinds of frustration-induced ordered phases except a trivial FM phase. Two of the ordered phases are valence bond solid (VBS) with spin-singlet dimerization, which is a rather conventional order by disorder. Still, it is interesting to note that the VBS states possess an Affleck-Kennedy-Lieb-Tasaki-type topological hidden order. The remaining two phases are ferrimagnetic orders, each of which is distinguished by commensurate or incommensurate spin-spin correlation. It is striking that the ferrimagnetic orders are not associated with geometrical symmetry breaking; instead, the global spin-rotation symmetry is broken. In other words, the system lowers its energy via the FM inter-leg interaction by polarizing both of the AFM Heisenberg legs. This is a rare type of order by disorder. Besides, the incommensurate ferrimagnetic state appears as a consequence of the competition between a polarization and a critical Tomonaga-Luttinger-liquid behavior in the AFM Heisenberg legs.
\end{abstract}

\maketitle

\section{Introduction}
Low-dimensional frustrated quantum magnets, in which a macroscopic number of quasi-degenerate states compete with each other, provide an ideal playground for the emergence of exotic phenomena~\cite{Vasiliev18}. For instance, the interplay of frustration and fluctuations could lead to unexpected condensed matter orders at low temperatures by spontaneously breaking some sort of symmetry – {\it order by disorder} –~\cite{Villain80}. Long-range ordered (LRO) magnetic state with breaking a spatial symmetry as well as valence bond solid (VBS) state with a formation of disentangled-unit--like local spin-singlet pair are typical examples of order by disorder. Moreover, when quantum fluctuations between the quasi-degenerate states prevent a selection of particular order, one ends up with spin liquids. Modern theories have brought us new insight by identifying spin liquids as topological phases of matter~\cite{Oshikawa06,Jiang12}. In recent years the realization of topological phases in frustrated spin systems has been one of the central topics in condensed matter physics~\cite{Messio12,Bauer14,Lao18}.

In one-dimensional (1D) and spin-$\frac{1}{2}$ case, quantum fluctuations are maximized so that we may place more expectations on the discovery of novel ground states by the cooperative effects with magnetic frustration. A most simply-structured 1D frustrated system is the so-called spin-$\frac{1}{2}$ $J_1$-$J_2$ chain consisting of nearest-neighbor $J_1$ and next-nearest-neighbor $J_2$ couplings. When both $J_1$ and $J_2$ are antiferromagnetic (AFM), the ground state is a VBS, the nature of which can be grasped by the Majumdar-Ghosh (MG) model~\cite{Majumdar69}, at $J_2/J_1 \gtrsim 0.24$~\cite{Okamoto92,Eggert96}. The idea of MG model was generalized to the Affleck-Kennedy-Lieb-Tasaki (AKLT) model~\cite{Affleck87} exhibiting spin-$1$ VBS ground state with a symmetry protected topological order~\cite{Pollmann12}.

Meanwhile, the $J_1$-$J_2$ chain with ferromagnetic (FM) $J_1$ and AFM $J_2$, which is known as a standard magnetic model for quasi-1D edge-shared cuprates such as Li$_2$CuO$_2$~\cite{deGraaf2002}, LiCuSbO$_4$~\cite{Grafe17}, LiCuVO$_4$ \cite{Orlova17}, Li$_2$ZrCuO$_2$~\cite{Drechsler07}, Rb$_2$Cu$_2$Mo$_3$O$_{12}$~\cite{Ueda18}, and PbCuSO$_4$(OH)$_2$~\cite{Wolter12}, encloses wider array of states of matter. Theoretically, this model has been extensively studied: Other than a trivial FM state in the dominant $J_1$ region, the ground state is a topological VBS accompanied by spontaneous multiple dimerization orders~\cite{furukawa12,clio19} (More details are described in Sec.~\ref{J1J2GS} of this article). It is also intriguing that a vector chirality and multimagnon bound states are induced in the presence of magnetic field~\cite{Kecke2007,Sudan2009,Hikihara08,Sato09}. Especially, the detection of nematic or higher multipolar phases is one of the most exciting experimental current issues~\cite{Buttgen14,Nawa13,Nawa14,Pregelj15,Grafe17,Heinze19}. Sensitive features to even tiny interchain couplings are another characteristic of this system~\cite{Ueda09,Li2CuO2,J1J2IC}.

\begin{figure}[tbh]
\centering
\includegraphics[width=0.9\linewidth]{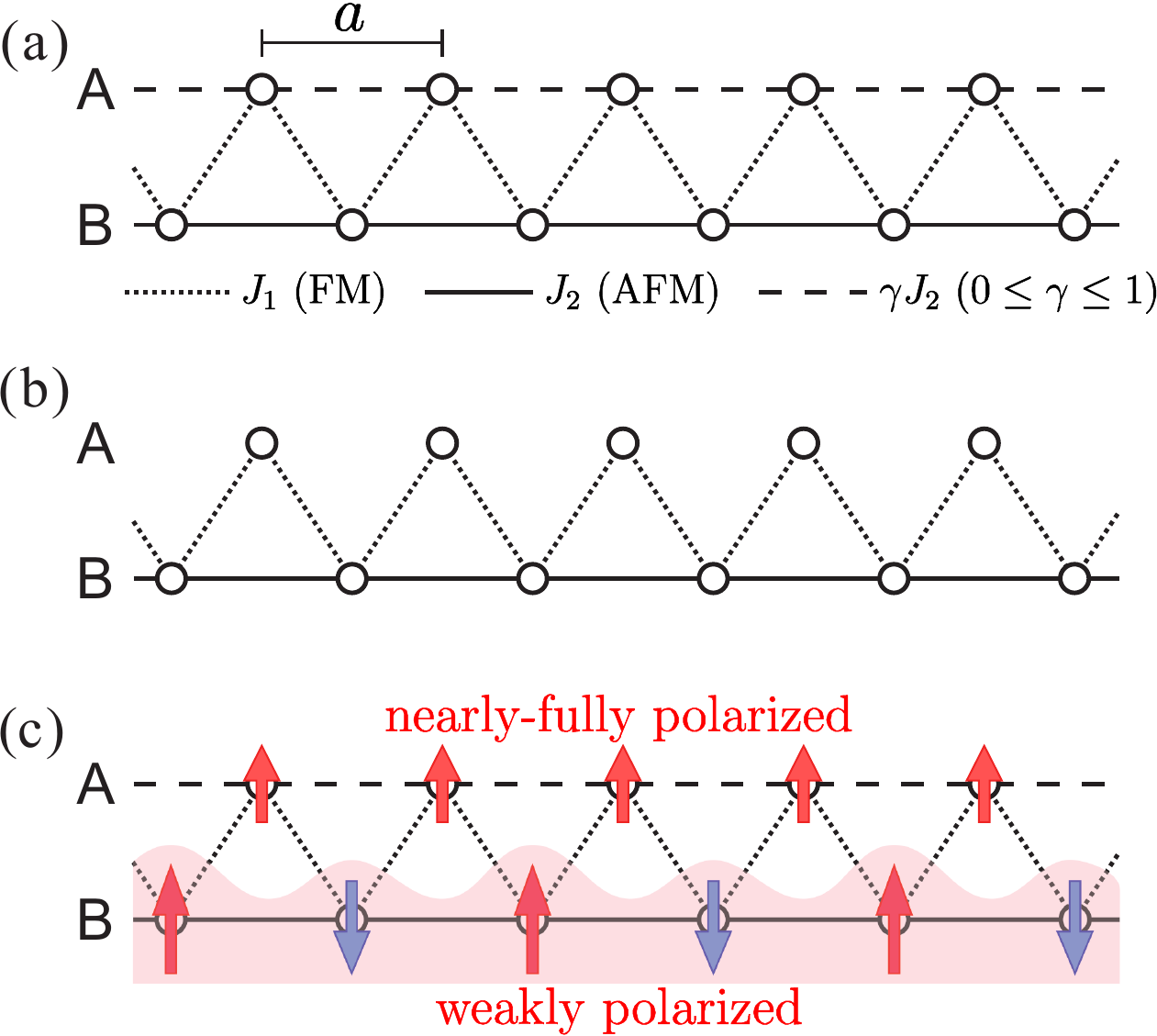}
\caption{
(a) Lattice structure of the asymmetric $J_1$-$J_2$ zigzag ladder. The indices `{\bf A}' and `{\bf B}' denote apical and basal chains, respectively. The lattice spacing $a$ is set as a distance between neighboring sites along the chains. The AFM interaction in the apical chain is controlled by $\gamma$. (b) Lattice structure of the so-called delta chain (or sawtooth chain) which is realized in the limit of $\gamma=0$. (c) Schematic representation of the ferrimagnetic state with global spin-rotation-symmetry breaking.
}
\label{lattice}
\end{figure}

Another typical example of frustrated 1D system is the delta chain (or sawtooth chain). The lattice structure is a series of triangles, as shown in Fig.~\ref{lattice}(b), which is similar to that of the $J_1$-$J_2$ chain but certain parts of $J_2$ bonds are missing. There have been several candidates for delta-chain and related materials:
YCuO$_{0.25}$~\cite{Sen96,LeBacq05},
[Cu({\it bpy})(H$_2$O)][Cu({\it bpy})({\it mal})–(H$_2$O)](ClO$_4$)$_2$~\cite{Ruiz-Perez00},
Zn{\it L}$_2$S$_4$ ({\it L} = Er, Tm, and Yb)~\cite{Lau06},
Cu(AsO$_4$)(OH)$\cdot$3H$_2$O~\cite{Kikuchi11},
Mn$_2$GeO$_4$~\cite{White12},
Rb$_2$Fe$_2$O(AsO$_4$)$_2$~\cite{Garlea14},
CuFe$_2$Ge$_2$~\cite{May16},
Fe$_{10}$Gd$_{10}$~\cite{Baniodeh18},
Cu$_2$Cl(OH)$_3$~\cite{Heinze19-2},
Fe$_2$O(SeO$_3$)$_2$~\cite{Gnezdilov19},
and V$_6$O$_{13}$~\cite{Toriyama14}.
In these materials, a wide array of complex phases has been experimentally observed. Delta-chain systems may offer an outlook towards promising prospects on novel magnetic phenomena.

The magnetic properties of delta-chain systems are totally different in different signs of $J_1$ and $J_2$. For the case of $J_1<0$ and $J_2>0$ (typically, referred as FM-AFM delta chain), only two corresponding materials have been recognized. One of them is malonatobridged copper complexes [Cu({\it bpy})(H$_2$O)][Cu({\it bpy})({\it mal})–(H$_2$O)](ClO$_4$)$_2$~\cite{Ruiz-Perez00}. The magnetic Cu$^{2+}$ ions with effective spin-$\frac{1}{2}$ form into a delta-chain network. The base and the other exchange couplings in a triangle made of malonate were estimated, respectively, as AFM ($J_2=6.0$K) and FM ($J_1=-6.6$K) from the analysis of magnetic susceptibility $\chi(T)$; and, as AFM ($J_2=10.9$K) and FM ($J_1=-12.0$K) from the fitting of magnetization curve (typically, referred as FM-AFM delta chain). In either case, the ratio of AFM and FM couplings is close to $1$. This means that the material would be in the region of strong frustration. Theoretically, the ground state was predicted to be a ferrimagnetic state but the detailed spin structures are less understood~\cite{tonegawa04}. In fact, only qualitative behavior of measured magnetization curve could be explained by assuming a ferrimagnetic ground state~\cite{Inagaki05,Kaburagi05}. A deeper understanding of the ferrimagnetic state is necessary to resolve the remaining discrepancy between experiment and theory.

The second candidate of the FM-AFM delta-chain materials is a mixed 3{\it d}$/4${\it f} cyclic coordination cluster system Fe$_{10}$Gd$_{10}$~\cite{Baniodeh18}. 
In these days, the delta-chain physics is increasingly attracting attention due to the synthesis of Fe$_{10}$Gd$_{10}$. This cluster consists of $10+10$ alternating Gd and Fe ions. The exchange couplings were estimated as FM ($J_1=-1.0$~K) between Fe and Gd ions, AFM ($J_2=0.65$~K) between Fe ions, and nearly zero between Gd ions; the magnetic ions form an FM-AFM delta chain short ring. The parameter ratio $J_2/|J_1|=0.65$ seems to be very close the FM quantum critical point $J_2/|J_1|=0.7$~\cite{krivnov14} Although the spin values of Fe and Gd ions are higher than $S=1/2$ ($S=5/2$ and $S=7/2$, respectively), quantum fluctuations would play important roles to determine the ground state because of the quantum criticality~\cite{Sachdev08}.
This means that the magnetic properties can be drastically changed upon even a small variation of external influences such as magnetic field, pressure, chemical means, and gating current. 
So, this delta chain material is drawing attention also from the perspective of controlling magnetic states in molecular spintronics~\cite{Roch08}.

For comparison, a few examples of delta-chain materials only with AFM interactions ($J_1, J_2>0$) have been also reported. With the help of MG-like projection method, the magnetic properties of the AFM-AFM delta-chain are better understood than those of the FM-AFM one~\cite{Monti91,Sen96,Nakamura96,Blundell03}. A peculiarly interesting feature is the dispersionless kink-antikink domain wall excitations to the dimerized VBS ground state. A kink is highly localized only in the range of one triangle. The first candidate of AFM-AFM delta-chain materials was the delafossite YCuO$_{2.5}$~\cite{Sen96}. However, a first-principle calculation revealed that the ratio of $J_2/J_1$ in YCuO$_{2.5}$ is out of the range of the dimerized VBS ground state and additional intrachain FM interaction is significantly large~\cite{LeBacq05}. Very recently, the other candidate materials Cu$_2$Cl(OH)$_3$ ($S=1/2$)~\cite{Heinze19-2} and Fe$_2$O(SeO$_3$)$_2$ ($S=5/2$) have been reported. They indeed exhibit characteristic features of AFM-AFM delta chain: a magnetization plateau at half-saturation~\cite{Inagaki05, Richter08} in Cu$_2$Cl(OH)$_3$ and an almost flat-band one-magnon excitation spectrum in Fe$_2$O(SeO$_3$)$_2$.

As mentioned above, the research of frustrated 1D systems with $J_1$-$J_2$ or delta-chain structures has become more and more active. Interestingly, each of the $J_1$-$J_2$ chain and the delta chain is expressed as a limiting case of an asymmetric $J_1$-$J_2$ zigzag ladder, defined as two different AFM Heisenberg chains coupled by zigzag-shaped interchain FM interaction [see Fig.~\ref{lattice}(a)]. When one of the Heisenberg chains vanishes, it is the delta chain; and, when the Heisenberg chains are equivalent, it is the $J_1$-$J_2$ chain. However, it is known that their ground states are completely different. Then, one may simply question how the two limiting cases are connected. Of particular interest is that the effect of exchange coupling towards the $J_1$-$J_2$ chain can be a likely perturbation in real delta-chain compounds, e.g., the effect of tiny coupling between Gd ions in Fe$_{10}$Gd$_{10}$.

In this paper, we therefore study an asymmetric FM-AFM $J_1$-$J_2$ zigzag ladder using the density-matrix renormalization group (DMRG) technique. We first clarify the detailed spin structure and low-energy excitations of ferrimagnetic state in the delta chain limit. We suggest that the ferrimagnetic state is a rare type of order by disorder, where the energy is lowered by FM fluctuation between two polarized AFM Heisenberg chains with spontaneous breaking of the global spin-rotation symmetry. Then, we examine how the ferrimagnetic state is collapsed and connected to the well-known incommensurate spiral state in the $J_1$-$J_2$ chain. We also find there exist two kinds of VBS phases in the spiral region. Finally, we obtain the ground-state phase diagram of asymmetric $J_1$-$J_2$ zigzag ladder with interpolating between the delta chain and $J_1$-$J_2$ chain.

The paper is organized as follows: In Sec. II our spin model is explained and the applied numerical methods are described. In Sec. III we briefly mention to-date knowledge on the ground state for two limiting cases of our spin model. In Sec. IV we present our numerical results and discuss how the two limiting cases are connected. Finally we end the paper with a summary in Sec. V.

\section{Model and Method}

\subsection{Model}

The asymmetric $J_1$-$J_2$ zigzag ladder is defined as two Heisenberg chains coupled by zigzag-shaped interchain interaction. The lattice structure is sketched in Fig.~\ref{lattice}(a). We call a leg chain with larger interaction ``basal chain'' and the other with smaller interaction ``apical chain''. The Hamiltonian is written as
\begin{align}
	\nonumber
	H = &J_1 \sum_{i} \bm{S}_{{\rm A},i} \cdot (\bm{S}_{{\rm B},i} + \bm{S}_{{\rm B},{i+1}})\\
	+ &J_2 \sum_{i} (\bm{S}_{{\rm B},i} \cdot \bm{S}_{{\rm B}, i+1}+\gamma\bm{S}_{{\rm A},i} \cdot \bm{S}_{{\rm A},i+1}),
	\label{ham}
\end{align}
where $\mathbf{S}_{{\rm B},i}$ is spin-$\frac{1}{2}$ operator at site $i$ on the basal chain and $\mathbf{S}_{{\rm A},i}$ is that on the apical chain. We focus on the case of FM interchain coupling ($J_1<0$) and AFM intrachain coupling ($J_2>0$). The intrachain interaction of apical chain is controlled by $\gamma$ ($0 \le \gamma \le 1$). The system \eqref{ham} corresponds to the so-called delta chain (or sawtooth chain) at $\gamma=0$ [Fig.~\ref{lattice}(b)] and the so-called $J_1$-$J_2$ chain at $\gamma=1$. The existing knowledge on the ground-state properties of these chains is briefly summarized in the next section. In our numerical calculations the chain lengths of basal and apical chains are denoted as $L_{\rm B}$ and $L_{\rm A}$, respectively. The total number of sites is $L=L_{\rm B}+L_{\rm A}$. In this paper, we call the system for $0 < \gamma < 1$ ``asymmetric $J_1$-$J_2$ zigzag ladder'', which has been little or not studied. In the case of $J_1>0$ and $J_2>0$, there are a few studies~\cite{Chen01,Chen03}.

\subsection{DMRG methods}

In order to examine the ground state and low-energy excitations of asymmetric $J_1$-$J_2$ zigzag ladder, we employ the DMRG techniques; namely, conventional DMRG (hereafter referred to simply as DMRG), dynamical DMRG (DDMRG), and matrix-product-state-based infinite DMRG (iDMRG) methods. They are used in a complementary fashion to further confirm our numerical results.

The DMRG method is a very powerful numerical method for various (quasi-)1D quantum systems~\cite{White92}. However, some difficulties are often involved in the DMRG analysis for strongly frustrated systems like Eq.~\eqref{ham}. First, the system size dependence of physical quantities is usually not straightforward. Therefore, relatively many data points are required to perform a reasonable finite-size scaling analysis. We thus study systems with length up to $L=161$ ($L_{\rm B}=81, L_{\rm A}=80$) under open boundary conditions (OBC) and systems with length up to $L=64$ ($L_{\rm B}=32, L_{\rm A}=32$) under periodic boundary conditions (PBC). Either OBC or PBC is chosen depending on the calculated quantity. Second, a lot of nearly-degenerate states are present around the ground state. To obtain results accurate enough, a relatively large number of density-matrix eigenstates $m$ must be kept in the renormalization procedure. In this paper, we keep up to $m=8000$ density-matrix eigenstates, which is much larger than that kept in usual DMRG calculations for 1D systems, and extrapolate the calculated quantities to the limit $m \to \infty$ if necessary. In this way, we can obtain quite accurate ground states within the error of $\Delta E/L=10^{-8}|J_1|$.

For the calculation of dynamical quantities, we use the DDMRG method which has been developed for calculating dynamical correlation functions at zero temperature in quantum lattice models~\cite{Jeckelmann02}. Since the DDMRG algorithm performs best for OBC, we study a open cluster with length up to $L=129$ ($L_{\rm B}=65, L_{\rm A}=64$). The DDMRG approach is based on a variational principle so that we have to prepare a `good trial function' of the ground state with the density-matrix eigenstates. Therefore, we keep $m=1200$ to obtain the ground state in the first ten DMRG sweeps and keep $m=600$ to calculate the excitation spectrum. In this way, the maximum truncation error, i.e., the discarded weight, is about $1 \times10^{-5}$, while the maximum error in the ground-state and low-lying excited states energies is about $10^{-4}|J_1|$.

The iDMRG method is very useful because it enables us to obtain the physical quantities directly in the thermodynamic limit~\cite{mcculloch08,Schollwock11}, if the matrix product state is not too complicated and the simulation can be performed accurately enough. In our iDMRG calculations, typical truncation errors are $10^{-8}$ using bond dimensions $\chi$ up to $6000$. In this way, the effective correlation length near criticality is less or at most equal to $500$, so that most of interesting parameter region of the system \eqref{ham} can be reasonably examined by our iDMRG simulations.

\section{Previous studies for limiting cases}

So far, our system for two limiting cases, namely, $J_1$-$J_2$ chain ($\gamma=1$) and delta chain ($\gamma=0$), has been extensively studied. In this section, we briefly summarize to-date knowledge on the ground state for the limiting cases.

\subsection{$J_1$-$J_2$ chain ($\gamma=1$)}\label{J1J2GS}

At $\gamma=1$, we are dealing with the $J_1$-$J_2$ chain, which may be also recognized as {\it symmetric} $J_1$-$J_2$ zigzag ladder. In the limit of $J_2/|J_1|=0$, the system is a simple FM Heisenberg chain with FM ordered ground state. Increasing $J_2/|J_1|$, the FM state persists up to $J_2/|J_1|=1/4$; then, a first-order phase transition from the FM to an incommensurate (``spiral'') state occurs~\cite{bursill95}. The total spin in the incommensurate phase is zero ($S_{\rm tot}=0$). Since quantum fluctuations disappear at the FM critical point, the critical value $J_2/|J_1|=1/4$ can be recognized similarly both in the quantum as well as in the classical model~\cite{bader79,hartel08}.

At $J_2/|J_1|>1/4$, the incommensurate correlations are short ranged in the quantum model~\cite{nersesyan98,sirker11}. Instead, the system exhibits a spontaneous nearest-neighbor FM dimerization with breaking of translation symmetry, as a consequence of the quantum fluctuations typical of magnetic frustration, i.e, order by disorder. By regarding the ferromagnetically dimerized spin-1/2 pair as a spin-$1$ site, the system is effectively mapped onto a spin-$1$ Heisenberg chain and an Affleck-Kennedy-Lieb-Tasaki (AKLT)-like hidden topological order protected by global $Z_2 \times Z_2$ symmetry is naively expected as a Haldane state~\cite{Affleck87,oshikawa92} (also see Sec.~\ref{stop}). In fact, the hidden order has been numerically confirmed~\cite{furukawa12,clio19}.

Furthermore, the existence of (exponentially small) singlet-triplet gap at $J_2/|J_1| \gtrsim 3.3$ was predicted by the field-theory analysis~\cite{itoi01} and its verification had been a longstanding issue. Only recently, the gap was numerically estimated: with increasing $J_2/|J_1|$ it starts to open at $J_2/|J_1|=1/4$, reaches its maximum $\sim 0.007|J_1|$ around $J_2/|J_1|=0.65$, and exponentially decreases~\cite{clio19}. The ground state is a kind of VBS state with spin-singlet formations between third-neighbor sites. Therefore, the magnitude of gap basically scales to the strength of third-neighbor valence bond.

\subsection{Delta chain ($\gamma=0$)}

At $\gamma=0$, the system is the delta chain consisting of a linear chain of corner-sharing triangles [Fig.~\ref{lattice}(b)]. The ground state properties are less understood than those of the $J_1$-$J_2$ chain. One main reason is that numerical investigation of the delta chain is particularly difficult due to the strong magnetic frustration and a number of nearly-degenerate states near the ground state. Especially at the FM critical point is macroscopically degenerate and consists of multi-magnon configurations formed by independent localized magnons and the special localized multi-magnon complexes~\cite{krivnov14}.

Nonetheless, a numerical study could identify the ground state at $J_2/|J_1|<1/2$ to be FM; that at $J_2/|J_1|>1/2$ to be ferrimagnetic~\cite{tonegawa04}. The total spins of the ferromagnetic and ferrimagnetic phases are $L/2$ and $L/4$, respectively. To understand the origin and the properties of this ferrimagnetic state, the delta chain in the large limit of easy-axis exchange anisotropy was studied~\cite{Dmitriev16}. In this limit the system can be reduced to a 1D XXZ basal chain under a static magnetic field depending on the magnetic structure of apical chain. The ground state was identified as ferrimagnetic with fully polarized apical spins and weakly polarized basal spins. It is expected that some essential features may be inherent in the isotropic SU(2) limit. In fact, the spin structure agrees qualitatively to the ferrimagnetic state determined in this paper [see Fig.~\ref{lattice}(c)].

\section{Results}

\begin{figure}[t]
\centering
\includegraphics[width=0.8\linewidth]{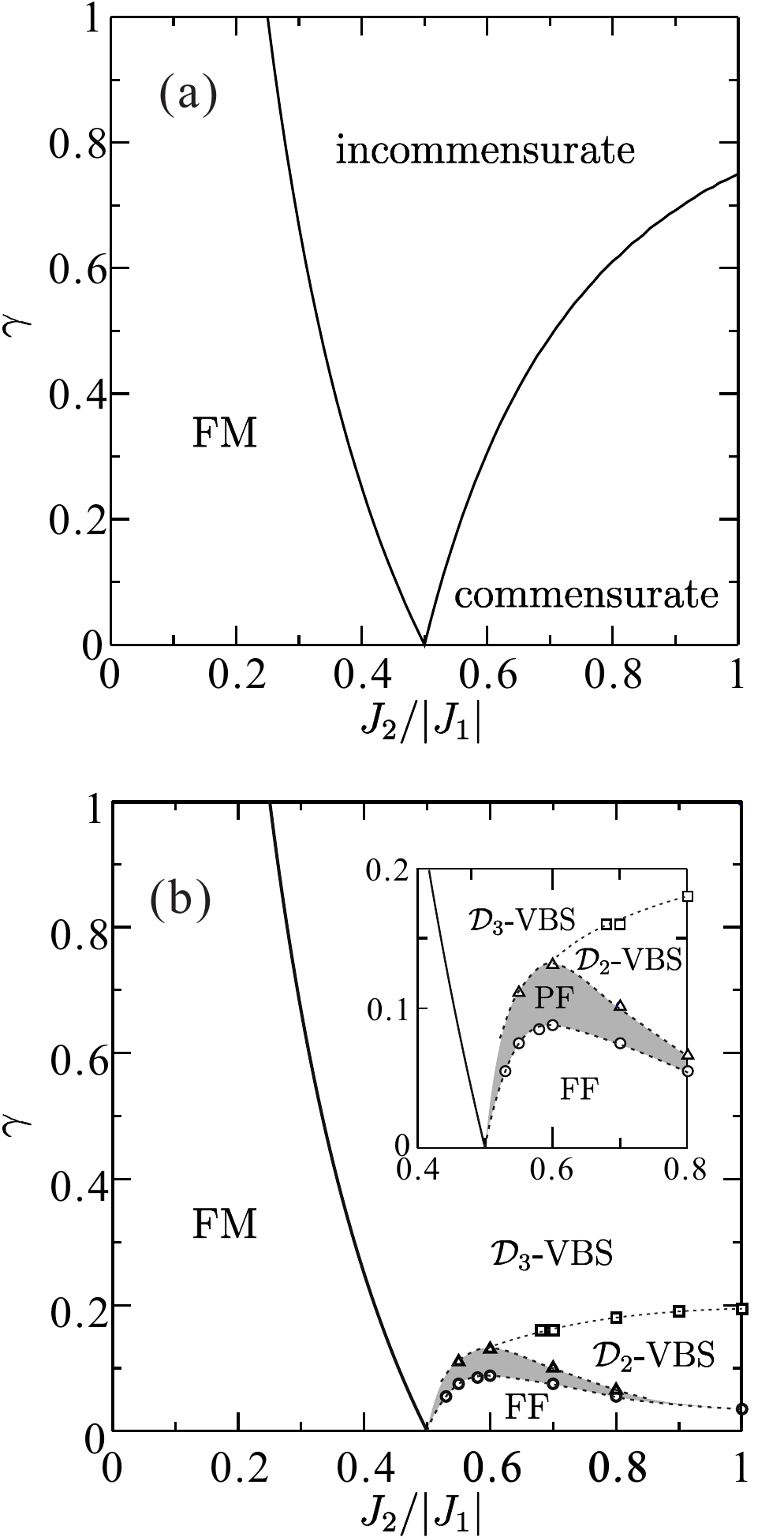}
\caption{
(a) Classical ground-state phase diagram of the asymmetric $J_1$-$J_2$ zigzag ladder. The phases are characterized by propagation vector $q=q_{\rm min}$ minimizing $J_q$ [Eq.~\eqref{SWT_Jq}]: FM ($q_{\rm min}=0$); commensurate ($q_{\rm min}=\pi$); incommensurate ($0<q_{\rm min}<\pi$). (b) Ground-state phase diagram of the asymmetric $J_1$-$J_2$ zigzag chain [Eq.~\eqref{ham}] determined by DMRG calculations. Inset: enlarged view around the PF phase.
}
\label{fig_pd}
\end{figure}

\subsection{classical limit}

As mentioned above, the FM critical point is known to be $J_2/|J_1|=1/4$ for the $J_1$-$J_2$ chain ($\gamma=1$) and $J_2/|J_1|=1/2$ for the delta chain ($\gamma=0$). To be examined first is how the critical point changes between the limiting cases, i.e., $0<\gamma<1$. Since the quantum fluctuations vanish at the FM critical point, the critical value can be exactly estimated by the classical spin wave theory (SWT). The Fourier transform of our Hamiltonian \eqref{ham} reads
\begin{align}
	H = \frac{1}{2}\sum_q J_q \vec{S}_q \cdot \vec{S}_{-q}
	\label{hamq}
\end{align}
with
\begin{align}
	\nonumber
	J_q=&(1+\gamma)J_2\cos q\\
	&\pm \sqrt{(1-\gamma)^2J_2^2(1-\cos q)^2+4J_1^2\cos^2(q/2)}.
	\label{SWT_Jq}
\end{align}
If Eq.~\eqref{SWT_Jq} has a minimum at $q=0$, the system is in an FM ground state. The FM critical point is thus derived as
\begin{align}
	\frac{J_{2,{\rm c}}}{|J_1|}=\frac{1}{2(1+\gamma)}.
	\label{FM_QCP}
\end{align}
As shown in Fig.~\ref{fig_pd}(a), the FM region is simply shrunk with increasing $\gamma$ because the AFM interaction is increased in the apical chain. We have confirmed this FM critical boundary numerically by calculating the total spin $S_{\rm tot}$ of the whole system, which is defined as
\begin{align}
	\langle \vec{S}^2 \rangle = S_{\rm tot}(S_{\rm tot}+1) = \sum_{i,j} \langle \vec{S}_i \cdot \vec{S}_j \rangle.
	\label{totalS}
\end{align}
It can be also verified by finding the absence of LRO FM state in the spin-spin correlation functions. These results are shown in Appendix A.

By evaluating $q$ ($\equiv q_{\rm min}$) value to minimize Eq.~\eqref{SWT_Jq}, a classical ground-state phase diagram is obtained as Fig.~\ref{fig_pd}(a). There are three kinds of LRO phases: FM phase with $q_{\rm min}=0$, incommensurate phase $0<q_{\rm min}<\pi$, and commensurate phase with $q_{\rm min}=\pi$. Since the ferrimagnetic state in Fig.~\ref{lattice}(c) is of commensurate with $q=\pi$ and the propagation number of the $J_1$-$J_2$ chain is incommensurate, the SWT results are consistent with those of the quantum system \eqref{ham} in the two limiting cases $\gamma=0$ and $\gamma=1$. Therefore, even in the quantum system an incommensurate-commensurate phase transition is naively expected at finite $\gamma$ with $J_2/|J_1|$ fixed.

\subsection{$\gamma=0$: delta chain}

Although the ground state of the delta chain is most probably ferrimagnetic at $J_2/|J_1|>1/2$, the detailed magnetic structure and properties have not been fully settled. To gain further insight into them, we here calculate the total spin, spin-spin correlation functions, and stabilization energy of ferrimagnetic state for the delta chain. We need to pick through the system-size dependence of the quantities to deal with nontrivial finite-size effects of the delta chain.

\subsubsection{total spin}

\begin{figure}[t]
\centering
\includegraphics[width=0.7\linewidth]{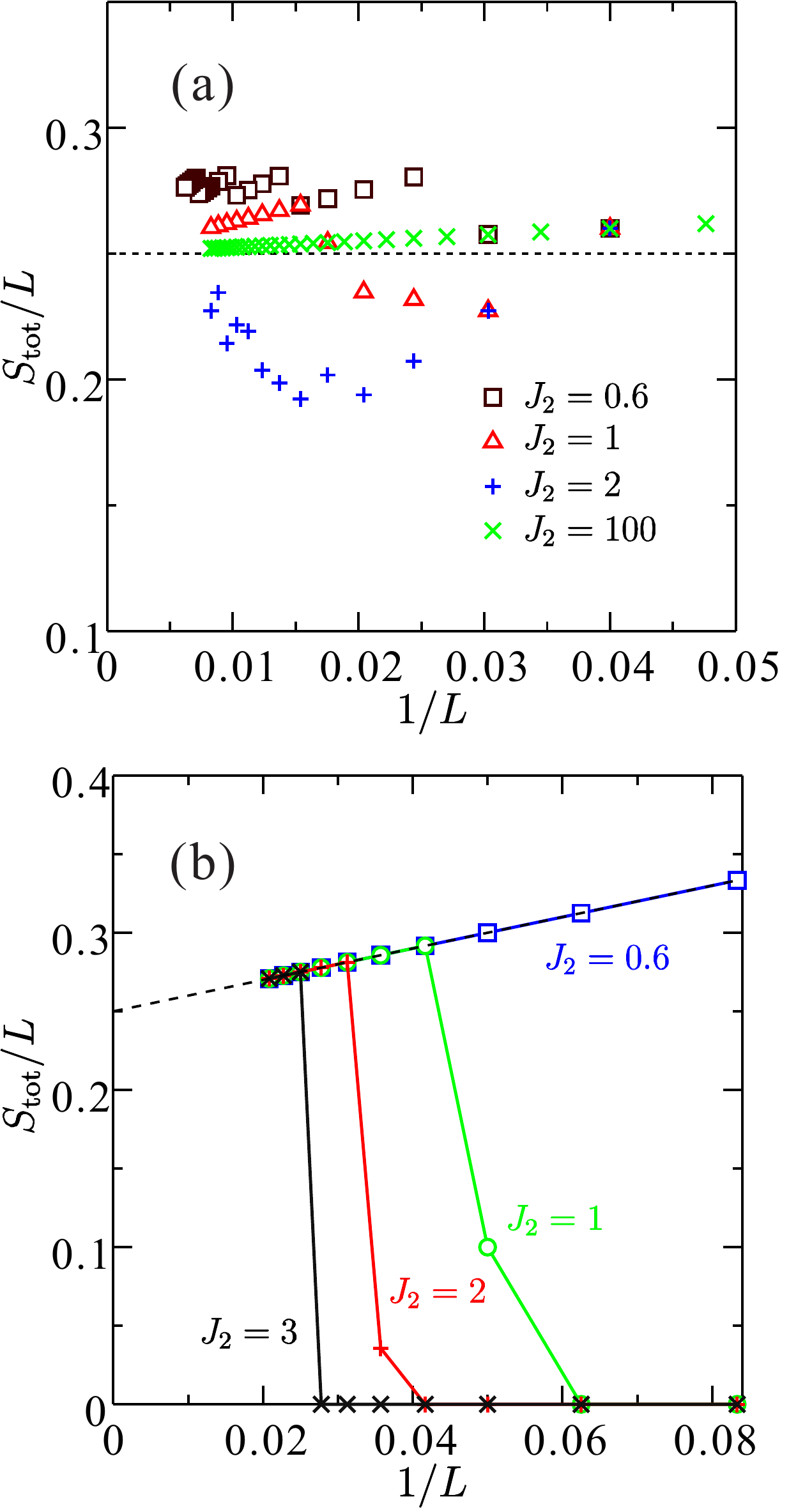}
\caption{
Finite-size scaling analysis of the total spin per site for the delta chain ($\gamma=0$), where (a) OBC and (b) PBC are applied. The dotted lines are guide for eyes.
}
\label{S_delta_chain}
\end{figure}

Investigating the total spin $S_{\rm tot}$ is a simple way to explore the possibility of ferrimagnetic state. As shown below, the system-size dependence of $S_{\rm tot}$ is significantly different between applying OBC and PBC. This obviously implies the difficulty of performing numerical calculations for this system. Nevertheless, the fact that they should coincide in the thermodynamic limit $L\to\infty$ makes the finite-scaling analysis even more reliable. We here use the DMRG method.

Let us first see the case of OBC. In Fig.~\ref{S_delta_chain}(a) the total spin per site $S_{\rm tot}/L$ is plotted as a function of inverse system size $1/L$ for several values of $J_2/|J_1|$. When $J_2/|J_1|$ is order of $1$, the effect of strong frustration is explicitly embedded in the finite-size scaling; namely, due to the Friedel oscillation, $S_{\rm tot}/L$ awkwardly oscillates with $1/L$. However, as expected, such an oscillation no longer appears in a case of large $J_2/|J_1|=100$. When $J_2/|J_1| \gg 1$, a straightforward scaling is allowed since the frustration is much weaker. Eventually, for all the $J_2/|J_1|$ values, $S_{\rm tot}/L$ seems to be extrapolated to $1/4$ in the thermodynamic limit. Although one may think that the PBC should be applied if the Friedel oscillation is problematic, the situation is not so simple as explained below.

We then turn to the case of PBC. In Fig.~\ref{S_delta_chain}(b) $S_{\rm tot}/L$ is plotted as a function of $1/L$ for several values of $J_2/|J_1|$. Unlike in the case of OBC, the oscillation is not seen in $S_{\rm tot}/L$ vs. $1/L$. Instead, there exists a `critical' system size to achieve finite $S_{\rm tot}/L$ in the ground state. This is caused by a kind of typical finite-size effects: Under the PBC, the basal chain forms a plaquette singlet and the basal spins are more or less screened. This screening leads to the reduction of FM interaction between the basal and apical chains although the FM fluctuations between the two chains are essential to stabilize the ferrimagnetic state (see Sec.~\ref{SS_delta}). Besides, only a small screening may be sufficient to collapse the ferrimagnetic state because the stabilization gap of ferrimagnetic state is very small (see Sec.~\ref{gap_delta}). Consequently, the ferrimagnetic state can be readily prevented under the PBC. For reference, the system-size dependence of energies for spin-singlet and ferrimagnetic states is shown in Appendix B. Since the triplet excitation gap of plaquette singlet roughly scales to $J_2/|J_1|$ with a fixed system size, the critical system size is larger for larger $J_2/|J_1|$ as seen in Fig.~\ref{S_delta_chain}(b). Once the system size goes beyond the critical one, $S_{\rm tot}/L$ approaches smoothly to $1/4$ at $L\to\infty$. The fitting function is $S_{\rm tot}/L=1/4+1/L$ for any $J_2/|J_1|$.

Thus, we have confirmed that the total spin of the delta chain at $J_2/|J_1|>1/2$ is $S_{\rm tot}=L/4$, indicating the ferrimagnetic state, and the ferrimagnetic phase persists up to the large $J_2/|J_1|$ limit.

\subsubsection{spin-spin correlation}\label{SS_delta}

\begin{figure}[t]
\centering
\includegraphics[width=1.0\linewidth]{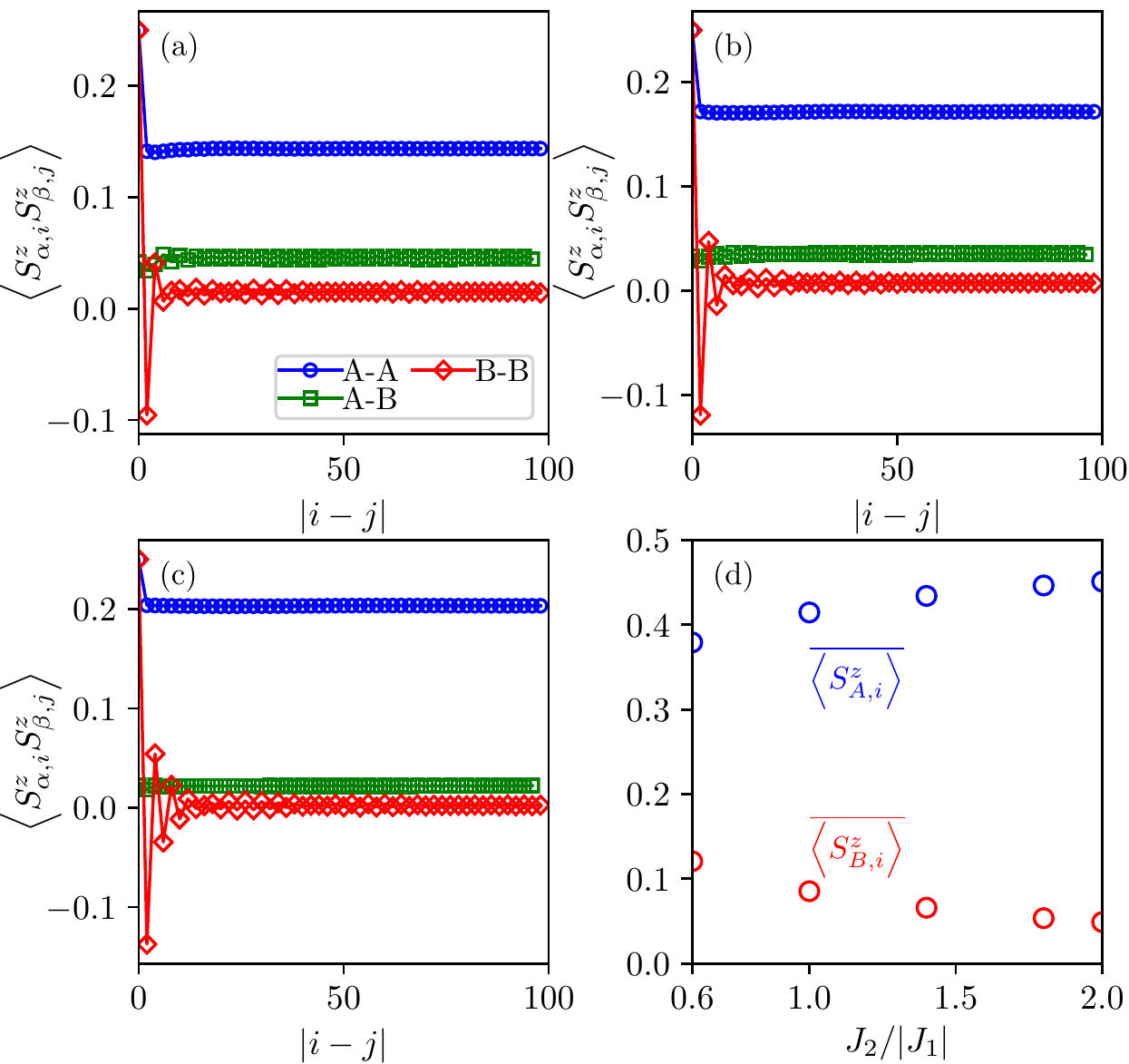}
\caption{
Spin-spin correlation function $\langle S^z_{\alpha,i} S^z_{\beta,j} \rangle$ for the delta chain ($\gamma=0$) as a function of distance $|i-j|$ at (a) $J_2=0.6$, (b) $1$, and (c) $2$. The total $S^z$ sector is set to be $S^z_{\rm tot}=L/4$. The legends denote as A-A: $(\alpha,\beta)=({\rm A},{\rm A})$, B-B: $(\alpha,\beta)=({\rm B},{\rm B})$, and A-B: $(\alpha,\beta)=({\rm A},{\rm B})$. (d) Averaged values of $\langle S^z_{\alpha,i} \rangle$ on the apical and basal sites as a function of $J_2/|J_1|$.
}
\label{corr_delta}
\end{figure}

Then, in order to determine the magnetic structure of the ferrimagnetic state, we examine spin-spin correlation functions $\langle \bm{S}_{\alpha,i} \cdot \bm{S}_{\beta, j} \rangle$ between apical sites: $(\alpha,\beta)=({\rm A},{\rm A})$, between basal sites: $(\alpha,\beta)=({\rm B},{\rm B})$, and between apical and basal sites: $(\alpha,\beta)=({\rm A},{\rm B})/({\rm B},{\rm A})$. For convenience, we fix the z-component of total spin at the total spin retained in the ferrimagnetic ground state, i.e., $S^z_{\rm tot}=S_{\rm tot}=L/4$. It lifts the ground-state degeneracy due to the SU(2) symmetry breaking and the (spontaneous) magnetization direction is restricted to the $z$-axis direction. Hence, we need only to see $\langle S^z_{\alpha,i} S^z_{\beta,j} \rangle$ instead of $\langle \bm{S}_{\alpha,i} \cdot \bm{S}_{\beta, j} \rangle$.

In Fig.~\ref{corr_delta}(a)-(c) iDMRG results of the correlation function $\langle S^z_{\alpha,i} S^z_{\beta,j} \rangle$ are plotted as a function of distance $|i-j|$ for $J_2/|J_1|=0.6$, $1$, and $2$. We clearly see long-ranged correlations indicating a magnetic order for all the $J_2/|J_1|$ values. If no magnetic order exists, all the correlation functions converge to $(S^z_{\rm tot}/L)^2=1/16$ at long distance limit $|i-j|\to\infty$. The intrachain correlations $\langle S^z_{A,i} S^z_{A,j} \rangle$ and $\langle S^z_{B,i} S^z_{B,j} \rangle$ are both positive at the long distance, and the former correlation is much larger than the latter one. This means that the apical spins are nearly-fully polarized and the basal spins are only weakly polarized. In addition, a N\'eel-like staggered oscillation is clearly seen leastwise at the short distance in $\langle S^z_{B,i} S^z_{B,j} \rangle$. These correlations immediately correspond to a ferrimagnetic state, which is schematically sketched in Fig.~\ref{lattice}(c). This picture seems to be valid in the whole region of $J_2/|J_1|>0.5$.

Now, another question may arise: Is the N\'eel-like staggered spin alignment on the basal chain LRO? The answer is NO. Although $\langle S^z_{B,i} S^z_{B,j} \rangle$ indeed exhibits an AFM oscillation at short distances $|i-j|$, it vanishes at the long distance limit. In fact, the oscillating part of $\langle S^z_{B,i} S^z_{B,j} \rangle$ exhibits a power-law decay indicating a critical behavior of the Tomonaga-Luttinger liquid (TLL) (see Appendix C). This also excludes the possibility of VBS formation in the basal chain because an exponential decay of the spin-spin correlation should be found in a VBS state. It means that, very surprisingly, the present ferrimagnetic order is an order by disorder but without any geometrical symmetry breaking. Instead, the magnetic frustration is relaxed by a spontaneous breaking of the global spin-rotation symmetry.  In other words, the system can gain energy from the FM interaction between the apical and basal chains by polarizing both of the chains. This is a very rare type of order by disorder.

This picture of order by disorder may be further convinced by looking at the relation between the polarization level of basal spins and the stabilization of ferrimagnetic order. Thus, to see the $J_2/|J_1|$-dependence of polarization level in more detail, we examine the expectation values of local spin moment $S^z_{A,i}$ and $S^z_{B,i}$. In Fig.~\ref{corr_delta}(d) the averaged values of $\langle S^z_{A,i} \rangle$ and $\langle S^z_{B,i} \rangle$ are plotted as a function of $J_2/|J_1|$. We find that with increasing $J_2/|J_1|$, $\overline{\langle S^z_{A,i} \rangle}$ increases and saturates at $1/2$; while, $\overline{\langle S^z_{B,i} \rangle}$ decreases and goes down to $0$ in the large $J_2/|J_1|$ limit. The decrease of $\overline{\langle S^z_{B,i} \rangle}$ may be naively expected by the fact that the basal chain approaches a 1D SU(2) Heisenberg model at large $J_2/|J_1|$; while, the increase of $\overline{\langle S^z_{B,i} \rangle}$ is a simple consequence of the condition $\overline{\langle S^z_{A,i} \rangle}+\overline{\langle S^z_{B,i} \rangle}=1/2$. If the order-by-disorder picture associated by interchain FM interaction is true, the ferrimagnetic state at larger $J_2/|J_1|$ should be fragile because of smaller polarization of basal spins. Actually, the stabilization energy of ferrimagnetic order decreases with increasing $J_2/|J_1|$ as described in the following subsection.

\subsubsection{stabilization gap}\label{gap_delta}

\begin{figure}[t]
\centering
\includegraphics[width=0.7\linewidth]{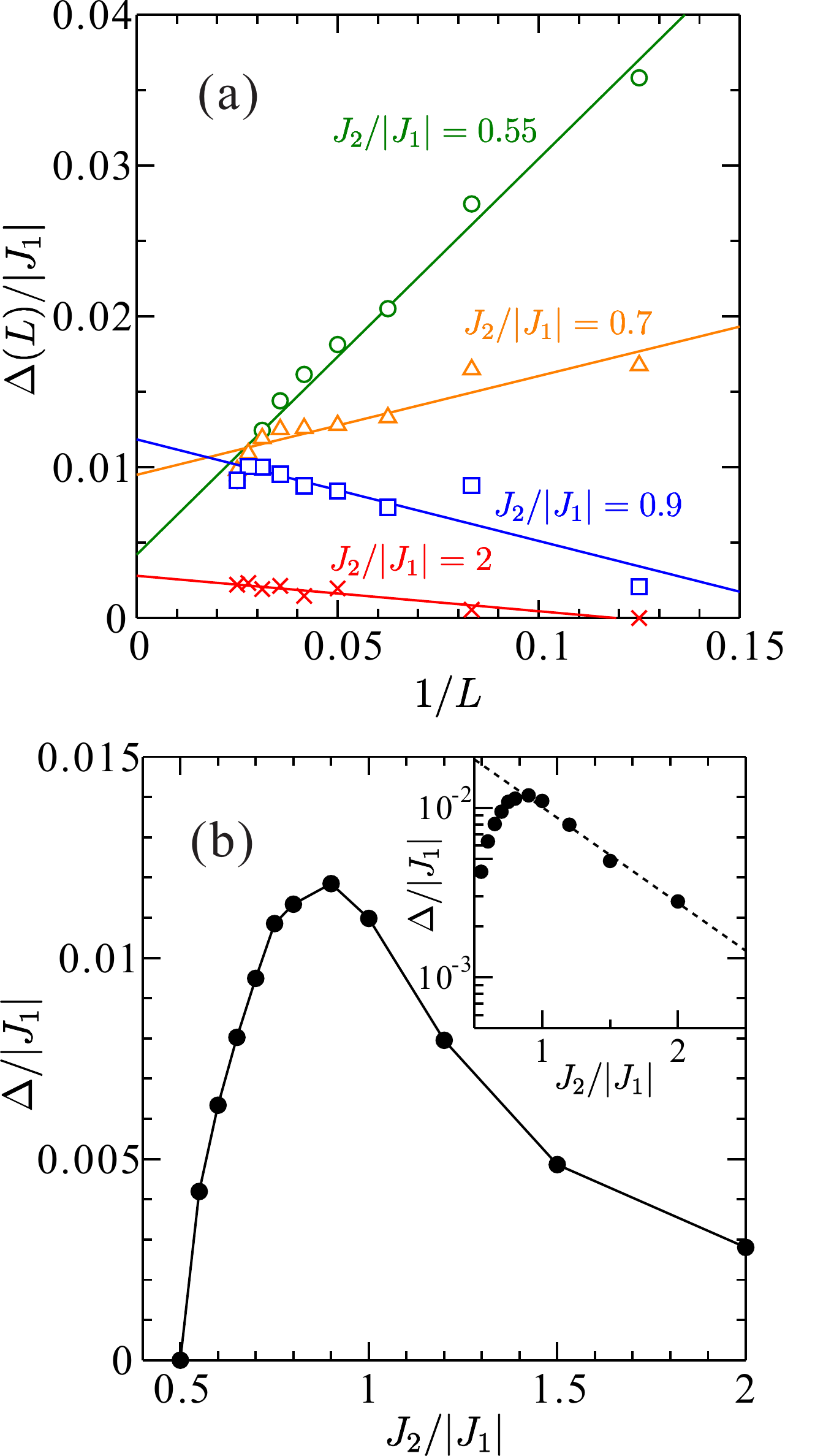}
\caption{
Energy difference between the lowest $S_{\rm tot}=0$ state and ferrimagnetic ground state for the delta chain, as a stabilization gap of the ferrimagnetic state. (a) Finite-size scaling and (b) the extrapolated values $\Delta/|J_1|$ as a function of $J_2/|J_1|$. Inset: Semi-log plot of $\Delta/|J_1|$ as a function of $J_2/|J_1|$. The dotted line is a fit for the large $J_2/|J_1|$ region: $\Delta/|J_1|=0.037\exp(-1.3J_2/|J_1|)$.
}
\label{gap_delta}
\end{figure}

To figure out the stability of the ferrimagnetic state, we calculate a stabilization gap defined by energy difference between the ferrimagnetic state and a lowest state in the $S_{\rm tot}=0$ sector:
\begin{align}
	\Delta(L)=E_0(0,L)-E_0(S_{\rm g.s.},L), \ \ \ \Delta=\lim_{L\to\infty}\Delta(L),
	\label{eq_gap}
\end{align}
where $E_0(S,L)$ is the lowest-state energy of $L$-site periodic system in the $S_{\rm tot}=S$ sector, and $S_{\rm g.s.}$ is a total spin of the ground state. The energy of ferrimagnetic state can be simply obtained as a ground-state energy, though, as mentioned above, the total spin $S_{\rm g.s.}/L$ in the ferrimagnetic state can deviate from $1/4$ for a finite cluster. Whereas, it is nontrivial to estimate the energy of a lowest state in the $S_{\rm tot}=0$ sector because it is an excited state and there are many quasi-degenerate states near the ferrimagnetic ground state. A proper way needs to be provided to extract the $S_{\rm tot}=0$ state. We therefore consider the following augmented Hamiltonian
\begin{align}
	H^\prime = H + \lambda \vec{S}^2,
	\label{ham_S0}
\end{align}
where $H$ is our original Hamiltonian~\eqref{ham} and the additional term corresponds to the total spin operator ($\lambda>0$). By setting $\lambda$ to be large enough, all the states with $S_{\rm tot}>0$ are lifted. Eventually, we can find a state with $S_{\rm tot}=0$ as the lowest state.

In Fig.~\ref{gap_delta}(a) the energy difference $\Delta(L)$ is plotted as a function of $1/L$ for several $J_2/|J_1|$ values. Since the magnetic frustration causes a sine-like oscillation in $\Delta(L)$ vs. $1/L$~\cite{clio19}, the finite-size scaling analysis is not very straightforward. Still, the plotted values show the sine-like oscillation with roughly more than one period, an acceptable scaling with linear function $\Delta(L)/|J_1|=\Delta/|J_1|+{\rm A}/L$ (${\rm A}$ is fitting parameter) may be possible. Actually, even if we assume a more general fitting function $\Delta(L)/|J_1|=\Delta/|J_1|+{\rm A}/L^\eta$, the extrapolated values of $\Delta/|J_1|$ are almost unchanged because $\eta\approx1$ is always achieved. In Fig.~\ref{gap_delta}(b) the extrapolated values of $\Delta$ are plotted as a function of $J_2/|J_1|$. At the FM critical point $J_2/|J_1|=1/2$, the lowest energies for all $S_{\rm tot}$ sectors are degenerate and it leads to $\Delta=0$. As soon as the system goes into the ferrimagnetic phase, the stabilization gap $\Delta$ steeply increases, reaches a maximum around $J_2/|J_1|=1$, and decreases with further increasing $J_2/|J_1|$. The magnetic frustration would be strongest at the maximum position $J_2/|J_1|\sim1$ because each triangle is fully frustrated. This is another indication of the fact that the ferrimagnetic state is originated from order by disorder. As shown in the inset of Fig.~\ref{gap_delta}(b), the stabilization gap seems to decay exponentially with $J_2/|J_1|$ in the large $J_2/|J_1|$ region. It means that the ferrimagnetic state is rapidly destabilized in the large $J_2/|J_1|$ regime although it persists up to $J_2/|J_1|=\infty$ in a precise sense. This is consistent with the rapid decrease of $\overline{\langle S^z_{B,i} \rangle}$ with $J_2/|J_1|$. The system can gain only little energy from the interchain FM interaction in case where the basal spins are not really polarized. In short, the quantum fluctuations between the apical and the basal chains play an essential role to stabilize the ferrimagnetic state.

\subsubsection{dynamical spin structure factor}

\begin{figure}[t]
\centering
\includegraphics[width=1.0\linewidth]{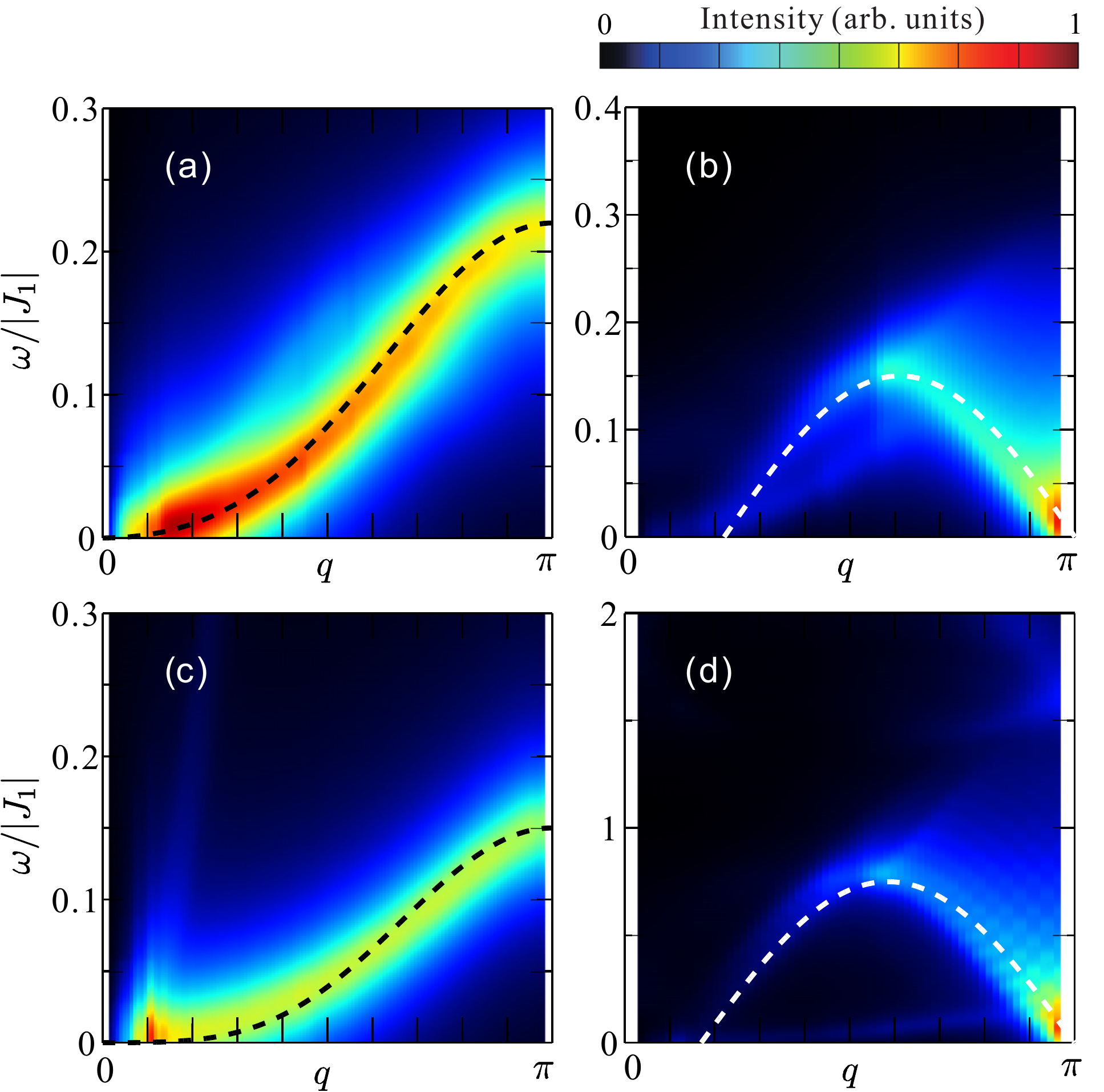}
\caption{
Dynamical spin structure factors for (a) apical and (b) basal chains at $J_2/|J_1|=0.6$. (c)(d) The same spectra at $J_2/|J_1|=1$. Finite broadening $\eta$ is introduced: $\eta=0.03|J_1|$ in (a) and (c), $\eta=0.02|J_1|$ in (b), and $\eta=0.05|J_1|$ in (d). The dotted lines are approximate analytical expressions of the main dispersions (see text).
}
\label{fig_Sqw}
\end{figure}

In order to provide further insight into the ferrimagnetic structure, we investigate the low-energy excitations of the delta chain. We calculate dynamical spin structure factors for both the apical and basal chains with using the DDMRG method. The dynamical spin structure factor is defined as
\begin{eqnarray}
\nonumber
S_\alpha(q,\omega) &=& \frac{1}{\pi}{\rm Im} \langle \psi_0 | S^z_{\alpha,q} \frac{1}{\hat{H}+\omega-E_0-{\rm i}\eta} S^z_{\alpha,q} | \psi_0 \rangle\\
&=& \sum_\nu |\langle \psi_\nu |S^z_{\alpha,q}| \psi_0 \rangle|^2 \delta(\omega-E_\nu+E_0),
\label{spec}
\end{eqnarray}
where $\alpha$ denotes either apical (A) or basal (B) chain, $|\psi_\nu \rangle$ and $E_\nu$ are the $\nu$-th eingenstate and the eigenenergy of the system, respectively ($\nu=0$ corresponds to the ground state). Under OBC, we define the momentum-dependent spin operators as
\begin{align}
S^z_{\alpha,q} = \sqrt{\frac{2}{L_\alpha+1}} \sum_l e^{iqr_i} S^z_{\alpha,i},
\label{operator_obc}
\end{align}
with (quasi-)momentum $q=\pi Z_x/(L_\alpha+1)$ for integers $1 \le Z_x \le L$. We use open clusters with $L_{\rm A}=31, L_{\rm B}=32$ for $S_B(q,\omega)$ and with $L_{\rm A}=64, L_{\rm B}=65$ for $S_A(q,\omega)$. In Fig.~\ref{fig_Sqw}, DDMRG results of the dynamical spin structure factors are shown for $J_2/|J_1|=0.6$ and $1$.

Let us see the spectrum for the apical chain. Since the spins are fully polarized on the apical chain, the main dispersion of $S_A(q,\omega)$ is basically described by that of the 1D FM Heisenberg chain. For $J_2/|J_1|=0.6$, a precise fitting leads to $\omega_q/|J_1|=J^\prime(\cos(q)-1)+J^{''}(\cos(2q)-1)$ with nearest-neighbor FM coupling $J^\prime=-0.11$ and next-nearest-neighbor AFM coupling $J^{''}=0.016$. Interestingly, we find that the dominant FM coupling is effectively induced on the apical chain in spite of no direct interaction between apical sites. A similar fitting for $J_2/|J_1|=1$ gives $J^\prime=-0.075$ $J^{''}=0.018$. The reduction of $|J^\prime|$ with increasing $J_2/|J_1|$ is naturally expected because the apical spins are completely free in the large $J_2/|J_1|$ limit. This reduction also reflects the weakening of ferrimagnetic state.

We then turn to the spectrum for the basal chain. Although the basal chain is weakly polarized, we expect that the fundamental excitations could be at least qualitatively described by those of the 1D SU(2) Heisenberg chain because the dominant short-range correlation is AFM. For $J_2/|J_1|=0.6$, the magnon dispersion (lower bound of the continuum) of $S_B(q,\omega)$ is certainly sine-like function and the well-known shaped two-spinon continuum is seen. However, surprisingly, such the weak spin polarization ($S_{\rm B, tot}/L_{\rm B}=0.11$) drastically suppresses the dispersion width down to $0.15|J_1|$ from that of the 1D SU(2) Heisenberg chain $\pi|J_1|/2$~\cite{Karbach97}. It is interesting that the dispersion width is rapidly recovered to $0.75|J_1|$ when the spin polarization is slightly reduced to $S_{\rm B, tot}/L_{\rm B}=0.085$ at $J_2/|J_1|=1$. Another effect of the weak polarization on the dispersion is a shift of node. Due to the dominant AFM fluctuation on the basal chain in the whole region of ferrimagnetic phase, a largest peak always appears at $(q,\omega)=(\pi,0)$. If there is no spin polarization, the other node should be at $q=0$ but it is actually shifted to higher $q$ value as seen in Fig.~\ref{fig_Sqw}(b)(d). This behavior is similar to the case in the presence of magnetic field. Namely, the node position can be expressed as $q=2\pi \langle S_{\rm B,tot} \rangle/L_{\rm B}$ where $\langle S_{\rm B,tot} \rangle$ is the total spin of the basal chain with length $L_{\rm B}$.

\subsection{Finite $\gamma$: asymmetric $J_1$-$J_2$ zigzag ladder}

As described above, we have confirmed that the ferrimagnetic state is indeed stabilized at $J_2/|J_1|>1/2$ in the delta chain ($\gamma=0$). Then, let us see what happens when the apical sites are connected by AFM interaction, the strength of which can be controlled by $\gamma$. Since the system is in a singlet ground state, i.e., $S_{\rm tot}=0$, in the $J_1$-$J_2$ chain ($\gamma=1$), the collapse of ferrimagnetic state is naively expected at some $\gamma (<1)$. Incidentally, the ferrimagnetic state is trivially enhanced if an FM interaction is introduced between apical sites.

\subsubsection{total spin}

\begin{figure}[t]
\centering
\includegraphics[width=0.7\linewidth]{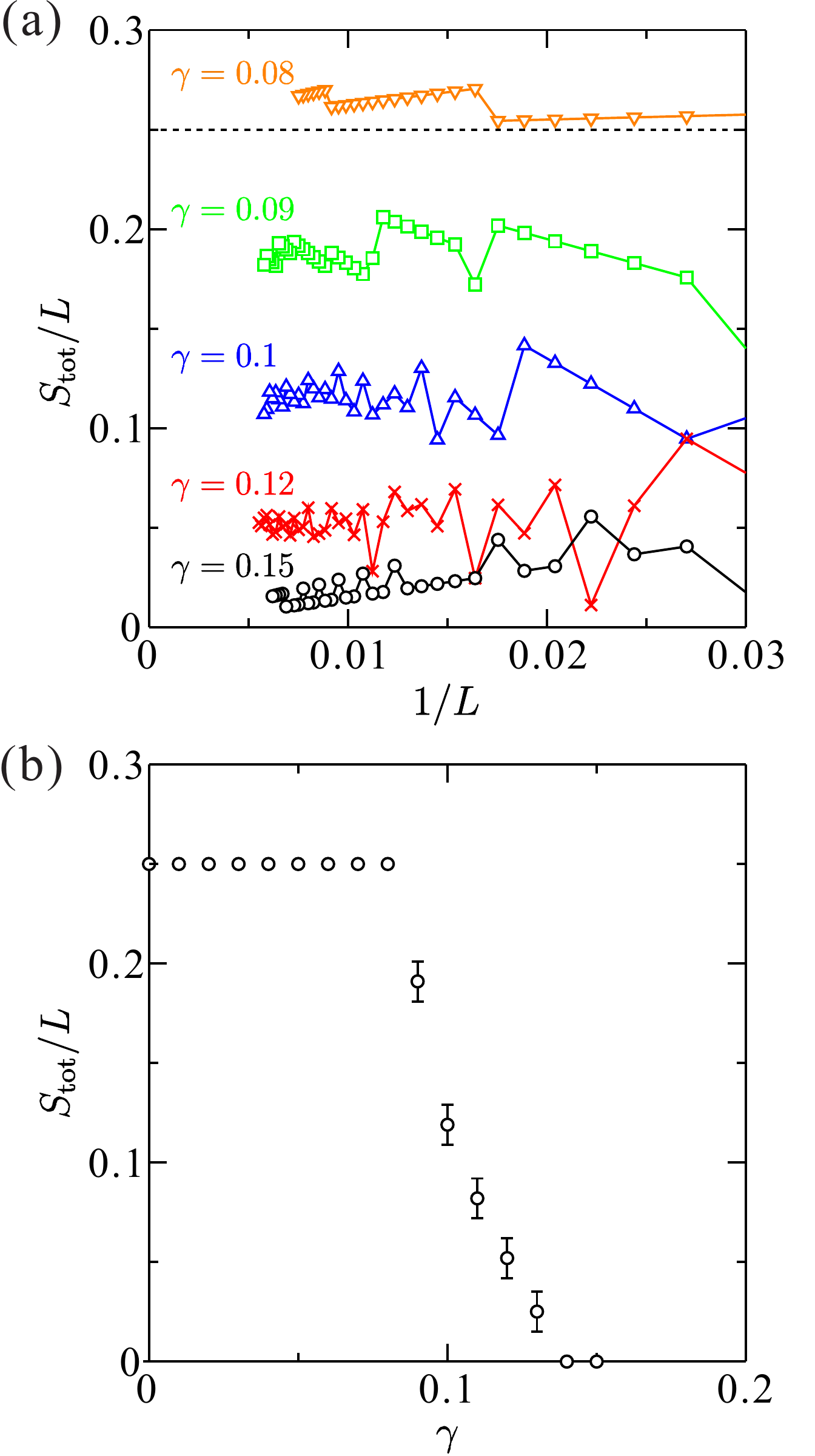}
\caption{
(a) System-size dependence of total spin per site $S_{\rm tot}/L$ for several $\gamma$ values with $J_2/|J_1|=0.6$ fixed. The dashed line indicates the value of $S_{\rm tot}/L$ for the full ferrimagnetic state. (b) The $L \to \infty$ extrapolated values of $S_{\rm tot}/L$ as a function of $\gamma$.
}
\label{S_gamma}
\end{figure}

Simply, we examine the $\gamma$-dependence of the total spin to identify when and how the ferrimagnetic state is destabilized. In Fig.~\ref{S_gamma}(a) the total spin per site $S_{\rm tot}/L$ is plotted as a function of $1/L$ for several $\gamma$ values with $J_2/|J_1|=0.6$ fixed. Due to the strong frustration, the value of $S_{\rm tot}/L$ oscillates with $1/L$; however, we may perform a reasonable finite-size scaling analysis with finer data points using large enough clusters. We here use open clusters with length up to $L=L_{\rm A}+L_{\rm B}=100+101=201$. The $L\to\infty$ extrapolated value of $S_{\rm tot}/L$ is plotted as a function of $\gamma$ in Fig.~\ref{S_gamma}(b).

At $\gamma=0$, the ground state is in the ferrimagnetic state with nearly-fully polarized apical spins and the total spin per site is $S_{\rm tot}=L/4$. Hereafter, we call this state ``{\it full} ferrimagnetic (FF) state'' to discriminate it from another ferrimagnetic state with $S_{\rm tot}<L/4$ (denoted as ``{\it partial} ferrimagnetic (PF) state'') which appears below. When the AFM interaction between apical spins is switched on, one may naively expect a collapse or weakening of the full spin polarization in the apical chain; nevertheless, interestingly, the FF condition $S_{\rm tot}=L/4$ survives up to $\gamma \approx 0.08$. This can be interpreted as follows: Roughly speaking, since the system can gain more energy from interchain FM interaction than AFM interaction between apical spins, the FM LRO in the apical chain is still maintained at $\gamma \lesssim 0.08$.

With increasing $\gamma$ from $0.08$, the competition between interchain FM interaction and apical intrachain AFM interaction derives a new state. As seen in Fig.~\ref{S_gamma}(a), at $\gamma=0.09$ the value of $S_{\rm tot}/L$ seems to converge at a finite but smaller value than $1/4$ as $1/L$ decreases. This clearly suggests that some sort of collapse of the FF state happens around $\gamma=0.09$. With further increasing $\gamma$, the value of $S_{\rm tot}/L$ appears to be continuously reduced and reaches zero around $\gamma=0.14$ [see Fig.~\ref{S_gamma}(b)]. Surprisingly, we find that there exists a finite $\gamma$-range exhibiting $0<S_{\rm tot}/L<1/4$. Since the basal chain basically keeps its weakly polarized or nearly singlet state, it would be a good guess that spin polarization on the apical sites is gradually collapsed with increasing $\gamma$ in this PF phase ($0.08 \lesssim \gamma \lesssim 0.14$). In other words, the ferrimagnetic order by disorder in association with the global spin-rotation-symmetry breaking disappears around $\gamma=0.14$. Actually, as stated below, the system has the other order by disorder, i.e., dimerization order, at $\gamma \gtrsim 0.14$. The region of the PF phase is shown as a shaded area in the ground-state phase diagram [Fig.~\ref{fig_pd}(b)].

We make some remarks on the existence of PF phase. Such a `halfway' magnetization $0<S_{\rm tot}/L<1/4$ in a ferrimagnetic state is generally prohibited by the Marshall-Lieb-Mattis (MLM) theorem~\cite{Marshall55,Lieb62}. There is an exception to this, however, when the ferrimagnetic order and a quasi-long-range order of TLL compete~\cite{Furuya14}. This corresponds to the competition between small FM polarization and dominant AFM fluctuations in the basal chain of our system. As confirmed in Appendix C, the basal chain in the FF state indeed exhibits a TLL behavior.

\subsubsection{spin-spin correlation}

\begin{figure}[t]
\centering
\includegraphics[width=1.0\linewidth]{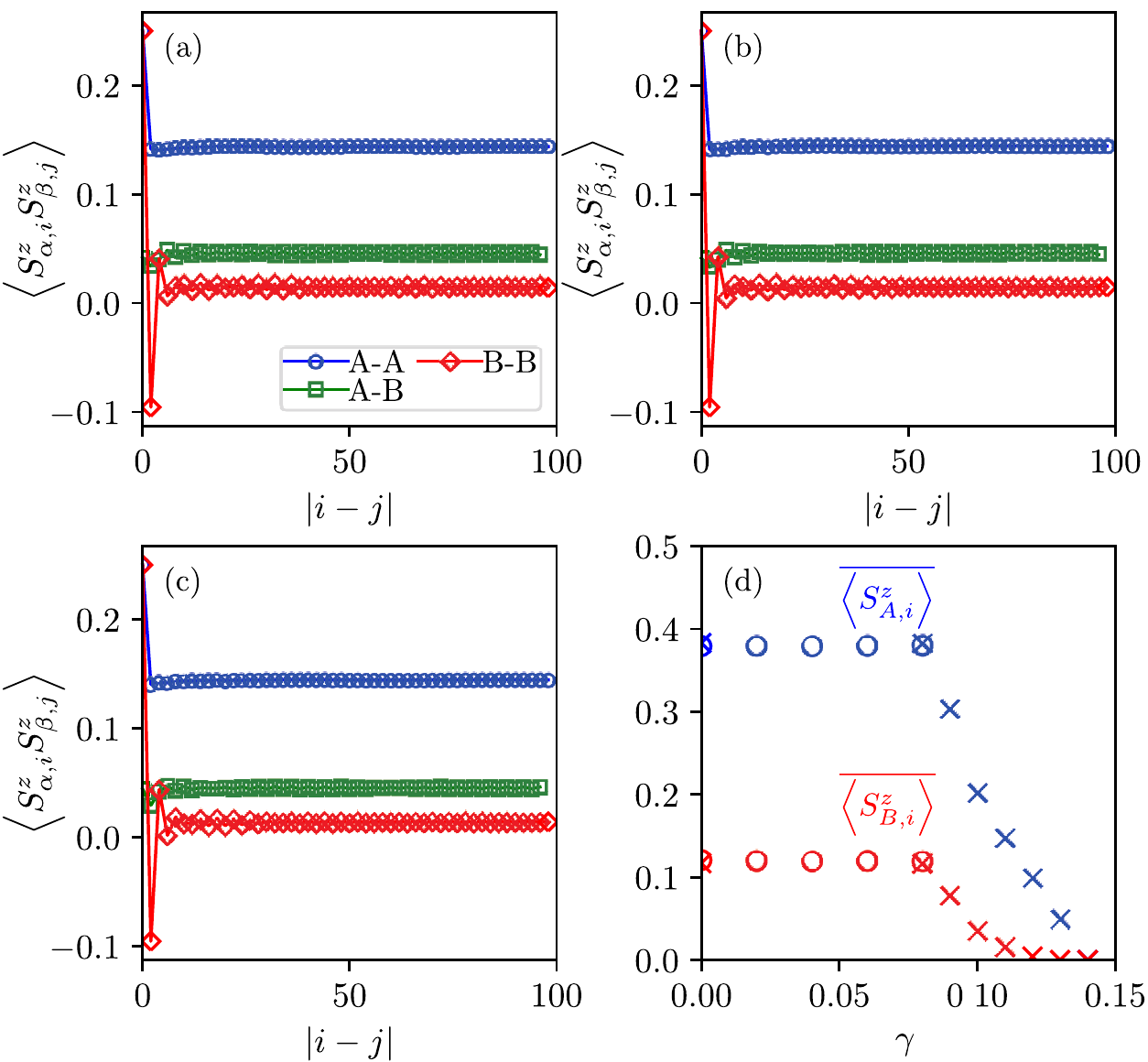}
\caption{
Spin-spin correlation function $\langle S^z_i S^z_j \rangle$ of the asymmetric $J_1$-$J_2$ zigzag ladder as a function of $|i-j|$ with fixed $S^z_{\rm tot}/L=1/4$ and $J_2/|J_1|=0.6$ for (a) $\gamma=0$, (b) $0.04$, and (c) $0.08$. (d) Averaged values of $\langle S^z_i \rangle$ on the apical and basal sites as a function of $\gamma$ for $J_2/|J_1|=0.6$. The circles and crosses denote iDMRG and DMRG results, respectively.
}
\label{corr_gamma}
\end{figure}

We then consider the evolution of spin-spin correlation function with $\gamma$ in the FF phase. In Fig.~\ref{corr_gamma}(a)-(c) iDMRG results of the spin-spin correlation function $\langle S^z_{\alpha,i} S^z_{\beta,j} \rangle$ for $\gamma=0$, $0.04$, and $0.08$ with fixed $J_2/|J_1|=0.6$ are plotted as a function of distance $|i-j|$. We keep $S^z_{\rm tot}=4/L$ as done in Fig.~\ref{corr_delta}.

As far as the FF state is maintained up to $\gamma \approx 0.08$, the correlation functions seem to be almost independent of $\gamma$. Accordingly, the expectation values of $S^z_{A,i}$ and $S^z_{B,i}$ are unchanged up to $\gamma \approx 0.08$, as shown in Fig.~\ref{corr_gamma}(d). Perhaps, one may naively expect the reduction of $\overline{\langle S^z_{A,i} \rangle}$ with increasing $\gamma$, i.e., with increasing AFM coupling between apical spins. However, this is not actually the case. This can be understood as follows: as estimated by the fitting of low-energy excitation spectrum, an FM coupling with the magnitude $\sim 0.11|J_1|$ is effectively induced between neighboring apical sites at $\gamma=0$. Thus, the apical chain may be effectively mapped onto an FM chain with $J_{\rm eff}=-0.11|J_1|$. Therefore, it would be rather natural that the (nearly) full polarization is free of the influence of additional AFM coupling $\gamma J_2$ until it reaches around $\sim0.11|J_1|$.

No $\gamma$-dependence of the spin structure up to $\gamma \approx 0.08$ then indicates that the total energy of FF state is simply lifted by the AFM interaction $\gamma J_2$ between nearly-fully polarized apical spins; it switches into the energy level with a metastable PF state at $\gamma \approx 0.09$. Hence, the FF to PF phase transition is of the first order. It can be also confirmed by a steep (or almost discontinuous) change of $S_{\rm tot}$ and $\overline{\langle S^z_{\alpha,i} \rangle}$ at $\gamma \approx 0.09$. On the other hand, both of $\overline{\langle S^z_{A,i} \rangle}$ and $\overline{\langle S^z_{B,i} \rangle}$ smoothly approach to zero around $\gamma \approx 0.14$. Thus, the transition from PF to the spiral singlet ($S_{\rm tot}=0$) region is of the second order or continuous.

\subsubsection{stabilization gap}

\begin{figure}[t]
\centering
\includegraphics[width=0.7\linewidth]{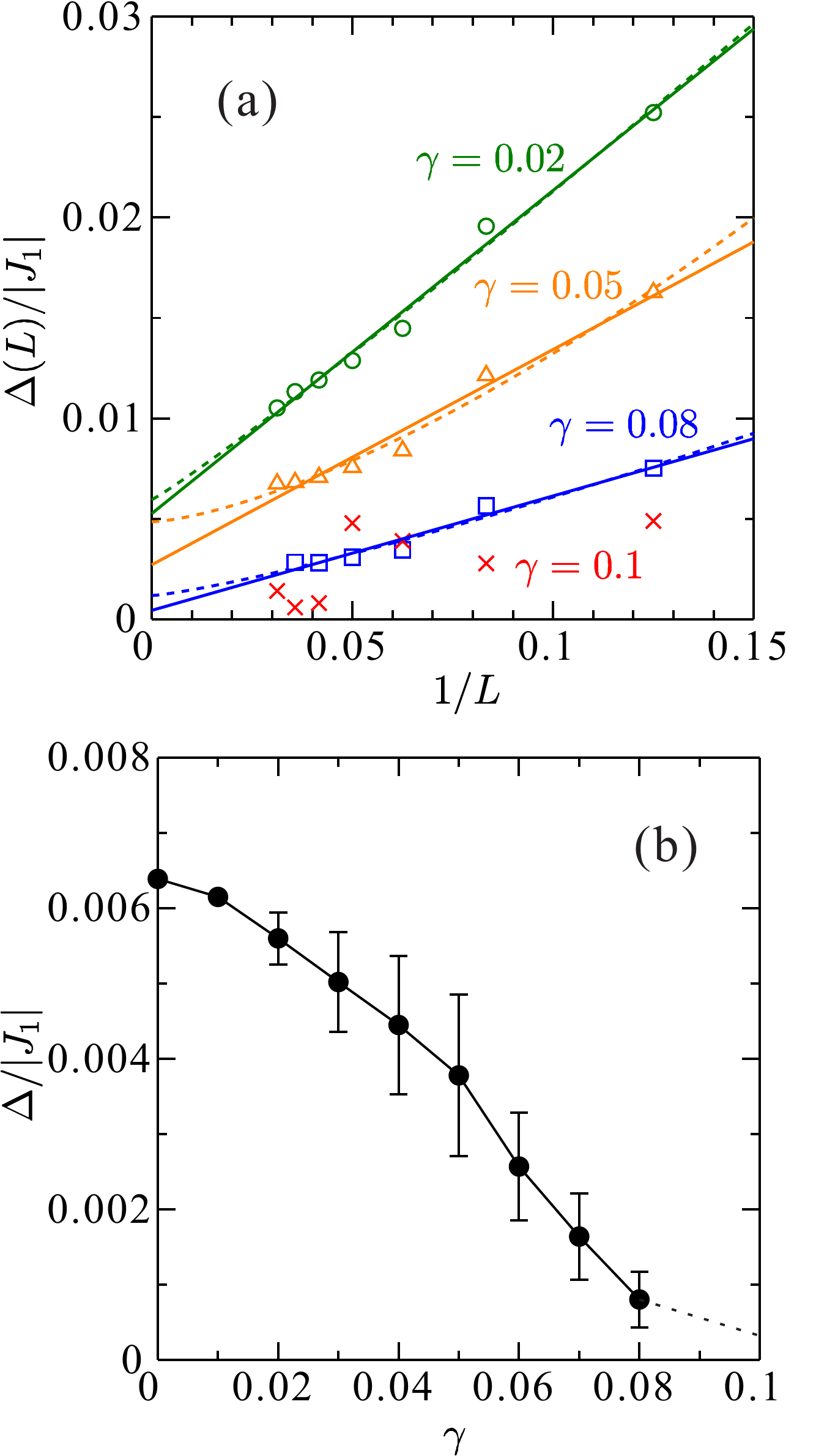}
\caption{
(a) Finite-size scaling of energy difference between  lowest $S_{\rm tot}=0$ state and ferrimagnetic ground state for the asymmetric $J_1$-$J_2$ zigzag ladder with fixed $J_2/|J_1|=0.6$. The solid and dotted lines show the fitting results with $\Delta(L)/|J_1|=\Delta/|J_1|+{\rm A}/L$ and $\Delta(L)/|J_1|=\Delta/|J_1|+{\rm A}/L^\eta$, respectively. (b) Extrapolated values of $\Delta/|J_1|$ as a function of $\gamma$. The width of error bar means the difference of $\Delta/|J_1|$ obtained by the two fitting functions.
}
\label{gap_gamma}
\end{figure}

To quantify the stability of ferrimagnetic state at finite $\gamma$, we calculate the stabilization energy $\Delta$ [Eq.\eqref{eq_gap}] as done in the case of delta chain. In Fig.~\ref{gap_gamma}(a) the finite-size scaling analysis of $\Delta(L)$ is performed for several $\gamma$ values with $J_2/|J_1|=0.6$ fixed. Although a sine-like oscillation is present as in the case of delta chain, an acceptable scaling may be possible in the FF phase ($\gamma \lesssim 0.09$). We here employ two kinds of fitting functions: $\Delta(L)/|J_1|=\Delta/|J_1|+{\rm A}/L$ and $\Delta(L)/|J_1|=\Delta/|J_1|+{\rm A}/L^\eta$. Since the former (latter) function seems to underestimate (overestimate) the extrapolated value of $\Delta/|J_1|$, their averaged value is plotted as a function of $\gamma$ in Fig.~\ref{gap_gamma}(b). The width of error bar means the differnce between two values of $\Delta/|J_1|$ obtained by the two fitting functions. The stabilization gap is approximately linearly reduced by $\gamma$ up to the critical point $\gamma \approx 0.08$. This also supports the above speculation that the total energy of FF state is simply lifted by $\gamma J_2$.

In the PF phase ($\gamma \gtrsim 0.09$), however, the scaling analysis of $\Delta(L)$ is virtually impossible. As an example, $\Delta(L)/|J_1|$ vs. $1/L$ for $\gamma=0.1$ is shown in Fig.~\ref{gap_gamma}(a). This difficulty is caused by the following several factors: (i) As shown in the next subsection, an incommensurate oscillation is involved in the PF state. (ii) The total spin per site $S_{\rm tot}/L$ is strongly dependent on system size since the states in different $S_{\rm tot}$ sectors are extremely quasi-degenerate around the ground state. (iii) The available system size is strictly limited because the second term of Eq.\eqref{ham_S0} includes long-range interactions and a periodic cluster must be used. Nevertheless, the stabilization gap should be positive due to the nonzero total spin of the ground state [Fig.~\ref{S_gamma}(b)]. We can, at least, confirm that the PF order is very fragile with the stabilization gap $\Delta<4.3\times10^{-4}|J_1|$ at $\gamma=0.08$. This small stabilization gap also tells us that there are a macroscopic number of quasi-degenerate states belonging to different $S_{\rm tot}$ sectors, since the total spin is continuously varied from $L/4$ to $0$ with $\gamma$ in the PF phase.

\subsubsection{Static spin structure factor}

\begin{figure}[t]
\centering
\includegraphics[width=0.9\linewidth]{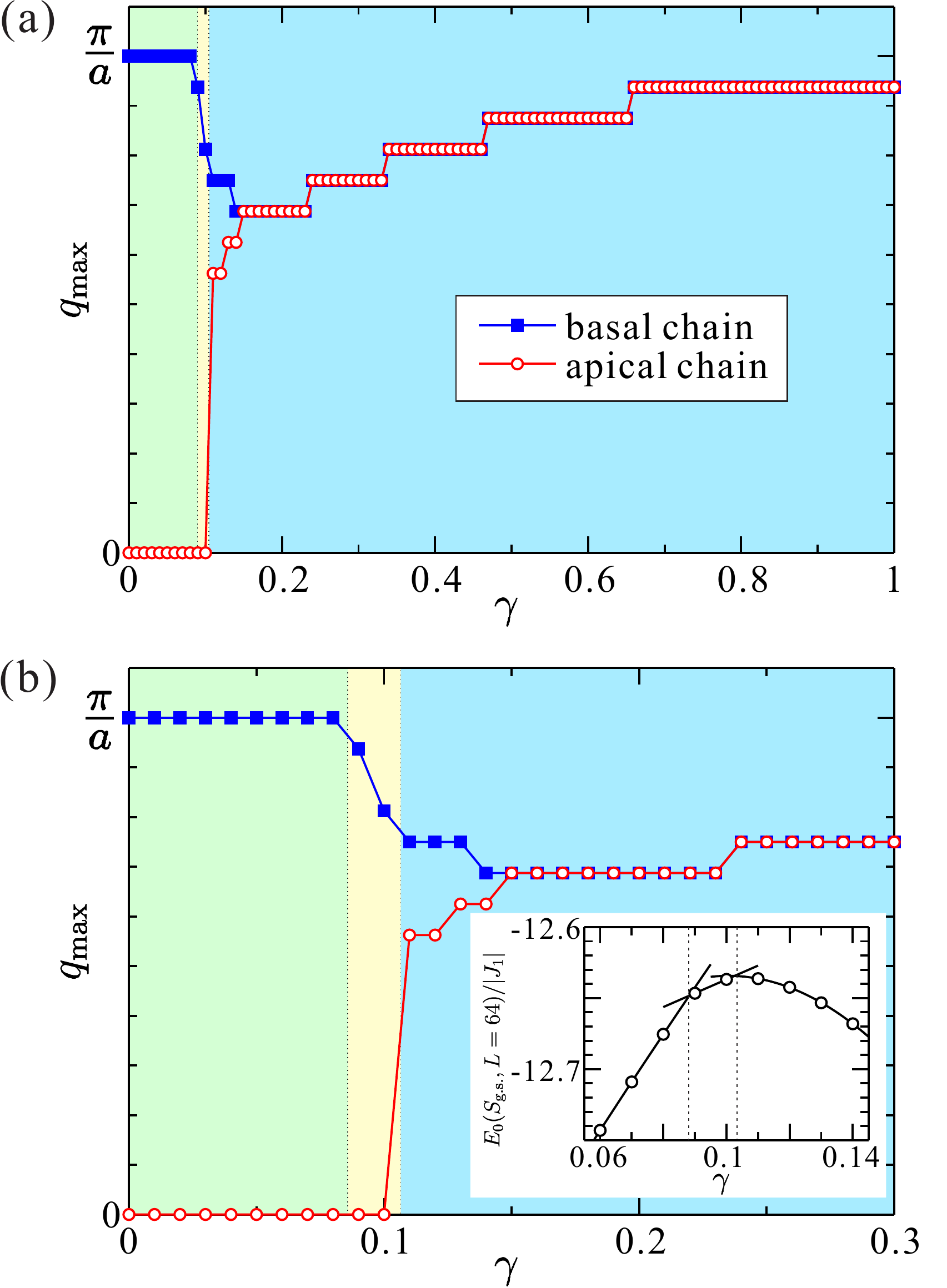}
\caption{
(a) Static spin structure factor for $J_2/|J_1|=0.6$ as a function of $\gamma$. The lattice spacing $a$ is set as shown in Fig.~\ref{lattice}(a). (b) Enlarged figure of (a) for $0 \le\gamma \le 0.3$. Inset: Ground-state energy as a function of $\gamma$.
}
\label{fig_Sq}
\end{figure}

It is important to see how the intrachain spin modulation changes with $\gamma$. A most significant quantity to know it would be the propagation number which can be extracted as a maximum position of static spin structure factor. Therefore, we calculate the static spin structure factor for each of the apical and basal chains. It is defined as
\begin{align}
S_\alpha(q) = \frac{1}{L_\alpha^2} \sum_{i.j=1}^{L_\alpha} e^{iq(r_i-r_j)} \langle \bm{S}_{\alpha,i} \cdot \bm{S}_{\alpha,j} \rangle
\label{Sq}
\end{align}
for the apical ($\alpha={\rm A}$) or basal ($\alpha={\rm B}$) chains. The lattice spacing $a$ is set as shown in Fig.~\ref{lattice}(a). In Fig.~\ref{fig_Sq} the propagation numbers $q_{\rm max}$ for $J_2/|J_1|=0.6$ using a 64-site periodic cluster are plotted as a function of $\gamma$.

At $\gamma=0$, the system is in the FF state whose spin structure is commensurate and simple as shown in Fig.~\ref{lattice}(c). The apical spins are nearly-fully polarized and the propagation number is $q_{\rm max}=0$, whereas for basal chain the dominant correlation is AFM ($q_{\rm max}=\pi/a$) although it is slightly polarized. It is obvious that this spin structure persists in the whole region of the FF phase at $0 \lesssim \gamma \lesssim 0.08$.

In the singlet ($S_{\rm tot}=0$) phase at $\gamma \gtrsim 0.14$, the propagation numbers of apical and basal chains are both incommensurate, i.e., $0<q_{\rm max}<\pi/a$. At $\gamma=1$, they coincide and it is estimated as $q_{\rm max}=0.958$. This value is in good agreement with our previous estimation $q_{\rm max}=0.958$~\cite{linarite_2012} in the thermodynamic limit. With decreasing $\gamma$, the propagation number is reduced because short-range FM correlation is relatively enhanced. Interestingly, they are equal or very close down to the critical point at $\gamma \approx 0.14$ and obviously split for smaller $\gamma$. It would be a good guess that the short-ranged spiral structure of the $J_1$-$J_2$ chain ($\gamma=1$) is approximately maintained down to $\gamma \approx 0.14$. In a broad sense, this incommensurate region can be referred to as a spiral singlet phase.

In Fig.~\ref{fig_Sq}(b) an enlarged figure around the PF phase ($0.08 \lesssim \gamma \lesssim 0.14$) is shown. As shown in the inset of Fig.~\ref{fig_Sq}(b) the phase boundaries are recognized by level crossing of the ground state energies. It clearly indicates the existence of an intermediate phase between the FF and spiral singlet phases, though the region ($0.08 \gtrsim \gamma \gtrsim 0.11$) is a bit narrower than that in the thermodynamic limit ($0.08 \gtrsim \gamma \gtrsim 0.14$) due to finite size effects. The intermediate phase is the PF phase as described above.

In the PF phase, the dominant correlation of the basal chain seems to be incommensurate and the propagation number of the apical chain keeps $q_{\rm max}=0$. It is a natural consequence of the global spin-rotation-symmetry breaking because the total spin can be no longer finite if both of the propagation numbers are nonzero. This incommensurate propagation is a consequence of the {\it halfway} magnetization so that the prohibition by the MLM theorem is also avoided by the TLL characteristic of basal chain. Similar incommensurate features have been reported for PF state in frustrated systems~\cite{Hida07,Hida08}.

\subsubsection{Dimerization order}

\begin{figure}[t]
\centering
\includegraphics[width=0.6\linewidth]{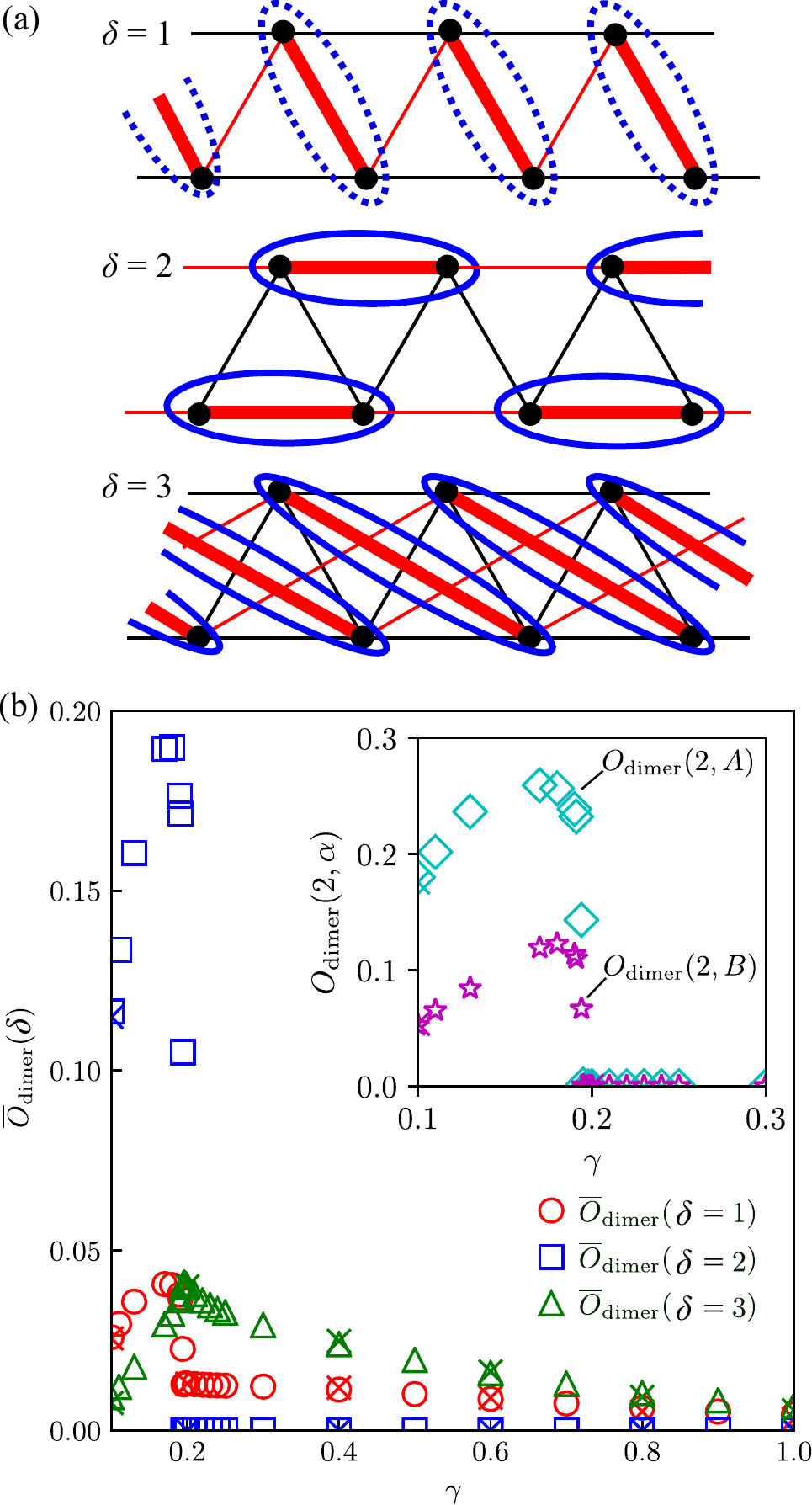}
\caption{
(a) Schematic pictures of possible dimerization order. The states for $\delta=2$ and $\delta=3$ are characterized as VBS. A solid (dotted) ellipse denotes a spin-singlet (spin-triplet) dimer. (b) Averaged dimerization order parameters $\mathcal O_{\rm dimer}(\delta)$ as a function of $\gamma$ at $J_2/|J_1|=1$. Inset: each contribution to $\mathcal O_{\rm dimer}(2)$ from the apical and basal chains.
}
\label{fig_SdS}
\end{figure}

So, let us see more about the magnetic properties of the spiral singlet ($S_{\rm tot}=0$) phase at larger $\gamma$. It is known that the system has LRO with spontaneous dimerization in the $J_1$-$J_2$ chain ($\gamma=1$)~\cite{furukawa12,clio19}. Therefore, as a starting point it would be reasonable to examine the evolution of dimerization order parameters with decreasing $\gamma$ from $1$. The dimerization order parameter between sites distant $\delta$ along the $J_1$ zigzag chain is defined as
\begin{align}
	\nonumber
	\mathcal O_{\rm dimer}(\delta)=\lim_{L\to\infty}|\langle \bm S_{{\rm A},i} &\cdot \bm S_{{\rm B},i-(\delta-1)/2} \rangle \\
	&- \langle \bm S_{{\rm A},i} \cdot \bm S_{{\rm B},i+(\delta+1)/2} \rangle|,
	\label{dimerodd}
\end{align}
for odd $\delta$, and
\begin{align}
	\mathcal O_{\rm dimer}(\delta)=\lim_{L\to\infty}|\langle \bm S_{\alpha,i} \cdot \bm S_{\alpha,i+\delta/2} \rangle - \langle \bm S_{\alpha,i} \cdot \bm S_{\alpha,i+\delta/2} \rangle|,
	\label{dimereven}
\end{align}
for even $\delta$. If $\mathcal O_{\rm dimer}(\delta)$ is finite for $\delta$, it signifies a long-range dimerization order associated with mirror-symmetry breaking for odd $\delta$ or translation-symmetry breaking for even $\delta$. We here study the case of $\delta=1$, $2$, and $3$. Schematic pictures of the possible dimerization orders are shown in Fig.~\ref{fig_SdS}(a). A finite $\mathcal O_{\rm dimer}(\delta)$ for $\delta=2$ and $3$ indicates a valence bond formation, i.e., spin-singlet formation, between two sites on the dimerized bond. In Fig.~\ref{fig_SdS}(b) the iDMRG results of dimerization order parameter $\mathcal O_{\rm dimer}(\delta)$ for $\delta=1$, $2$, and $3$ are plotted as a function of $\gamma$ at $J_2/|J_1|=1$. For confirmation, we also estimate $\mathcal O_{\rm dimer}(\delta)$ in the thermodynamic limit for some $\gamma$ values using DMRG under OBC. We can find their excellent agreement with the iDMRG results. Note that the FM interaction $J_1$ at both edges in the open clusters is set to be zero. It enables us to perform the finite-size scaling analysis more easily because the competing two translation-symmetry breaking states are explicitly separated~\cite{clio19}. For confirmation, we have checked that the ground state in the thermodynamic limit does not depend on the choice of boundary conditions.

At $\gamma=1$, two dimerization orders with $\delta=1$ and $\delta=3$ coexist~\cite{clio19}. With decreasing $\gamma$, interestingly, $\mathcal O_{\rm dimer}(3)$ is significantly enhanced and $\mathcal O_{\rm dimer}(1)$ is slightly increased down to $\gamma \approx 0.195$. With fixed $J_2/|J_1|=1$, we may deduce that the magnetic frustration is largest in the limit of $\gamma=0$ where the system is a series of isotropic triangles with uniform magnitude of interactions. The valence bond pair may be strengthened to screen spins more strongly for the relaxation of larger magnetic frustration at smaller $\gamma$. Since $\mathcal O_{\rm dimer}(2)=0$ down to $\gamma \approx 0.195$, this state is dominantly characterized as a VBS state with $\delta=3$ dimerization order (we call it ``$\mathcal D_3$-VBS'' state).

Even more surprisingly, $\mathcal O_{\rm dimer}(2)$ exhibits a steep increase (almost jump) at $\gamma \approx 0.195$. The other order parameters $\mathcal O_{\rm dimer}(1)$ and $\mathcal O_{\rm dimer}(3)$ are also not differentiable with $\gamma$ at this point. This clearly indicates another first-order transition at $\gamma \approx 0.195$. We note that both of the apical and basal chains are spontaneously dimerized along the chain direction in the region of $\mathcal O_{\rm dimer}(2)$. The order parameter for each chain is plotted in the inset of Fig.~\ref{fig_SdS}(b) (The main figure shows the averaged value). Since the value of $\mathcal O_{\rm dimer}(2)$ is much larger than the other dimerization order parameters at $0.035 \lesssim \gamma \lesssim 0.195$, the state is recognized as a VBS one with $\delta=2$ dimerization order (we call it ``$\mathcal D_2$-VBS'' state). Since this $\delta=2$ dimerization order is associated with a translation-symmetry breaking, the magnetic structure consists of a supercell with four sites. More detailed analysis is given in Appendix D.

\begin{figure}[tb]
  \includegraphics[width=1.0\columnwidth]{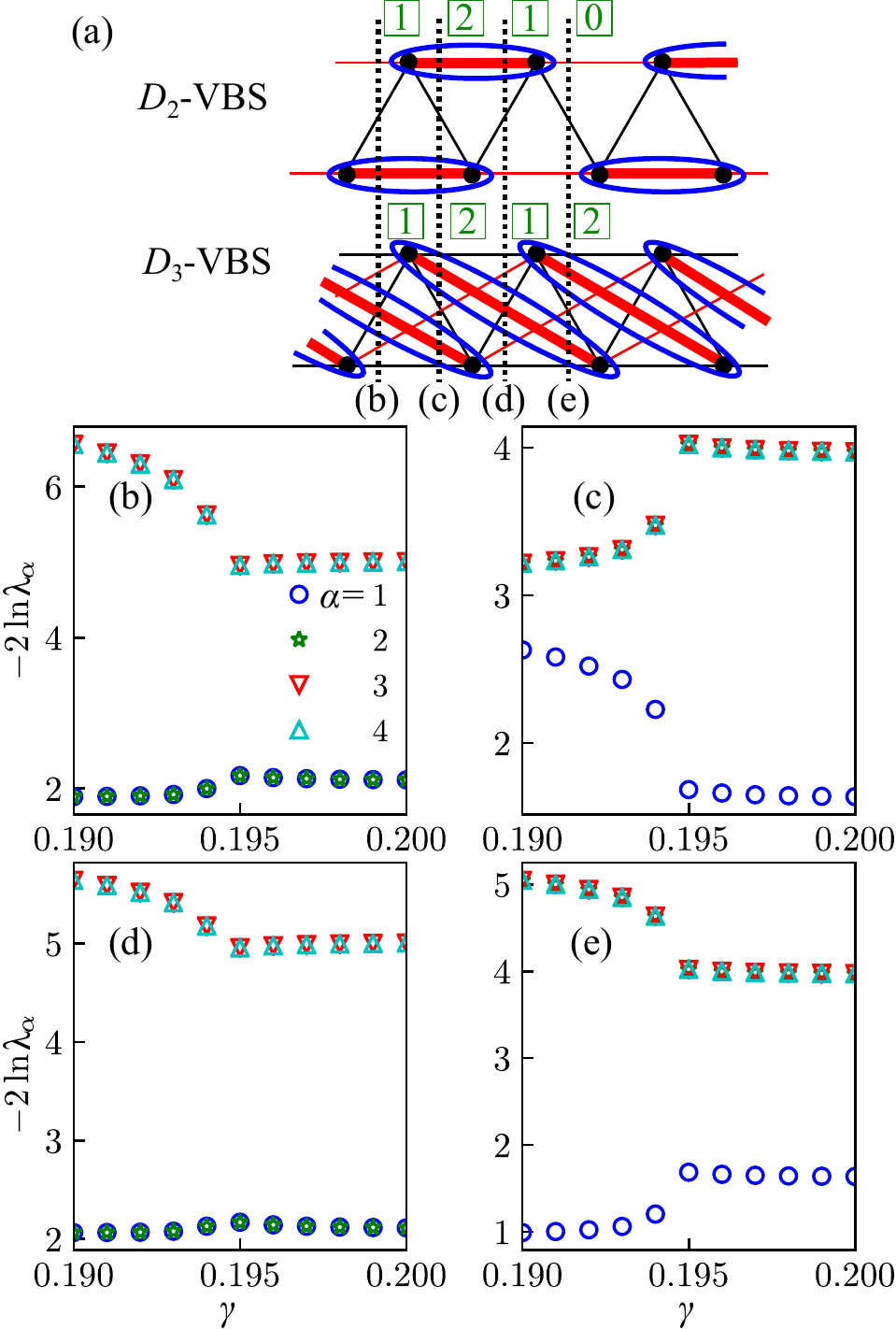}
  \caption{
    (a) Schematic pictures of considered splitting of the system into two subsystems in the $\mathcal D_2$-VBS and $\mathcal D_3$-VBS state. A solid ellipse denotes a spin-singlet pair. The number of singlet pair crossing with each cut is shown in the green square.
  (b)-(e) Entanglement spectrum for the corresponding splitting as a function of $\gamma$ at $J_2/|J_1|=1$.}
  \label{fig_espectrum}
\end{figure}

The dimerization order can be also detected by studying the topological properties of the system. Then, let us see the entanglement spectrum (ES)~\cite{li08} which can be obtained by a canonical representation of the an infinite matrix-product-state in the iDMRG calculations~\cite{Pollmann12}. Using Schmidt decomposition, the ground-state wave function can be expressed as
\begin{equation}
|\psi\rangle=\sum_i e^{-\xi_i/2}|\phi_i^{\rm A}\rangle \otimes |\phi_i^{\rm B}\rangle,
\label{xi}
\end{equation}
where the states $|\phi_i^S\rangle$  correspond to an orthonormal basis for the subsystem $S$ (either A or B). We study the ES for several kinds of splitting pattern between subsystems A and B. The splitting patterns are sketched in Fig.~\ref{fig_espectrum}(a). In our iDMRG calculations, the ES $\{\xi_\alpha\}$ is simply obtained as $\xi_\alpha=-2\ln \lambda_\alpha$, where $\{\lambda_\alpha^2\}$ are the singular values of the reduced density matrices after the bipartite splitting. The low-lying four ES levels are plotted as function of $\gamma$ in Fig.~\ref{fig_espectrum}(b)-(e).

  When the one dimerized singlet pair straddles the subsystems A and B [Fig.~\ref{fig_espectrum}(b) and (d)], the lowest entanglement level is doubly degenerate as a reflection of the edge state. On the other hand, when this is not the case [Fig.~\ref{fig_espectrum}(c) and (e)], the lowest entanglement level is non-degenerate. Our result for the $\mathcal D_3$-VBS state is consistent with previous research on the symmetric case~\cite{furukawa12}. These facts would strongly support the formation of long-range dimerization order. We can also find a discontinuous change of $\partial (-2\ln \lambda_\alpha)/\partial \gamma$ at $\gamma\sim0.195$, which seems to correspond to the transition point from the $\mathcal D_2$-VBS to $\mathcal D_3$-VBS state. We note that the difference between (b) and (d) as well as (c) and (e) in Fig.~\ref{fig_espectrum} comes from the asymmetric nature of our system. More details about the asymmetric nature are discussed in Appendix D.

In the spiral singlet phase, the system is in either $\mathcal D_2$-VBS or $\mathcal D_3$-VBS state. The phase boundary between them is shown in Fig.~\ref{fig_pd}(b). In general, the spin gap, namely, energy difference between spin-singlet ground state and spin-triplet first excited state, is expected to be finite when the system is in a VBS state. In the $\mathcal D_3$-VBS phase, the spin gap simply scales to an energy to break a valence bond for $\delta=3$. In the $\mathcal D_2$-VBS phase, each of the apical and basal chains has a different valence bond. Nevertheless, it is easy to imagine that the valence bond in the apical chain is more fragile because of smaller AFM interaction, although $\mathcal O_{\rm dimer}(2)$ for the apical chain is larger than that for the basal chain. Thus, the spin gap in the $\mathcal D_2$-VBS phase scales to an energy to break a valence bond in the apical chain. This means that a larger energy than the spin gap is needed to break a valence bond in the basal chain. It would provide a 1/2-plateau in the magnetization process with magnetic field.

\subsubsection{String order}\label{stop}

\begin{figure}[t]
\centering
\includegraphics[width=0.7\linewidth]{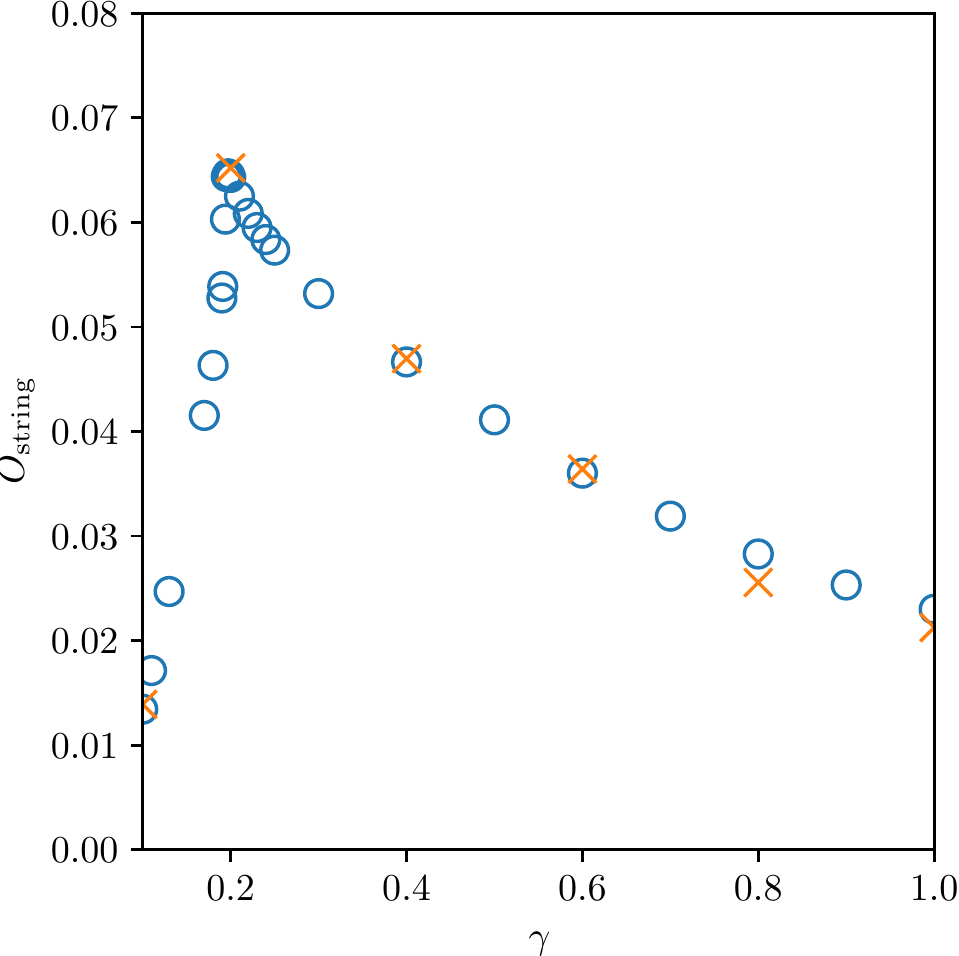}
\caption{
String order parameter as a function of $\gamma$ at $J_2/|J_1|=1$ using iDMRG (circles) and DMRG (crosses) methods. The DMRG results are extrapolated values to the thermodynamic limit.
}
\label{fig_SOPz}
\end{figure}

We have confirmed the existence of nearest-neighbor ($\delta=1$) FM dimerization order in the whole spiral singlet region. A spin-triplet pair may be effectively formed in the each ferromagnetically dimerized bond: By relating three states $|\uparrow\uparrow\rangle$, $|\uparrow\downarrow\rangle+|\downarrow\uparrow\rangle)/\sqrt{2}$, and $|\downarrow\downarrow\rangle$ to $S^z=1$, $0$, and $-1$ states, respectively, the resultant spin on the dimerized bond can be reduced to a spin-$1$ degree of freedom. Consequently, the system could be mapped onto a $S=1$ Heisenberg chain accompanied by the emergent effective spin-$1$ degrees of freedom with the dimerized two spin-$\frac{1}{2}$'s~\cite{furukawa12}. Furthermore, the presence of third-neighbor AFM dimerization order ensures a valence bond formation between the neighboring effective $S=1$ sites~\cite{clio19}. It leads to finite spin gap as a Haldane gap in symmetry-protected VBS state~\cite{Affleck87,Pollmann12}. Although the spin gap is a good indicator to measure the stability of VBS state, it would be too small to correctly estimate with DMRG method in most of the parameter region of system \eqref{ham}. Alternatively, the stability of VBS state associated with the Haldane picture can be evaluated by examining the string order parameter~\cite{deNijs89}:
\begin{align}
	\nonumber
	\mathcal O_\mathrm{string}^z = - &\lim_{|k-j| \to \infty} \langle
	(S^z_{{\rm A},k}+S^z_{{\rm B},k})\\
	&\exp[i\pi\sum_{l=k+1}^{j-1} (S^z_{{\rm A},l}+S^z_{{\rm B},l})]
	(S^z_{{\rm A},j}+S^z_{{\rm B},j})\rangle.
	\label{stringorder1}
\end{align}
For our system \eqref{ham}, Eq.~\eqref{stringorder1} can be simplified as
\begin{align}
	\nonumber
	\mathcal O_\mathrm{string}^z = - &\lim_{|k-j| \to \infty} (-4)^{j-k-2} \langle
	(S^z_{{\rm A},k}+S^z_{{\rm B},k})\\
	&\prod_{l=k+1}^{j-1} S^z_{{\rm A},l}S^z_{{\rm B},l}
	(S^z_{{\rm A},j}+S^z_{{\rm B},j})\rangle.
	\label{stringorder2}
 \rangle
\end{align}
The finite value of $\mathcal O_\mathrm{string}^z$ suggests the formation of a VBS state having a hidden topological long-range string order. Since two-fold degeneracy of the ground state due to the FM dimerization is lifted in our numerical calculations, $|\mathcal O_\mathrm{string}^z|$ can have two different values depending on how to select $k$ and $j$. We then take their average.

In Fig.~\ref{fig_SOPz} iDMRG and DMRG results for the string order parameter are plotted as a function of $\gamma$ at $J_2/|J_1|=1$. We can see a good agreement between the iDMRG and DMRG values. With decreasing $\gamma$ from $1$, $\mathcal O_\mathrm{string}^z$ is significantly increased and has a pointed top at $\gamma \approx 0.195$, which is the first-order transition point between $\mathcal D_3$-VBS and $\mathcal D_2$-VBS phases. With further decreasing $\gamma$, it decreases and vanishes at $\gamma=0.035$, which is the second-order transition point between $\mathcal D_2$-VBS and ferrimagnetic phases. We notice that the overall trend of $\mathcal O_\mathrm{string}^z$ is similar to that of the $\delta=3$ dimerization order parameter $\mathcal O_{\rm dimer}(3)$. It means that the stability of string order is dominated by the strength of valence bond with $\delta=3$. In other words, a Haldane state is produced as a structure where all the neighboring effective spin-$1$ sites are bridged by the $\delta=3$ valence bonds. Even though it is a Haldane state, the maximum value $\mathcal O_\mathrm{string}^z\sim0.065$ is much smaller than $\mathcal O_\mathrm{string}^z=\frac{4}{9}\simeq0.4444$ for the {\it perfect} VBS state for the AKLT model~\cite{Affleck87} and $\mathcal O_\mathrm{string}^z\simeq0.3743$ for the $S=1$ Heisenberg chain~\cite{White93}. This small value of $\mathcal O_\mathrm{string}^z$ is interpreted as a sign of fragility of the $\mathcal D_3$-VBS state, which is however comparable with the maximum value for the $J_1$-$J_2$ chain at $J_2/|J_1| \approx 0.6$ ($\mathcal O_\mathrm{string}^z\sim0.06$)~\cite{furukawa12,clio19}.

\section{Summary}

We studied the asymmetric $S=\frac{1}{2}$ $J_1$-$J_2$ zigzag ladder, defined as two different AFM Heisenberg chains coupled by zigzag-shaped interchain FM interaction, using the DMRG-based techniques. The AFM chain with larger (smaller) interaction is referred to as apical (basal) chain.

First, a classical phase diagram was obtained by the spin-wave theory. It contains three phases: FM, commensurate, and incommensurate phases. It offers the possibility of commensurate-incommensurate phase transition by tuning the ratio of AFM interaction of the apical and basal chains in the quantum case.

Next, we revisited the ferrimagnetism in the so-called delta chain as the vanishing limit of AFM interaction in the apical chain. The ferrimagnetic state is characterized by total spin $S_{\rm tot}=L/4$. By carefully examining the long-range spin-spin correlation functions and low-energy excitations, we pointed out that the origin of ferrimagnetic state is order by disorder without geometrical symmetry breaking but with a global spin-rotation-symmetry breaking. Accordingly, the system can gain energy from the FM interaction between the polarized apical and basal spins. So to speak, FM fluctuations play an essential role to lower the ground state energy against the magnetic frustration. This is a rare type of order by disorder. And yet the basal chain is essentially a critical AFM Heisenberg chain as a TLL and its polarization is rather ill-conditioned as a state of chain itself. In this regards, one could interpret this to mean that the ferrimagnetic order competes with a quasi-long-range AFM order of TLL.

Then, we examined how the ferrimagnetic state is affected by AFM interaction of the apical chain, which is controlled by $\gamma$. We found that the ferrimagnetic state with $S_{\rm tot}=L/4$ is maintained up to a finite value of $\gamma$; and with further increasing $\gamma$ the system goes into spiral singlet ($S_{\rm tot}=0$) phase at a certain amount of $\gamma$. Of particular interest is the appearance of another ferrimagnetic phase characterized by $0<S_{\rm tot}<L/4$ between the $S_{\rm tot}=L/4$ ferrimagnetic and $S_{\rm tot}=0$ phases. Such a `halfway' magnetization, which is generally prohibited by the MLM theorem, is now allowed under the competition between ferrimagnetic and quasi-long-range order of TLL.

Finally, the detailed magnetic properties of spiral singlet phase was investigated. With evaluating various dimerization order parameters, we confirmed that the $S_{\rm tot}=0$ region is covered by two kinds of VBS phases depending on the parameter values. One is the $\mathcal D_2$-VBS phase where the valence bond is formed along the apical and basal chains; the other is the $\mathcal D_3$-VBS phase where the valence bond is formed between the apical and basal spins. Interestingly, a hidden topological long-range string order exists in the whole region of VBS phases and its strength scales only to the stability of $\mathcal D_3$-VBS state.

We summarized the above numerical results as a ground-state phase diagram, which contains a variety of frustration-induced ordered phases. Especially, it is quite curious that a simple frustrated system, like the asymmetric zigzag ladder, exhibits very different kinds of order by disorder -- associated with geometrical symmetry and global spin-rotation symmetry.

\section*{Acknowledgements}

We thank S. Ejima for fruitful discussions and U. Nitzsche for technical assistance. This work was supported by Grants-in-Aid for Scientific Research from JSPS (Projects Nos. JP17K05530 and JP19J10805) and by SFB 1143 of the Deutsche Forschungsgemeinschaft (project-id 247310070). The iDMRG simulations were performed using the ITensor library~\cite{ITensor}.

\appendix

\section{Numerical confirmation of ferromagnetic critical point}

\begin{figure}[tbh]
  \includegraphics[width=0.7\columnwidth]{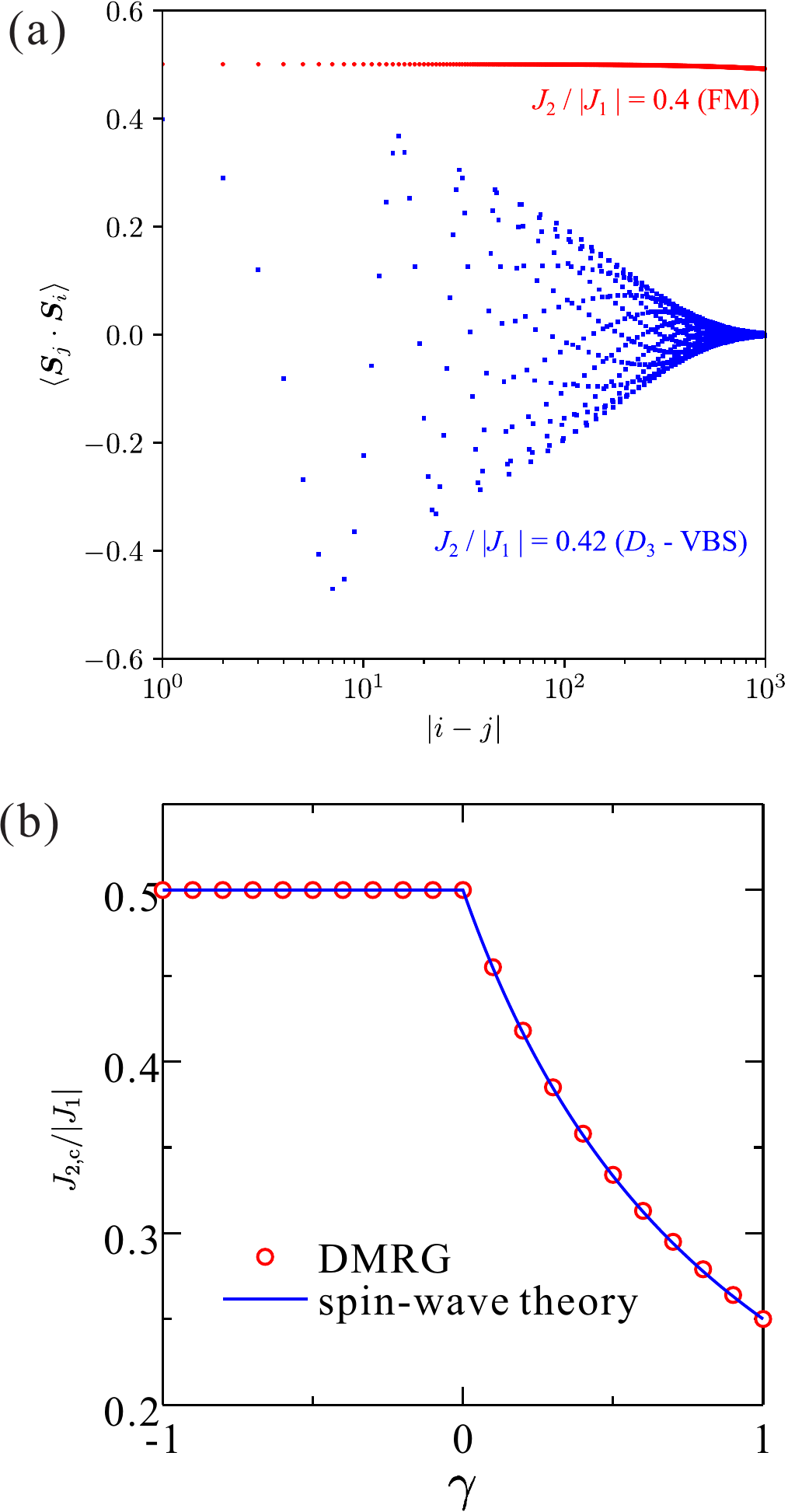}
  \caption{(a) iDMRG results for spin-spin correlation function $\langle \bm{S}_i\cdot\bm{S}_j \rangle$ as a function of distance $|i-j|$ with $\gamma=0.2$ fixed. (b) Comparison of the $\gamma$-dependence of FM critical points $J_{2,{\rm c}}/|J_1|$ estimated by DMRG and spin-wave theory.
  }
  \label{fig_SS_FM}
\end{figure}

In the main text, we have obtained the $\gamma$-dependence of FM critical point by classical SWT. It is given as
\begin{align}
	\frac{J_{2,{\rm c}}}{|J_1|}=\frac{1}{2(1+\gamma)}
	\label{FM_QCP_appendix1}
\end{align}
for $\gamma>0$. This derivation can be easily extended to negative $\gamma$. Then, when $\gamma<0$ we obtain the FM critical point as
\begin{align}
	\frac{J_{2,{\rm c}}}{|J_1|}=\frac{1}{2},
	\label{FM_QCP_appendix2}
\end{align}
which is independent of $\gamma$.

Since the quantum fluctuations vanish at the FM critical point, Eqs.~\eqref{FM_QCP_appendix1} and \eqref{FM_QCP_appendix2} should be exact. We here confirm them numerically. A simplest way is to check the presence or absence of long-range FM order. In Fig.~\ref{fig_SS_FM}(a) we show iDMRG result for the spin-spin correlation function $\langle \bm{S}_i\cdot\bm{S}_j \rangle$ at $\gamma=0.2$ as a function of distance $|i-j|$. For convenience, the site indices $i$ and $j$ are taken along the $J_1$ zigzag chain. At $J_2/|J_1|=0.4$ we can clearly see an FM long-range order with $\langle \bm{S}_i\cdot\bm{S}_j \rangle=\frac{1}{4}$ for any pair of $i$ and $j$ at the large distance. However, the FM long-range order seems to be already destroyed at only a slightly larger $J_2/|J_1|=0.42$. In fact, this iDMRG result is consistent with the SWT estimation $J_{2,{\rm c}}/|J_1|=0.417$ for $\gamma=0.2$. Additionally, since the system is in the $\mathcal D_3$-VBS state at $J_2/|J_1|=0.42$, an exponential decay of $\langle \bm{S}_i\cdot\bm{S}_j \rangle$ is thus expected.

Furthermore, we also numerically estimate the FM critical point for $\gamma<0$. In Fig.~\ref{fig_SS_FM}(b) we compare the $\gamma$-dependence of $J_{2,{\rm c}}/|J_1|$ values estimated by DMRG and SWT. We can find a good agreement between them. The reason why the FM critical point does not depend on $\gamma$ is as follows: For any negative $\gamma$ the apical chain is fully polarized and this FM order affects the basal chain as an external field via the interchain coupling $J_1$. Therefore, the FM phase transition boils down to the question of a competition between FM interchain interaction $J_1$ and AFM interaction $J_2$ in the basal chain, namely, independent of $\gamma$. Even intuitively, Eq.~\eqref{FM_QCP_appendix2} can be obtained by comparing the numbers of $J_1$ and $J_2$ bonds.

\section{Artificial enhancement of spin-singlet state with short chains}
\begin{figure}[tbh]
  \includegraphics[width=0.6\columnwidth]{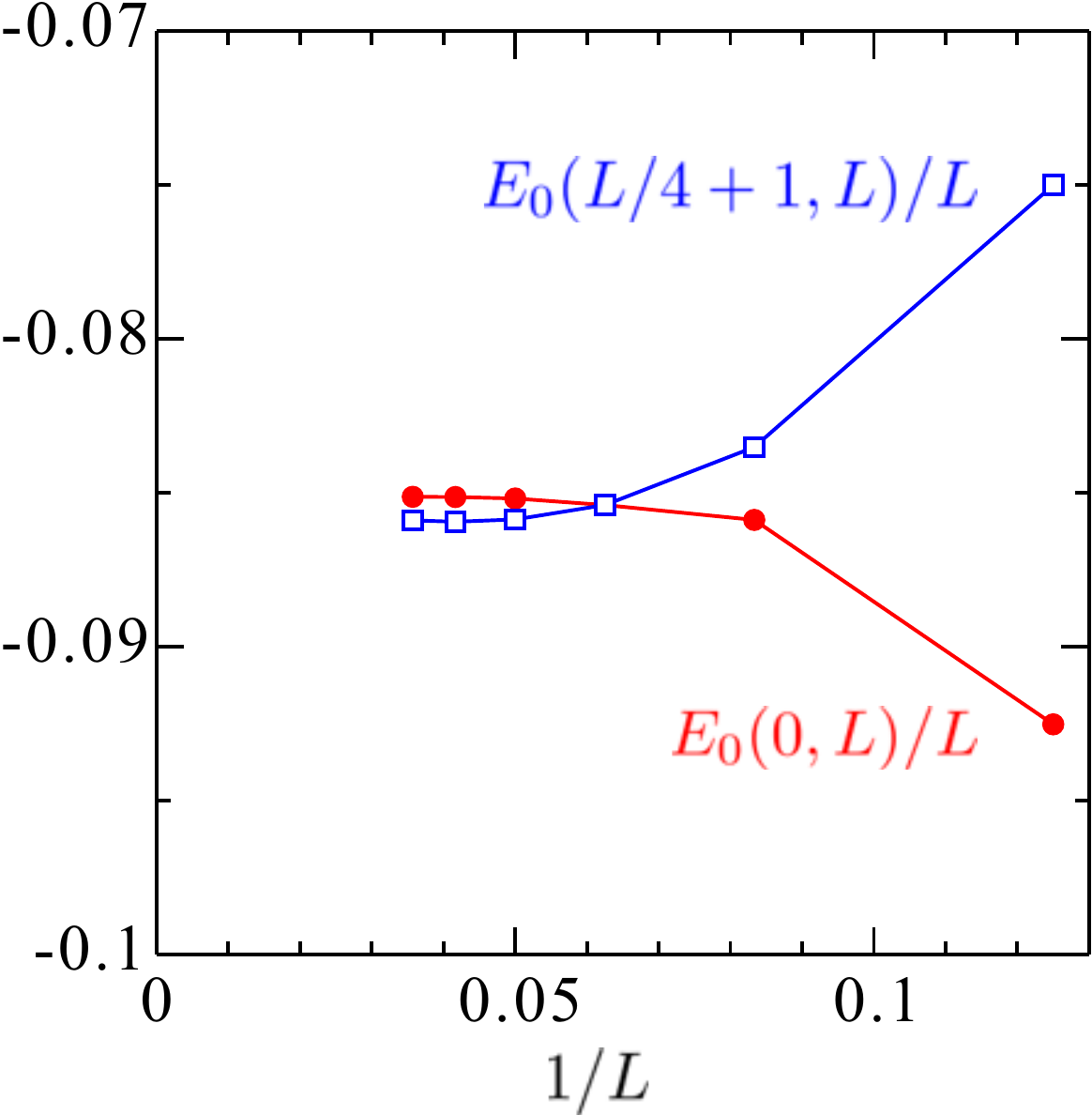}
  \caption{
Energies per site for lowest-lying spin-singlet and ferrimagnetic states at $J_1=-1$ and $J_2=0.8$ as a function of inverse system size. The PBC is used.
  }
  \label{fig_totalS_appendix}
\end{figure}

In Fig.~\ref{S_delta_chain}(b) of the main text, the total spin per site $S_{\rm tot}/L$ is plotted as a function of $1/L$ under PBC. With decreasing $1/L$, we see a drastic change of the total spin from $S_{\rm tot}/L=0$ to $S_{\rm tot}/L=1/4+1/L$ at some value of $1/L$. This is the consequence of a typical finite-size effect: If the system size is sufficiently small, the basal chain forms a short AFM ring under PBC and a plaquette singlet is highly stabilized. On the other hand, the ferrimagnetic state may be readily prevented due to the screening of basal spins. This finite-size effect could be eliminated by increasing the system size because the plaquette singlet is rapidly destabilized for larger systems. To confirm it, we compare the energies for spin-singlet and ferrimagnetic states in Fig.~\ref{fig_totalS_appendix}. When the system size is small, the energy for spin-singlet state is much lower than that for ferrimagnetic state. They approach with increasing system size and cross at some value of $L(\equiv L_{\rm c})$. Roughly speaking, one could interpret this to mean that the energy gain from ferrimagnetic formation becomes larger than that from the plaquette singlet formation of basal chain around the critical system size $L=L_{\rm c}$. Therefore, we take $S_{\rm tot}/L=1/4+1/L$ for the finite-size scaling and obtain $S_{\rm tot}/L=1/4$ in the thermodynamic limit $L\to\infty$.

\section{Tomonaga-Luttinger-liquid behavior of basal chain in full ferrimagnetic phase}

\begin{figure}[tbh]
  \includegraphics[width=0.7\columnwidth]{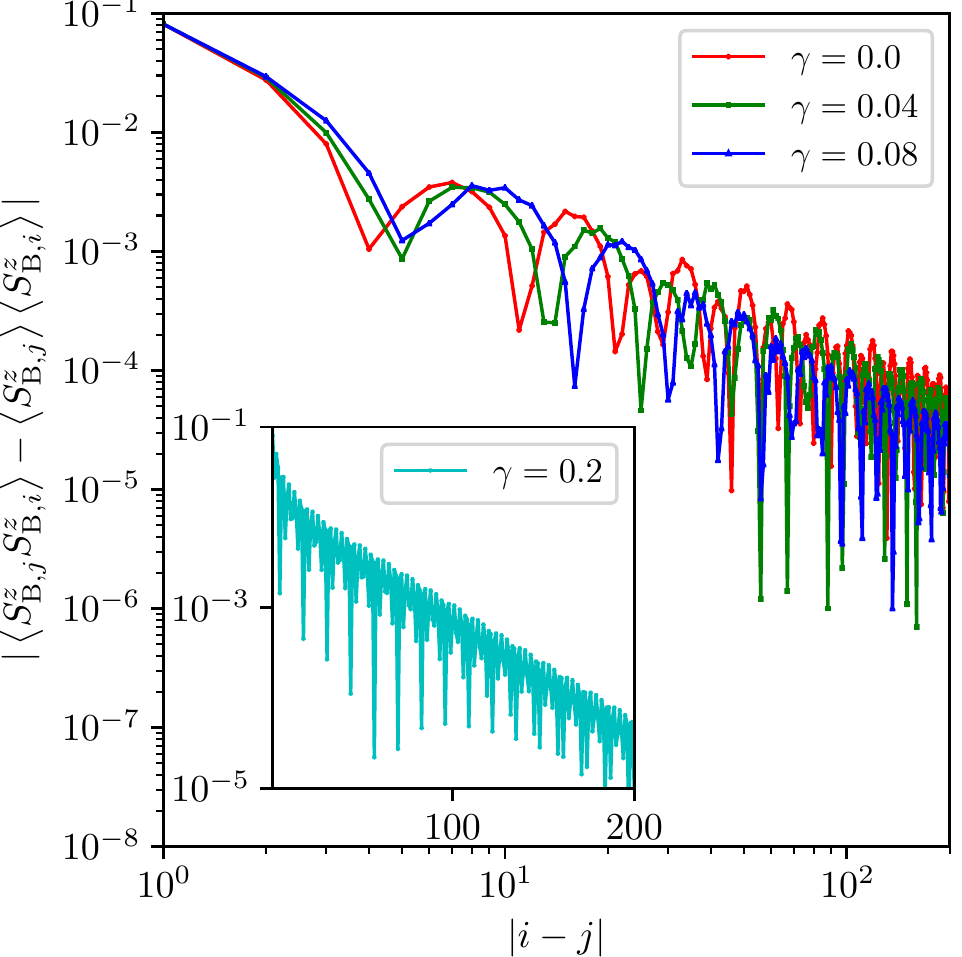}
  \caption{Spin-spin correlation function as a function of distance $|i-j|$ with $J_2/|J_1|=0.6$.
  }
  \label{fig_SzSz_appendix}
\end{figure}

As mentioned in the main text, the basal chain in the ferrimagnetic states should be a TLL because it only allows the system to have an incommensurate `halfway' magnetization other than the commensurate one. We can check it by looking at the spin-spin correlation in the basal chain. We plot iDMRG results of the spin-spin correlations $|\langle S^z_{{\rm B},i} S^z_{{\rm B},j} \rangle-\langle S^z_{{\rm B},i} \rangle\langle S^z_{{\rm B},j} \rangle|$ in the FF state are plotted as a function of distance $|i-j|$ in Fig.~\ref{fig_SzSz_appendix}. The $z$-component of total spin is fixed at $S_{\rm tot}^z=L/4$. We can see a power-law decay with distance, i.e., $|\langle S^z_{{\rm B},i} S^z_{{\rm B},j} \rangle-\langle S^z_{{\rm B},i} \rangle\langle S^z_{{\rm B},j} \rangle| \sim 1/|i-j|$, as a signature of TLL. For comfirmation, we also plot the spin-spin correlation for the $\mathcal D_3$-VBS state in the inset of Fig.~\ref{fig_SzSz_appendix}. It clearly exhibits an exponential decay indicating a gapped VBS state.

\section{Dimerization order with translation symmetry breaking}

\begin{figure}[tbh]
  \includegraphics[width=1.0\columnwidth]{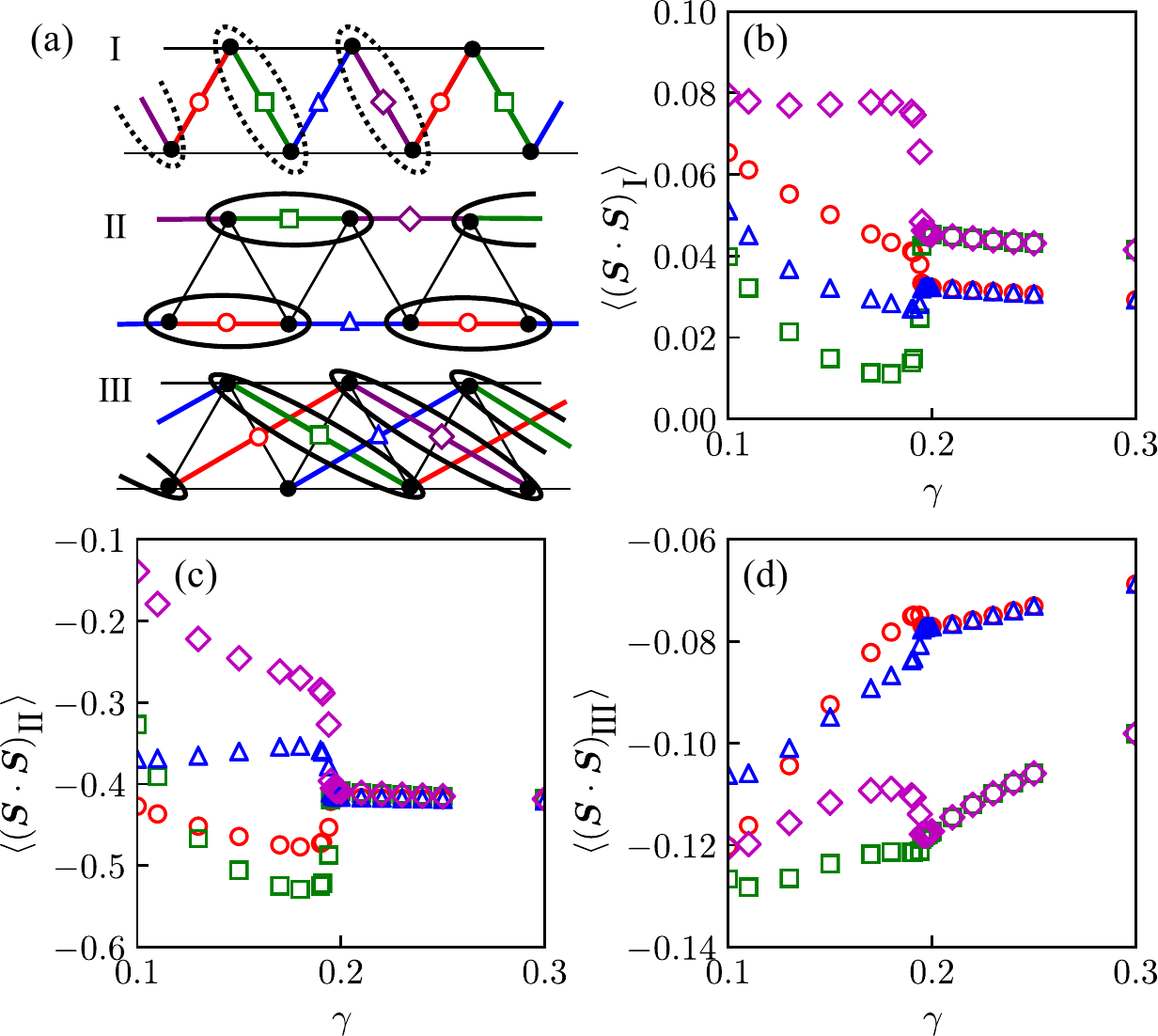}
  \caption{(a) Schematic pictures of possible dimerization order. The pictures I-III are related to the dimerization order parameter with $\delta=1-3$, respectively. A solid (dotted) ellipse denotes a spin-singlet (spin-triplet) dimer. The symbols on bond correspond to those used in the following plots (b)-(d). (b) Spin-spin correlation function for the bonds marked with the symbols in I. (c)(d) Similar plots to (a) for II and III.
  }
  \label{fig_SdS_appendix}
\end{figure}

In the main text, we have estimated the dimerization order parameters defined as the difference between spin-spin correlations for the `neighboring' bonds. To obtain further insight into the dimerization structure, we here look at the bare spin-spin correlations. In Fig.~\ref{fig_SdS_appendix}(b)-(d) we plot the spin-spin correlations for various pairs of spins as a function of $\gamma$. Each symbol corresponds to a bond marked by the same symbol in Fig.~\ref{fig_SdS_appendix}(a).

At $\gamma \gtrsim 0.195$, the valence bond structure is relatively simple. The correlation $\langle (\bm{S}_i\cdot\bm{S}_j)_{\rm I} \rangle$ has two different values and the difference between them provides the dimerization order parameter. This is the same for $\langle (\bm{S}_i\cdot\bm{S}_j)_{\rm III} \rangle$. No splitting of the correlation $\langle (\bm{S}_i\cdot\bm{S}_j)_{\rm II} \rangle$ indicates the absence of dimerization order parameter for $\delta=2$.

At $\gamma \lesssim 0.195$, the situation is a bit more complicated. The translation symmetry is broken due to the presence of dimerization order for $\delta=2$ and all the three dimerization orders coexist. As a result, the two values of $\langle (\bm{S}_i\cdot\bm{S}_j)_{\rm I} \rangle$ and $\langle (\bm{S}_i\cdot\bm{S}_j)_{\rm III} \rangle$ are further split into four values. Likewise, the correlation $\langle (\bm{S}_i\cdot\bm{S}_j)_{\rm II} \rangle$ is also split into four values because the strength of dimerization order can be different between the apical and basal chains. Therefore, the magnetic structure consists of a supercell with four sites in the $\mathcal D_2$-VBS phase.

\bibliography{asymj1j2}

\begin{thebibliography}{79}%
\makeatletter
\providecommand \@ifxundefined [1]{%
 \@ifx{#1\undefined}
}%
\providecommand \@ifnum [1]{%
 \ifnum #1\expandafter \@firstoftwo
 \else \expandafter \@secondoftwo
 \fi
}%
\providecommand \@ifx [1]{%
 \ifx #1\expandafter \@firstoftwo
 \else \expandafter \@secondoftwo
 \fi
}%
\providecommand \natexlab [1]{#1}%
\providecommand \enquote  [1]{``#1''}%
\providecommand \bibnamefont  [1]{#1}%
\providecommand \bibfnamefont [1]{#1}%
\providecommand \citenamefont [1]{#1}%
\providecommand \href@noop [0]{\@secondoftwo}%
\providecommand \href [0]{\begingroup \@sanitize@url \@href}%
\providecommand \@href[1]{\@@startlink{#1}\@@href}%
\providecommand \@@href[1]{\endgroup#1\@@endlink}%
\providecommand \@sanitize@url [0]{\catcode `\\12\catcode `\$12\catcode
  `\&12\catcode `\#12\catcode `\^12\catcode `\_12\catcode `\%12\relax}%
\providecommand \@@startlink[1]{}%
\providecommand \@@endlink[0]{}%
\providecommand \url  [0]{\begingroup\@sanitize@url \@url }%
\providecommand \@url [1]{\endgroup\@href {#1}{\urlprefix }}%
\providecommand \urlprefix  [0]{URL }%
\providecommand \Eprint [0]{\href }%
\providecommand \doibase [0]{https://doi.org/}%
\providecommand \selectlanguage [0]{\@gobble}%
\providecommand \bibinfo  [0]{\@secondoftwo}%
\providecommand \bibfield  [0]{\@secondoftwo}%
\providecommand \translation [1]{[#1]}%
\providecommand \BibitemOpen [0]{}%
\providecommand \bibitemStop [0]{}%
\providecommand \bibitemNoStop [0]{.\EOS\space}%
\providecommand \EOS [0]{\spacefactor3000\relax}%
\providecommand \BibitemShut  [1]{\csname bibitem#1\endcsname}%
\let\auto@bib@innerbib\@empty
\bibitem [{\citenamefont {Vasiliev}\ \emph {et~al.}(2018)\citenamefont
  {Vasiliev}, \citenamefont {Volkova}, \citenamefont {Zvereva},\ and\
  \citenamefont {Markina}}]{Vasiliev18}%
  \BibitemOpen
  \bibfield  {author} {\bibinfo {author} {\bibfnamefont {A.}~\bibnamefont
  {Vasiliev}}, \bibinfo {author} {\bibfnamefont {O.}~\bibnamefont {Volkova}},
  \bibinfo {author} {\bibfnamefont {E.}~\bibnamefont {Zvereva}},\ and\ \bibinfo
  {author} {\bibfnamefont {M.}~\bibnamefont {Markina}},\ }\bibfield  {title}
  {\bibinfo {title} {Milestones of low-{D} quantum magnetism},\ }\href@noop {}
  {\bibfield  {journal} {\bibinfo  {journal} {npj Quantum Materials}\ }\textbf
  {\bibinfo {volume} {3}},\ \bibinfo {pages} {18} (\bibinfo {year}
  {2018})}\BibitemShut {NoStop}%
\bibitem [{\citenamefont {Villain}\ \emph {et~al.}(1980)\citenamefont
  {Villain}, \citenamefont {Bidaux}, \citenamefont {Carton},\ and\
  \citenamefont {Conte}}]{Villain80}%
  \BibitemOpen
  \bibfield  {author} {\bibinfo {author} {\bibfnamefont {J.}~\bibnamefont
  {Villain}}, \bibinfo {author} {\bibfnamefont {R.}~\bibnamefont {Bidaux}},
  \bibinfo {author} {\bibfnamefont {J.-P.}\ \bibnamefont {Carton}},\ and\
  \bibinfo {author} {\bibfnamefont {R.}~\bibnamefont {Conte}},\ }\bibfield
  {title} {\bibinfo {title} {Order as an effect of disorder},\ }\href
  {https://doi.org/10.1051/jphys:0198000410110126300} {\bibfield  {journal}
  {\bibinfo  {journal} {J. Phys. (France)}\ }\textbf {\bibinfo {volume} {41}},\
  \bibinfo {pages} {1263} (\bibinfo {year} {1980})}\BibitemShut {NoStop}%
\bibitem [{\citenamefont {Oshikawa}\ and\ \citenamefont
  {Senthil}(2006)}]{Oshikawa06}%
  \BibitemOpen
  \bibfield  {author} {\bibinfo {author} {\bibfnamefont {M.}~\bibnamefont
  {Oshikawa}}\ and\ \bibinfo {author} {\bibfnamefont {T.}~\bibnamefont
  {Senthil}},\ }\bibfield  {title} {\bibinfo {title} {Fractionalization,
  topological order, and quasiparticle statistics},\ }\href
  {https://doi.org/10.1103/PhysRevLett.96.060601} {\bibfield  {journal}
  {\bibinfo  {journal} {Phys. Rev. Lett.}\ }\textbf {\bibinfo {volume} {96}},\
  \bibinfo {pages} {060601} (\bibinfo {year} {2006})}\BibitemShut {NoStop}%
\bibitem [{\citenamefont {Jiang}\ \emph {et~al.}(2012)\citenamefont {Jiang},
  \citenamefont {Wang},\ and\ \citenamefont {Balents}}]{Jiang12}%
  \BibitemOpen
  \bibfield  {author} {\bibinfo {author} {\bibfnamefont {H.-C.}\ \bibnamefont
  {Jiang}}, \bibinfo {author} {\bibfnamefont {Z.}~\bibnamefont {Wang}},\ and\
  \bibinfo {author} {\bibfnamefont {L.}~\bibnamefont {Balents}},\ }\bibfield
  {title} {\bibinfo {title} {Identifying topological order by entanglement
  entropy},\ }\href@noop {} {\bibfield  {journal} {\bibinfo  {journal} {Nature
  Physics}\ }\textbf {\bibinfo {volume} {8}},\ \bibinfo {pages} {902} (\bibinfo
  {year} {2012})}\BibitemShut {NoStop}%
\bibitem [{\citenamefont {Messio}\ \emph {et~al.}(2012)\citenamefont {Messio},
  \citenamefont {Bernu},\ and\ \citenamefont {Lhuillier}}]{Messio12}%
  \BibitemOpen
  \bibfield  {author} {\bibinfo {author} {\bibfnamefont {L.}~\bibnamefont
  {Messio}}, \bibinfo {author} {\bibfnamefont {B.}~\bibnamefont {Bernu}},\ and\
  \bibinfo {author} {\bibfnamefont {C.}~\bibnamefont {Lhuillier}},\ }\bibfield
  {title} {\bibinfo {title} {Kagome antiferromagnet: A chiral topological spin
  liquid?},\ }\href {https://doi.org/10.1103/PhysRevLett.108.207204} {\bibfield
   {journal} {\bibinfo  {journal} {Phys. Rev. Lett.}\ }\textbf {\bibinfo
  {volume} {108}},\ \bibinfo {pages} {207204} (\bibinfo {year}
  {2012})}\BibitemShut {NoStop}%
\bibitem [{\citenamefont {Bauer}\ \emph {et~al.}(2014)\citenamefont {Bauer},
  \citenamefont {Cincio}, \citenamefont {Keller}, \citenamefont {Dolfi},
  \citenamefont {Vidal}, \citenamefont {Trebst},\ and\ \citenamefont
  {Ludwig}}]{Bauer14}%
  \BibitemOpen
  \bibfield  {author} {\bibinfo {author} {\bibfnamefont {B.}~\bibnamefont
  {Bauer}}, \bibinfo {author} {\bibfnamefont {L.}~\bibnamefont {Cincio}},
  \bibinfo {author} {\bibfnamefont {B.}~\bibnamefont {Keller}}, \bibinfo
  {author} {\bibfnamefont {M.}~\bibnamefont {Dolfi}}, \bibinfo {author}
  {\bibfnamefont {G.}~\bibnamefont {Vidal}}, \bibinfo {author} {\bibfnamefont
  {S.}~\bibnamefont {Trebst}},\ and\ \bibinfo {author} {\bibfnamefont
  {A.}~\bibnamefont {Ludwig}},\ }\bibfield  {title} {\bibinfo {title} {Chiral
  spin liquid and emergent anyons in a kagome lattice mott insulator},\ }\href
  {https://doi.org/10.1038/ncomms6137} {\bibfield  {journal} {\bibinfo
  {journal} {Nat. Commun.}\ }\textbf {\bibinfo {volume} {5}},\ \bibinfo {pages}
  {5137} (\bibinfo {year} {2014})}\BibitemShut {NoStop}%
\bibitem [{\citenamefont {Lao}\ \emph {et~al.}(2018)\citenamefont {Lao},
  \citenamefont {Caravelli}, \citenamefont {Sheikh}, \citenamefont {Sklenar},
  \citenamefont {Gardeazabal}, \citenamefont {Watts}, \citenamefont {Albrecht},
  \citenamefont {Scholl}, \citenamefont {Dahmen}, \citenamefont {Nisoli} \emph
  {et~al.}}]{Lao18}%
  \BibitemOpen
  \bibfield  {author} {\bibinfo {author} {\bibfnamefont {Y.}~\bibnamefont
  {Lao}}, \bibinfo {author} {\bibfnamefont {F.}~\bibnamefont {Caravelli}},
  \bibinfo {author} {\bibfnamefont {M.}~\bibnamefont {Sheikh}}, \bibinfo
  {author} {\bibfnamefont {J.}~\bibnamefont {Sklenar}}, \bibinfo {author}
  {\bibfnamefont {D.}~\bibnamefont {Gardeazabal}}, \bibinfo {author}
  {\bibfnamefont {J.~D.}\ \bibnamefont {Watts}}, \bibinfo {author}
  {\bibfnamefont {A.~M.}\ \bibnamefont {Albrecht}}, \bibinfo {author}
  {\bibfnamefont {A.}~\bibnamefont {Scholl}}, \bibinfo {author} {\bibfnamefont
  {K.}~\bibnamefont {Dahmen}}, \bibinfo {author} {\bibfnamefont
  {C.}~\bibnamefont {Nisoli}}, \emph {et~al.},\ }\bibfield  {title} {\bibinfo
  {title} {Classical topological order in the kinetics of artificial spin
  ice},\ }\href@noop {} {\bibfield  {journal} {\bibinfo  {journal} {Nature
  Physics}\ }\textbf {\bibinfo {volume} {14}},\ \bibinfo {pages} {723}
  (\bibinfo {year} {2018})}\BibitemShut {NoStop}%
\bibitem [{\citenamefont {Majumdar}\ and\ \citenamefont
  {Ghosh}(1969)}]{Majumdar69}%
  \BibitemOpen
  \bibfield  {author} {\bibinfo {author} {\bibfnamefont {C.~K.}\ \bibnamefont
  {Majumdar}}\ and\ \bibinfo {author} {\bibfnamefont {D.~K.}\ \bibnamefont
  {Ghosh}},\ }\bibfield  {title} {\bibinfo {title} {On
  next‐nearest‐neighbor interaction in linear chain. i},\ }\href
  {https://doi.org/10.1063/1.1664978} {\bibfield  {journal} {\bibinfo
  {journal} {Journal of Mathematical Physics}\ }\textbf {\bibinfo {volume}
  {10}},\ \bibinfo {pages} {1388} (\bibinfo {year} {1969})}\BibitemShut
  {NoStop}%
\bibitem [{\citenamefont {Okamoto}\ and\ \citenamefont
  {Nomura}(1992)}]{Okamoto92}%
  \BibitemOpen
  \bibfield  {author} {\bibinfo {author} {\bibfnamefont {K.}~\bibnamefont
  {Okamoto}}\ and\ \bibinfo {author} {\bibfnamefont {K.}~\bibnamefont
  {Nomura}},\ }\bibfield  {title} {\bibinfo {title} {Fluid-dimer critical point
  in {$S=1/2$} antiferromagnetic {H}eisenberg chain with next nearest neighbor
  interactions},\ }\href
  {https://doi.org/https://doi.org/10.1016/0375-9601(92)90823-5} {\bibfield
  {journal} {\bibinfo  {journal} {Physics Letters A}\ }\textbf {\bibinfo
  {volume} {169}},\ \bibinfo {pages} {433 } (\bibinfo {year}
  {1992})}\BibitemShut {NoStop}%
\bibitem [{\citenamefont {Eggert}(1996)}]{Eggert96}%
  \BibitemOpen
  \bibfield  {author} {\bibinfo {author} {\bibfnamefont {S.}~\bibnamefont
  {Eggert}},\ }\bibfield  {title} {\bibinfo {title} {Numerical evidence for
  multiplicative logarithmic corrections from marginal operators},\ }\href
  {https://doi.org/10.1103/PhysRevB.54.R9612} {\bibfield  {journal} {\bibinfo
  {journal} {Phys. Rev. B}\ }\textbf {\bibinfo {volume} {54}},\ \bibinfo
  {pages} {R9612} (\bibinfo {year} {1996})}\BibitemShut {NoStop}%
\bibitem [{\citenamefont {Affleck}\ \emph {et~al.}(1987)\citenamefont
  {Affleck}, \citenamefont {Kennedy}, \citenamefont {Lieb},\ and\ \citenamefont
  {Tasaki}}]{Affleck87}%
  \BibitemOpen
  \bibfield  {author} {\bibinfo {author} {\bibfnamefont {I.}~\bibnamefont
  {Affleck}}, \bibinfo {author} {\bibfnamefont {T.}~\bibnamefont {Kennedy}},
  \bibinfo {author} {\bibfnamefont {E.~H.}\ \bibnamefont {Lieb}},\ and\
  \bibinfo {author} {\bibfnamefont {H.}~\bibnamefont {Tasaki}},\ }\bibfield
  {title} {\bibinfo {title} {Rigorous results on valence-bond ground states in
  antiferromagnets},\ }\href {https://doi.org/10.1103/PhysRevLett.59.799}
  {\bibfield  {journal} {\bibinfo  {journal} {Phys. Rev. Lett.}\ }\textbf
  {\bibinfo {volume} {59}},\ \bibinfo {pages} {799} (\bibinfo {year}
  {1987})}\BibitemShut {NoStop}%
\bibitem [{\citenamefont {Pollmann}\ \emph {et~al.}(2012)\citenamefont
  {Pollmann}, \citenamefont {Berg}, \citenamefont {Turner},\ and\ \citenamefont
  {Oshikawa}}]{Pollmann12}%
  \BibitemOpen
  \bibfield  {author} {\bibinfo {author} {\bibfnamefont {F.}~\bibnamefont
  {Pollmann}}, \bibinfo {author} {\bibfnamefont {E.}~\bibnamefont {Berg}},
  \bibinfo {author} {\bibfnamefont {A.}~\bibnamefont {Turner}},\ and\ \bibinfo
  {author} {\bibfnamefont {M.}~\bibnamefont {Oshikawa}},\ }\bibfield  {title}
  {\bibinfo {title} {Symmetry protection of topological phases in
  one-dimensional quantum spin systems},\ }\href
  {https://doi.org/10.1103/PhysRevB.85.075125} {\bibfield  {journal} {\bibinfo
  {journal} {Phys. Rev. B}\ }\textbf {\bibinfo {volume} {85}},\ \bibinfo
  {pages} {075125} (\bibinfo {year} {2012})}\BibitemShut {NoStop}%
\bibitem [{\citenamefont {de~Graaf}\ \emph {et~al.}(2002)\citenamefont
  {de~Graaf}, \citenamefont {de~P.~R.~Moreira}, \citenamefont {Illas},
  \citenamefont {Iglesias},\ and\ \citenamefont {Labarta}}]{deGraaf2002}%
  \BibitemOpen
  \bibfield  {author} {\bibinfo {author} {\bibfnamefont {C.}~\bibnamefont
  {de~Graaf}}, \bibinfo {author} {\bibfnamefont {I.}~\bibnamefont
  {de~P.~R.~Moreira}}, \bibinfo {author} {\bibfnamefont {F.}~\bibnamefont
  {Illas}}, \bibinfo {author} {\bibfnamefont {{\`{O}}.}~\bibnamefont
  {Iglesias}},\ and\ \bibinfo {author} {\bibfnamefont {A.}~\bibnamefont
  {Labarta}},\ }\bibfield  {title} {\bibinfo {title} {Magnetic structure of
  {L}i$_{2}${C}u{O}$_{2}$:{\hspace{1em}}fromab initiocalculations to
  macroscopic simulations},\ }\href
  {https://doi.org/10.1103/physrevb.66.014448} {\bibfield  {journal} {\bibinfo
  {journal} {Phys. Rev. B}\ }\textbf {\bibinfo {volume} {66}},\ \bibinfo
  {pages} {014448} (\bibinfo {year} {2002})}\BibitemShut {NoStop}%
\bibitem [{\citenamefont {Grafe}\ \emph {et~al.}(2017)\citenamefont {Grafe},
  \citenamefont {Nishimoto}, \citenamefont {Iakovleva}, \citenamefont
  {Vavilova}, \citenamefont {Spillecke}, \citenamefont {Alfonsov},
  \citenamefont {Sturza}, \citenamefont {Wurmehl}, \citenamefont {Nojiri},
  \citenamefont {Rosner}, \citenamefont {Richter}, \citenamefont
  {R\"{o}{\ss}ler}, \citenamefont {Drechsler}, \citenamefont {Kataev},\ and\
  \citenamefont {B\"{u}chner}}]{Grafe17}%
  \BibitemOpen
  \bibfield  {author} {\bibinfo {author} {\bibfnamefont {H.-J.}\ \bibnamefont
  {Grafe}}, \bibinfo {author} {\bibfnamefont {S.}~\bibnamefont {Nishimoto}},
  \bibinfo {author} {\bibfnamefont {M.}~\bibnamefont {Iakovleva}}, \bibinfo
  {author} {\bibfnamefont {E.}~\bibnamefont {Vavilova}}, \bibinfo {author}
  {\bibfnamefont {L.}~\bibnamefont {Spillecke}}, \bibinfo {author}
  {\bibfnamefont {A.}~\bibnamefont {Alfonsov}}, \bibinfo {author}
  {\bibfnamefont {M.-I.}\ \bibnamefont {Sturza}}, \bibinfo {author}
  {\bibfnamefont {S.}~\bibnamefont {Wurmehl}}, \bibinfo {author} {\bibfnamefont
  {H.}~\bibnamefont {Nojiri}}, \bibinfo {author} {\bibfnamefont
  {H.}~\bibnamefont {Rosner}}, \bibinfo {author} {\bibfnamefont
  {J.}~\bibnamefont {Richter}}, \bibinfo {author} {\bibfnamefont {U.~K.}\
  \bibnamefont {R\"{o}{\ss}ler}}, \bibinfo {author} {\bibfnamefont {S.-L.}\
  \bibnamefont {Drechsler}}, \bibinfo {author} {\bibfnamefont {V.}~\bibnamefont
  {Kataev}},\ and\ \bibinfo {author} {\bibfnamefont {B.}~\bibnamefont
  {B\"{u}chner}},\ }\bibfield  {title} {\bibinfo {title} {Signatures of a
  magnetic field-induced unconventional nematic liquid in the frustrated and
  anisotropic spin-chain cuprate {L}i{C}u{S}b{O}$_4$},\ }\href
  {https://doi.org/10.1038/s41598-017-06525-0} {\bibfield  {journal} {\bibinfo
  {journal} {Sci. Rep.}\ }\textbf {\bibinfo {volume} {7}},\ \bibinfo {pages}
  {6720} (\bibinfo {year} {2017})}\BibitemShut {NoStop}%
\bibitem [{\citenamefont {Orlova}\ \emph {et~al.}(2017)\citenamefont {Orlova},
  \citenamefont {Green}, \citenamefont {Law}, \citenamefont {Gorbunov},
  \citenamefont {Chanda}, \citenamefont {Kr\"amer}, \citenamefont
  {Horvati\ifmmode~\acute{c}\else \'{c}\fi{}}, \citenamefont {Kremer},
  \citenamefont {Wosnitza},\ and\ \citenamefont {Rikken}}]{Orlova17}%
  \BibitemOpen
  \bibfield  {author} {\bibinfo {author} {\bibfnamefont {A.}~\bibnamefont
  {Orlova}}, \bibinfo {author} {\bibfnamefont {E.~L.}\ \bibnamefont {Green}},
  \bibinfo {author} {\bibfnamefont {J.~M.}\ \bibnamefont {Law}}, \bibinfo
  {author} {\bibfnamefont {D.~I.}\ \bibnamefont {Gorbunov}}, \bibinfo {author}
  {\bibfnamefont {G.}~\bibnamefont {Chanda}}, \bibinfo {author} {\bibfnamefont
  {S.}~\bibnamefont {Kr\"amer}}, \bibinfo {author} {\bibfnamefont
  {M.}~\bibnamefont {Horvati\ifmmode~\acute{c}\else \'{c}\fi{}}}, \bibinfo
  {author} {\bibfnamefont {R.~K.}\ \bibnamefont {Kremer}}, \bibinfo {author}
  {\bibfnamefont {J.}~\bibnamefont {Wosnitza}},\ and\ \bibinfo {author}
  {\bibfnamefont {G.~L. J.~A.}\ \bibnamefont {Rikken}},\ }\bibfield  {title}
  {\bibinfo {title} {Nuclear magnetic resonance signature of the spin-nematic
  phase in ${\mathrm{licuvo}}_{4}$ at high magnetic fields},\ }\href
  {https://doi.org/10.1103/PhysRevLett.118.247201} {\bibfield  {journal}
  {\bibinfo  {journal} {Phys. Rev. Lett.}\ }\textbf {\bibinfo {volume} {118}},\
  \bibinfo {pages} {247201} (\bibinfo {year} {2017})}\BibitemShut {NoStop}%
\bibitem [{\citenamefont {Drechsler}\ \emph {et~al.}(2007)\citenamefont
  {Drechsler}, \citenamefont {Volkova}, \citenamefont {Vasiliev}, \citenamefont
  {Tristan}, \citenamefont {Richter}, \citenamefont {Schmitt}, \citenamefont
  {Rosner}, \citenamefont {M\'alek}, \citenamefont {Klingeler}, \citenamefont
  {Zvyagin},\ and\ \citenamefont {B\"uchner}}]{Drechsler07}%
  \BibitemOpen
  \bibfield  {author} {\bibinfo {author} {\bibfnamefont {S.-L.}\ \bibnamefont
  {Drechsler}}, \bibinfo {author} {\bibfnamefont {O.}~\bibnamefont {Volkova}},
  \bibinfo {author} {\bibfnamefont {A.~N.}\ \bibnamefont {Vasiliev}}, \bibinfo
  {author} {\bibfnamefont {N.}~\bibnamefont {Tristan}}, \bibinfo {author}
  {\bibfnamefont {J.}~\bibnamefont {Richter}}, \bibinfo {author} {\bibfnamefont
  {M.}~\bibnamefont {Schmitt}}, \bibinfo {author} {\bibfnamefont
  {H.}~\bibnamefont {Rosner}}, \bibinfo {author} {\bibfnamefont
  {J.}~\bibnamefont {M\'alek}}, \bibinfo {author} {\bibfnamefont
  {R.}~\bibnamefont {Klingeler}}, \bibinfo {author} {\bibfnamefont {A.~A.}\
  \bibnamefont {Zvyagin}},\ and\ \bibinfo {author} {\bibfnamefont
  {B.}~\bibnamefont {B\"uchner}},\ }\bibfield  {title} {\bibinfo {title}
  {Frustrated cuprate route from antiferromagnetic to ferromagnetic
  spin-$\frac{1}{2}$ {H}eisenberg chains: {L}i$_{2}${Z}r{C}u{O}$_{4}$ as a
  missing link near the quantum critical point},\ }\href
  {https://doi.org/10.1103/PhysRevLett.98.077202} {\bibfield  {journal}
  {\bibinfo  {journal} {Phys. Rev. Lett.}\ }\textbf {\bibinfo {volume} {98}},\
  \bibinfo {pages} {077202} (\bibinfo {year} {2007})}\BibitemShut {NoStop}%
\bibitem [{\citenamefont {Ueda}\ \emph {et~al.}(2018)\citenamefont {Ueda},
  \citenamefont {Onoda}, \citenamefont {Yamaguchi}, \citenamefont {Kimura},
  \citenamefont {Yoshizawa}, \citenamefont {Morioka}, \citenamefont {Hagiwara},
  \citenamefont {Hagihala}, \citenamefont {Soda}, \citenamefont {Masuda},
  \citenamefont {Sakakibara}, \citenamefont {Tomiyasu}, \citenamefont
  {Ohira-Kawamura}, \citenamefont {Nakajima}, \citenamefont {Kajimoto},
  \citenamefont {Nakamura}, \citenamefont {Inamura}, \citenamefont {Hase},\
  and\ \citenamefont {Yasui}}]{Ueda18}%
  \BibitemOpen
  \bibfield  {author} {\bibinfo {author} {\bibfnamefont {H.}~\bibnamefont
  {Ueda}}, \bibinfo {author} {\bibfnamefont {S.}~\bibnamefont {Onoda}},
  \bibinfo {author} {\bibfnamefont {Y.}~\bibnamefont {Yamaguchi}}, \bibinfo
  {author} {\bibfnamefont {T.}~\bibnamefont {Kimura}}, \bibinfo {author}
  {\bibfnamefont {D.}~\bibnamefont {Yoshizawa}}, \bibinfo {author}
  {\bibfnamefont {T.}~\bibnamefont {Morioka}}, \bibinfo {author} {\bibfnamefont
  {M.}~\bibnamefont {Hagiwara}}, \bibinfo {author} {\bibfnamefont
  {M.}~\bibnamefont {Hagihala}}, \bibinfo {author} {\bibfnamefont
  {M.}~\bibnamefont {Soda}}, \bibinfo {author} {\bibfnamefont {T.}~\bibnamefont
  {Masuda}}, \bibinfo {author} {\bibfnamefont {T.}~\bibnamefont {Sakakibara}},
  \bibinfo {author} {\bibfnamefont {K.}~\bibnamefont {Tomiyasu}}, \bibinfo
  {author} {\bibfnamefont {S.}~\bibnamefont {Ohira-Kawamura}}, \bibinfo
  {author} {\bibfnamefont {K.}~\bibnamefont {Nakajima}}, \bibinfo {author}
  {\bibfnamefont {R.}~\bibnamefont {Kajimoto}}, \bibinfo {author}
  {\bibfnamefont {M.}~\bibnamefont {Nakamura}}, \bibinfo {author}
  {\bibfnamefont {Y.}~\bibnamefont {Inamura}}, \bibinfo {author} {\bibfnamefont
  {M.}~\bibnamefont {Hase}},\ and\ \bibinfo {author} {\bibfnamefont
  {Y.}~\bibnamefont {Yasui}},\ }\bibfield  {title} {\bibinfo {title} {Emergent
  spin-1 {H}aldane gap and ferroelectricity in a frustrated spin-1/2 ladder},\
  }\href {http://arxiv.org/abs/1803.07081} {\bibfield  {journal} {\bibinfo
  {journal} {arXiv:1803.07081 [cond-mat]}\ } (\bibinfo {year} {2018})},\
  \bibinfo {note} {arXiv: 1803.07081}\BibitemShut {NoStop}%
\bibitem [{\citenamefont {Wolter}\ \emph {et~al.}(2012)\citenamefont {Wolter},
  \citenamefont {Lipps}, \citenamefont {Sch\"apers}, \citenamefont {Drechsler},
  \citenamefont {Nishimoto}, \citenamefont {Vogel}, \citenamefont {Kataev},
  \citenamefont {B\"uchner}, \citenamefont {Rosner}, \citenamefont {Schmitt},
  \citenamefont {Uhlarz}, \citenamefont {Skourski}, \citenamefont {Wosnitza},
  \citenamefont {S\"ullow},\ and\ \citenamefont {Rule}}]{Wolter12}%
  \BibitemOpen
  \bibfield  {author} {\bibinfo {author} {\bibfnamefont {A.~U.~B.}\
  \bibnamefont {Wolter}}, \bibinfo {author} {\bibfnamefont {F.}~\bibnamefont
  {Lipps}}, \bibinfo {author} {\bibfnamefont {M.}~\bibnamefont {Sch\"apers}},
  \bibinfo {author} {\bibfnamefont {S.-L.}\ \bibnamefont {Drechsler}}, \bibinfo
  {author} {\bibfnamefont {S.}~\bibnamefont {Nishimoto}}, \bibinfo {author}
  {\bibfnamefont {R.}~\bibnamefont {Vogel}}, \bibinfo {author} {\bibfnamefont
  {V.}~\bibnamefont {Kataev}}, \bibinfo {author} {\bibfnamefont
  {B.}~\bibnamefont {B\"uchner}}, \bibinfo {author} {\bibfnamefont
  {H.}~\bibnamefont {Rosner}}, \bibinfo {author} {\bibfnamefont
  {M.}~\bibnamefont {Schmitt}}, \bibinfo {author} {\bibfnamefont
  {M.}~\bibnamefont {Uhlarz}}, \bibinfo {author} {\bibfnamefont
  {Y.}~\bibnamefont {Skourski}}, \bibinfo {author} {\bibfnamefont
  {J.}~\bibnamefont {Wosnitza}}, \bibinfo {author} {\bibfnamefont
  {S.}~\bibnamefont {S\"ullow}},\ and\ \bibinfo {author} {\bibfnamefont
  {K.~C.}\ \bibnamefont {Rule}},\ }\bibfield  {title} {\bibinfo {title}
  {Magnetic properties and exchange integrals of the frustrated chain cuprate
  linarite {P}b{C}u{S}o$_{4}$({O}{H})$_{2}$},\ }\href
  {https://doi.org/10.1103/PhysRevB.85.014407} {\bibfield  {journal} {\bibinfo
  {journal} {Phys. Rev. B}\ }\textbf {\bibinfo {volume} {85}},\ \bibinfo
  {pages} {014407} (\bibinfo {year} {2012})}\BibitemShut {NoStop}%
\bibitem [{\citenamefont {Furukawa}\ \emph {et~al.}(2012)\citenamefont
  {Furukawa}, \citenamefont {Sato}, \citenamefont {Onoda},\ and\ \citenamefont
  {Furusaki}}]{furukawa12}%
  \BibitemOpen
  \bibfield  {author} {\bibinfo {author} {\bibfnamefont {S.}~\bibnamefont
  {Furukawa}}, \bibinfo {author} {\bibfnamefont {M.}~\bibnamefont {Sato}},
  \bibinfo {author} {\bibfnamefont {S.}~\bibnamefont {Onoda}},\ and\ \bibinfo
  {author} {\bibfnamefont {A.}~\bibnamefont {Furusaki}},\ }\bibfield  {title}
  {\bibinfo {title} {Ground-state phase diagram of a spin-$\frac{1}{2}$
  frustrated ferromagnetic {XXZ} chain: {H}aldane dimer phase and
  gapped/gapless chiral phases},\ }\href
  {https://doi.org/10.1103/PhysRevB.86.094417} {\bibfield  {journal} {\bibinfo
  {journal} {Phys. Rev. B}\ }\textbf {\bibinfo {volume} {86}},\ \bibinfo
  {pages} {094417} (\bibinfo {year} {2012})}\BibitemShut {NoStop}%
\bibitem [{\citenamefont {Agrapidis}\ \emph {et~al.}(2019)\citenamefont
  {Agrapidis}, \citenamefont {Drechsler}, \citenamefont {van~den Brink},\ and\
  \citenamefont {Nishimoto}}]{clio19}%
  \BibitemOpen
  \bibfield  {author} {\bibinfo {author} {\bibfnamefont {C.~E.}\ \bibnamefont
  {Agrapidis}}, \bibinfo {author} {\bibfnamefont {S.-L.}\ \bibnamefont
  {Drechsler}}, \bibinfo {author} {\bibfnamefont {J.}~\bibnamefont {van~den
  Brink}},\ and\ \bibinfo {author} {\bibfnamefont {S.}~\bibnamefont
  {Nishimoto}},\ }\bibfield  {title} {\bibinfo {title} {{Coexistence of
  valence-bond formation and topological order in the Frustrated Ferromagnetic
  $J_1$-$J_2$ Chain}},\ }\href {https://doi.org/10.21468/SciPostPhys.6.2.019}
  {\bibfield  {journal} {\bibinfo  {journal} {SciPost Phys.}\ }\textbf
  {\bibinfo {volume} {6}},\ \bibinfo {pages} {19} (\bibinfo {year}
  {2019})}\BibitemShut {NoStop}%
\bibitem [{\citenamefont {Kecke}\ \emph {et~al.}(2007)\citenamefont {Kecke},
  \citenamefont {Momoi},\ and\ \citenamefont {Furusaki}}]{Kecke2007}%
  \BibitemOpen
  \bibfield  {author} {\bibinfo {author} {\bibfnamefont {L.}~\bibnamefont
  {Kecke}}, \bibinfo {author} {\bibfnamefont {T.}~\bibnamefont {Momoi}},\ and\
  \bibinfo {author} {\bibfnamefont {A.}~\bibnamefont {Furusaki}},\ }\bibfield
  {title} {\bibinfo {title} {Multimagnon bound states in the frustrated
  ferromagnetic one-dimensional chain},\ }\href
  {https://doi.org/10.1103/physrevb.76.060407} {\bibfield  {journal} {\bibinfo
  {journal} {Phys. Rev. B}\ }\textbf {\bibinfo {volume} {76}},\ \bibinfo
  {pages} {060407(R)} (\bibinfo {year} {2007})}\BibitemShut {NoStop}%
\bibitem [{\citenamefont {Sudan}\ \emph {et~al.}(2009)\citenamefont {Sudan},
  \citenamefont {L\"{u}scher},\ and\ \citenamefont {L\"{a}uchli}}]{Sudan2009}%
  \BibitemOpen
  \bibfield  {author} {\bibinfo {author} {\bibfnamefont {J.}~\bibnamefont
  {Sudan}}, \bibinfo {author} {\bibfnamefont {A.}~\bibnamefont {L\"{u}scher}},\
  and\ \bibinfo {author} {\bibfnamefont {A.~M.}\ \bibnamefont {L\"{a}uchli}},\
  }\bibfield  {title} {\bibinfo {title} {Emergent multipolar spin correlations
  in a fluctuating spiral: The frustrated ferromagnetic spin-1/2 {H}eisenberg
  chain in a magnetic field},\ }\href
  {https://doi.org/10.1103/physrevb.80.140402} {\bibfield  {journal} {\bibinfo
  {journal} {Phys. Rev. B}\ }\textbf {\bibinfo {volume} {80}},\ \bibinfo
  {pages} {140402(R)} (\bibinfo {year} {2009})}\BibitemShut {NoStop}%
\bibitem [{\citenamefont {Hikihara}\ \emph {et~al.}(2008)\citenamefont
  {Hikihara}, \citenamefont {Kecke}, \citenamefont {Momoi},\ and\ \citenamefont
  {Furusaki}}]{Hikihara08}%
  \BibitemOpen
  \bibfield  {author} {\bibinfo {author} {\bibfnamefont {T.}~\bibnamefont
  {Hikihara}}, \bibinfo {author} {\bibfnamefont {L.}~\bibnamefont {Kecke}},
  \bibinfo {author} {\bibfnamefont {T.}~\bibnamefont {Momoi}},\ and\ \bibinfo
  {author} {\bibfnamefont {A.}~\bibnamefont {Furusaki}},\ }\bibfield  {title}
  {\bibinfo {title} {Vector chiral and multipolar orders in the
  spin-$\frac{1}{2}$ frustrated ferromagnetic chain in magnetic field},\ }\href
  {https://doi.org/10.1103/PhysRevB.78.144404} {\bibfield  {journal} {\bibinfo
  {journal} {Phys. Rev. B}\ }\textbf {\bibinfo {volume} {78}},\ \bibinfo
  {pages} {144404} (\bibinfo {year} {2008})}\BibitemShut {NoStop}%
\bibitem [{\citenamefont {Sato}\ \emph {et~al.}(2009)\citenamefont {Sato},
  \citenamefont {Momoi},\ and\ \citenamefont {Furusaki}}]{Sato09}%
  \BibitemOpen
  \bibfield  {author} {\bibinfo {author} {\bibfnamefont {M.}~\bibnamefont
  {Sato}}, \bibinfo {author} {\bibfnamefont {T.}~\bibnamefont {Momoi}},\ and\
  \bibinfo {author} {\bibfnamefont {A.}~\bibnamefont {Furusaki}},\ }\bibfield
  {title} {\bibinfo {title} {Nmr relaxation rate and dynamical structure
  factors in nematic and multipolar liquids of frustrated spin chains under
  magnetic fields},\ }\href {https://doi.org/10.1103/PhysRevB.79.060406}
  {\bibfield  {journal} {\bibinfo  {journal} {Phys. Rev. B}\ }\textbf {\bibinfo
  {volume} {79}},\ \bibinfo {pages} {060406} (\bibinfo {year}
  {2009})}\BibitemShut {NoStop}%
\bibitem [{\citenamefont {B\"uttgen}\ \emph {et~al.}(2014)\citenamefont
  {B\"uttgen}, \citenamefont {Nawa}, \citenamefont {Fujita}, \citenamefont
  {Hagiwara}, \citenamefont {Kuhns}, \citenamefont {Prokofiev}, \citenamefont
  {Reyes}, \citenamefont {Svistov}, \citenamefont {Yoshimura},\ and\
  \citenamefont {Takigawa}}]{Buttgen14}%
  \BibitemOpen
  \bibfield  {author} {\bibinfo {author} {\bibfnamefont {N.}~\bibnamefont
  {B\"uttgen}}, \bibinfo {author} {\bibfnamefont {K.}~\bibnamefont {Nawa}},
  \bibinfo {author} {\bibfnamefont {T.}~\bibnamefont {Fujita}}, \bibinfo
  {author} {\bibfnamefont {M.}~\bibnamefont {Hagiwara}}, \bibinfo {author}
  {\bibfnamefont {P.}~\bibnamefont {Kuhns}}, \bibinfo {author} {\bibfnamefont
  {A.}~\bibnamefont {Prokofiev}}, \bibinfo {author} {\bibfnamefont {A.~P.}\
  \bibnamefont {Reyes}}, \bibinfo {author} {\bibfnamefont {L.~E.}\ \bibnamefont
  {Svistov}}, \bibinfo {author} {\bibfnamefont {K.}~\bibnamefont {Yoshimura}},\
  and\ \bibinfo {author} {\bibfnamefont {M.}~\bibnamefont {Takigawa}},\
  }\bibfield  {title} {\bibinfo {title} {Search for a spin-nematic phase in the
  quasi-one-dimensional frustrated magnet {L}i{C}u{V}{O}$_{4}$},\ }\href
  {https://doi.org/10.1103/PhysRevB.90.134401} {\bibfield  {journal} {\bibinfo
  {journal} {Phys. Rev. B}\ }\textbf {\bibinfo {volume} {90}},\ \bibinfo
  {pages} {134401} (\bibinfo {year} {2014})}\BibitemShut {NoStop}%
\bibitem [{\citenamefont {Nawa}\ \emph {et~al.}(2013)\citenamefont {Nawa},
  \citenamefont {Takigawa}, \citenamefont {Yoshida},\ and\ \citenamefont
  {Yoshimura}}]{Nawa13}%
  \BibitemOpen
  \bibfield  {author} {\bibinfo {author} {\bibfnamefont {K.}~\bibnamefont
  {Nawa}}, \bibinfo {author} {\bibfnamefont {M.}~\bibnamefont {Takigawa}},
  \bibinfo {author} {\bibfnamefont {M.}~\bibnamefont {Yoshida}},\ and\ \bibinfo
  {author} {\bibfnamefont {K.}~\bibnamefont {Yoshimura}},\ }\bibfield  {title}
  {\bibinfo {title} {Anisotropic spin fluctuations in the quasi one-dimensional
  frustrated magnet licuvo4},\ }\href {https://doi.org/10.7566/JPSJ.82.094709}
  {\bibfield  {journal} {\bibinfo  {journal} {J. Phys. Soc. Jpn.}\ }\textbf
  {\bibinfo {volume} {82}},\ \bibinfo {pages} {094709} (\bibinfo {year}
  {2013})}\BibitemShut {NoStop}%
\bibitem [{\citenamefont {Nawa}\ \emph {et~al.}(2014)\citenamefont {Nawa},
  \citenamefont {Okamoto}, \citenamefont {Matsuo}, \citenamefont {Kindo},
  \citenamefont {Kitahara}, \citenamefont {Yoshida}, \citenamefont {Ikeda},
  \citenamefont {Hara}, \citenamefont {Sakurai}, \citenamefont {Okubo},
  \citenamefont {Ohta},\ and\ \citenamefont {Hiroi}}]{Nawa14}%
  \BibitemOpen
  \bibfield  {author} {\bibinfo {author} {\bibfnamefont {K.}~\bibnamefont
  {Nawa}}, \bibinfo {author} {\bibfnamefont {Y.}~\bibnamefont {Okamoto}},
  \bibinfo {author} {\bibfnamefont {A.}~\bibnamefont {Matsuo}}, \bibinfo
  {author} {\bibfnamefont {K.}~\bibnamefont {Kindo}}, \bibinfo {author}
  {\bibfnamefont {Y.}~\bibnamefont {Kitahara}}, \bibinfo {author}
  {\bibfnamefont {S.}~\bibnamefont {Yoshida}}, \bibinfo {author} {\bibfnamefont
  {S.}~\bibnamefont {Ikeda}}, \bibinfo {author} {\bibfnamefont
  {S.}~\bibnamefont {Hara}}, \bibinfo {author} {\bibfnamefont {T.}~\bibnamefont
  {Sakurai}}, \bibinfo {author} {\bibfnamefont {S.}~\bibnamefont {Okubo}},
  \bibinfo {author} {\bibfnamefont {H.}~\bibnamefont {Ohta}},\ and\ \bibinfo
  {author} {\bibfnamefont {Z.}~\bibnamefont {Hiroi}},\ }\bibfield  {title}
  {\bibinfo {title} {{N}a{C}u{M}o{O}$_{4}$({O}{H}) as a candidate frustrated
  ${J}_{1}$-${J}_2$ chain quantum magnet},\ }\href
  {https://doi.org/10.7566/JPSJ.83.103702} {\bibfield  {journal} {\bibinfo
  {journal} {J. Phys. Soc. Jpn.}\ }\textbf {\bibinfo {volume} {83}},\ \bibinfo
  {pages} {103702} (\bibinfo {year} {2014})}\BibitemShut {NoStop}%
\bibitem [{\citenamefont {Pregelj}\ \emph {et~al.}(2015)\citenamefont
  {Pregelj}, \citenamefont {Zorko}, \citenamefont {Zaharko}, \citenamefont
  {Nojiri}, \citenamefont {Berger}, \citenamefont {Chapon},\ and\ \citenamefont
  {Ar{\v{c}}on}}]{Pregelj15}%
  \BibitemOpen
  \bibfield  {author} {\bibinfo {author} {\bibfnamefont {M.}~\bibnamefont
  {Pregelj}}, \bibinfo {author} {\bibfnamefont {A.}~\bibnamefont {Zorko}},
  \bibinfo {author} {\bibfnamefont {O.}~\bibnamefont {Zaharko}}, \bibinfo
  {author} {\bibfnamefont {H.}~\bibnamefont {Nojiri}}, \bibinfo {author}
  {\bibfnamefont {H.}~\bibnamefont {Berger}}, \bibinfo {author} {\bibfnamefont
  {L.}~\bibnamefont {Chapon}},\ and\ \bibinfo {author} {\bibfnamefont
  {D.}~\bibnamefont {Ar{\v{c}}on}},\ }\bibfield  {title} {\bibinfo {title}
  {Spin-stripe phase in a frustrated zigzag spin-1/2 chain},\ }\href@noop {}
  {\bibfield  {journal} {\bibinfo  {journal} {Nature communications}\ }\textbf
  {\bibinfo {volume} {6}},\ \bibinfo {pages} {7255} (\bibinfo {year}
  {2015})}\BibitemShut {NoStop}%
\bibitem [{\citenamefont {Heinze}\ \emph
  {et~al.}(2019{\natexlab{a}})\citenamefont {Heinze}, \citenamefont {Bastien},
  \citenamefont {Ryll}, \citenamefont {Hoffmann}, \citenamefont {Reehuis},
  \citenamefont {Ouladdiaf}, \citenamefont {Bert}, \citenamefont {Kermarrec},
  \citenamefont {Mendels}, \citenamefont {Nishimoto}, \citenamefont
  {Drechsler}, \citenamefont {R\"o\ss{}ler}, \citenamefont {Rosner},
  \citenamefont {B\"uchner}, \citenamefont {Studer}, \citenamefont {Rule},
  \citenamefont {S\"ullow},\ and\ \citenamefont {Wolter}}]{Heinze19}%
  \BibitemOpen
  \bibfield  {author} {\bibinfo {author} {\bibfnamefont {L.}~\bibnamefont
  {Heinze}}, \bibinfo {author} {\bibfnamefont {G.}~\bibnamefont {Bastien}},
  \bibinfo {author} {\bibfnamefont {B.}~\bibnamefont {Ryll}}, \bibinfo {author}
  {\bibfnamefont {J.-U.}\ \bibnamefont {Hoffmann}}, \bibinfo {author}
  {\bibfnamefont {M.}~\bibnamefont {Reehuis}}, \bibinfo {author} {\bibfnamefont
  {B.}~\bibnamefont {Ouladdiaf}}, \bibinfo {author} {\bibfnamefont
  {F.}~\bibnamefont {Bert}}, \bibinfo {author} {\bibfnamefont {E.}~\bibnamefont
  {Kermarrec}}, \bibinfo {author} {\bibfnamefont {P.}~\bibnamefont {Mendels}},
  \bibinfo {author} {\bibfnamefont {S.}~\bibnamefont {Nishimoto}}, \bibinfo
  {author} {\bibfnamefont {S.-L.}\ \bibnamefont {Drechsler}}, \bibinfo {author}
  {\bibfnamefont {U.~K.}\ \bibnamefont {R\"o\ss{}ler}}, \bibinfo {author}
  {\bibfnamefont {H.}~\bibnamefont {Rosner}}, \bibinfo {author} {\bibfnamefont
  {B.}~\bibnamefont {B\"uchner}}, \bibinfo {author} {\bibfnamefont {A.~J.}\
  \bibnamefont {Studer}}, \bibinfo {author} {\bibfnamefont {K.~C.}\
  \bibnamefont {Rule}}, \bibinfo {author} {\bibfnamefont {S.}~\bibnamefont
  {S\"ullow}},\ and\ \bibinfo {author} {\bibfnamefont {A.~U.~B.}\ \bibnamefont
  {Wolter}},\ }\bibfield  {title} {\bibinfo {title} {Magnetic phase diagram of
  the frustrated spin chain compound linarite
  {P}b{C}u{S}{O}$_{4}$({O}{H})$_{2}$ as seen by neutron diffraction and
  $^{1}${H}-{N}{M}{R}},\ }\href {https://doi.org/10.1103/PhysRevB.99.094436}
  {\bibfield  {journal} {\bibinfo  {journal} {Phys. Rev. B}\ }\textbf {\bibinfo
  {volume} {99}},\ \bibinfo {pages} {094436} (\bibinfo {year}
  {2019}{\natexlab{a}})}\BibitemShut {NoStop}%
\bibitem [{\citenamefont {Ueda}\ and\ \citenamefont {Totsuka}(2009)}]{Ueda09}%
  \BibitemOpen
  \bibfield  {author} {\bibinfo {author} {\bibfnamefont {H.~T.}\ \bibnamefont
  {Ueda}}\ and\ \bibinfo {author} {\bibfnamefont {K.}~\bibnamefont {Totsuka}},\
  }\bibfield  {title} {\bibinfo {title} {Magnon bose-einstein condensation and
  various phases of three-dimensonal quantum helimagnets under high magnetic
  field},\ }\href {https://doi.org/10.1103/PhysRevB.80.014417} {\bibfield
  {journal} {\bibinfo  {journal} {Phys. Rev. B}\ }\textbf {\bibinfo {volume}
  {80}},\ \bibinfo {pages} {014417} (\bibinfo {year} {2009})}\BibitemShut
  {NoStop}%
\bibitem [{\citenamefont {Nishimoto}\ \emph {et~al.}(2011)\citenamefont
  {Nishimoto}, \citenamefont {Drechsler}, \citenamefont {Kuzian}, \citenamefont
  {van~den Brink}, \citenamefont {Richter}, \citenamefont {Lorenz},
  \citenamefont {Skourski}, \citenamefont {Klingeler},\ and\ \citenamefont
  {B\"uchner}}]{Li2CuO2}%
  \BibitemOpen
  \bibfield  {author} {\bibinfo {author} {\bibfnamefont {S.}~\bibnamefont
  {Nishimoto}}, \bibinfo {author} {\bibfnamefont {S.-L.}\ \bibnamefont
  {Drechsler}}, \bibinfo {author} {\bibfnamefont {R.~O.}\ \bibnamefont
  {Kuzian}}, \bibinfo {author} {\bibfnamefont {J.}~\bibnamefont {van~den
  Brink}}, \bibinfo {author} {\bibfnamefont {J.}~\bibnamefont {Richter}},
  \bibinfo {author} {\bibfnamefont {W.~E.~A.}\ \bibnamefont {Lorenz}}, \bibinfo
  {author} {\bibfnamefont {Y.}~\bibnamefont {Skourski}}, \bibinfo {author}
  {\bibfnamefont {R.}~\bibnamefont {Klingeler}},\ and\ \bibinfo {author}
  {\bibfnamefont {B.}~\bibnamefont {B\"uchner}},\ }\bibfield  {title} {\bibinfo
  {title} {Saturation field of frustrated chain cuprates: Broad regions of
  predominant interchain coupling},\ }\href
  {https://doi.org/10.1103/PhysRevLett.107.097201} {\bibfield  {journal}
  {\bibinfo  {journal} {Phys. Rev. Lett.}\ }\textbf {\bibinfo {volume} {107}},\
  \bibinfo {pages} {097201} (\bibinfo {year} {2011})}\BibitemShut {NoStop}%
\bibitem [{\citenamefont {Nishimoto}\ \emph {et~al.}(2015)\citenamefont
  {Nishimoto}, \citenamefont {Drechsler}, \citenamefont {Kuzian}, \citenamefont
  {Richter},\ and\ \citenamefont {van~den Brink}}]{J1J2IC}%
  \BibitemOpen
  \bibfield  {author} {\bibinfo {author} {\bibfnamefont {S.}~\bibnamefont
  {Nishimoto}}, \bibinfo {author} {\bibfnamefont {S.-L.}\ \bibnamefont
  {Drechsler}}, \bibinfo {author} {\bibfnamefont {R.}~\bibnamefont {Kuzian}},
  \bibinfo {author} {\bibfnamefont {J.}~\bibnamefont {Richter}},\ and\ \bibinfo
  {author} {\bibfnamefont {J.}~\bibnamefont {van~den Brink}},\ }\bibfield
  {title} {\bibinfo {title} {Interplay of interchain interactions and exchange
  anisotropy: Stability and fragility of multipolar states in
  spin-$\frac{1}{2}$ quasi-one-dimensional frustrated helimagnets},\ }\href
  {https://doi.org/10.1103/PhysRevB.92.214415} {\bibfield  {journal} {\bibinfo
  {journal} {Phys. Rev. B}\ }\textbf {\bibinfo {volume} {92}},\ \bibinfo
  {pages} {214415} (\bibinfo {year} {2015})}\BibitemShut {NoStop}%
\bibitem [{\citenamefont {Sen}\ \emph {et~al.}(1996)\citenamefont {Sen},
  \citenamefont {Shastry}, \citenamefont {Walstedt},\ and\ \citenamefont
  {Cava}}]{Sen96}%
  \BibitemOpen
  \bibfield  {author} {\bibinfo {author} {\bibfnamefont {D.}~\bibnamefont
  {Sen}}, \bibinfo {author} {\bibfnamefont {B.~S.}\ \bibnamefont {Shastry}},
  \bibinfo {author} {\bibfnamefont {R.~E.}\ \bibnamefont {Walstedt}},\ and\
  \bibinfo {author} {\bibfnamefont {R.}~\bibnamefont {Cava}},\ }\bibfield
  {title} {\bibinfo {title} {Quantum solitons in the sawtooth lattice},\ }\href
  {https://doi.org/10.1103/PhysRevB.53.6401} {\bibfield  {journal} {\bibinfo
  {journal} {Phys. Rev. B}\ }\textbf {\bibinfo {volume} {53}},\ \bibinfo
  {pages} {6401} (\bibinfo {year} {1996})}\BibitemShut {NoStop}%
\bibitem [{\citenamefont {Le~Bacq}\ \emph {et~al.}(2005)\citenamefont
  {Le~Bacq}, \citenamefont {Pasturel}, \citenamefont {Lacroix},\ and\
  \citenamefont {{N{\'u}{\~n}ez-Regueiro}}}]{LeBacq05}%
  \BibitemOpen
  \bibfield  {author} {\bibinfo {author} {\bibfnamefont {O.}~\bibnamefont
  {Le~Bacq}}, \bibinfo {author} {\bibfnamefont {A.}~\bibnamefont {Pasturel}},
  \bibinfo {author} {\bibfnamefont {C.}~\bibnamefont {Lacroix}},\ and\ \bibinfo
  {author} {\bibfnamefont {M.~D.}\ \bibnamefont {{N{\'u}{\~n}ez-Regueiro}}},\
  }\bibfield  {title} {\bibinfo {title} {First-principles determination of
  exchange interactions in delafossite {Y}{C}u{O}$_{2.5}$},\ }\href
  {https://doi.org/10.1103/PhysRevB.71.014432} {\bibfield  {journal} {\bibinfo
  {journal} {Phys. Rev. B}\ }\textbf {\bibinfo {volume} {71}},\ \bibinfo
  {pages} {014432} (\bibinfo {year} {2005})}\BibitemShut {NoStop}%
\bibitem [{\citenamefont {Ruiz-P\'erez}\ \emph {et~al.}(2000)\citenamefont
  {Ruiz-P\'erez}, \citenamefont {Hernández-Molina}, \citenamefont
  {Lorenzo-Luis}, \citenamefont {Lloret}, \citenamefont {Cano},\ and\
  \citenamefont {Julve}}]{Ruiz-Perez00}%
  \BibitemOpen
  \bibfield  {author} {\bibinfo {author} {\bibfnamefont {C.}~\bibnamefont
  {Ruiz-P\'erez}}, \bibinfo {author} {\bibfnamefont {M.}~\bibnamefont
  {Hernández-Molina}}, \bibinfo {author} {\bibfnamefont {P.}~\bibnamefont
  {Lorenzo-Luis}}, \bibinfo {author} {\bibfnamefont {F.}~\bibnamefont
  {Lloret}}, \bibinfo {author} {\bibfnamefont {J.}~\bibnamefont {Cano}},\ and\
  \bibinfo {author} {\bibfnamefont {M.}~\bibnamefont {Julve}},\ }\bibfield
  {title} {\bibinfo {title} {Magnetic coupling through the carbon skeleton of
  malonate in two polymorphs of
  {[Cu(bpy)(H$_{2}$O)][Cu(bpy)(mal)(H$_{2}$O)]}({C}l{O}$_{4}$)$_{2}$
  ({H}$_{2}$mal = {M}alonic {A}cid; bpy = 2,2'-{B}ipyridine)},\ }\href
  {https://doi.org/10.1021/ic000314n} {\bibfield  {journal} {\bibinfo
  {journal} {Inorganic Chemistry}\ }\textbf {\bibinfo {volume} {39}},\ \bibinfo
  {pages} {3845} (\bibinfo {year} {2000})}\BibitemShut {NoStop}%
\bibitem [{\citenamefont {Lau}\ \emph {et~al.}(2006)\citenamefont {Lau},
  \citenamefont {Ueland}, \citenamefont {Freitas}, \citenamefont {Dahlberg},
  \citenamefont {Schiffer},\ and\ \citenamefont {Cava}}]{Lau06}%
  \BibitemOpen
  \bibfield  {author} {\bibinfo {author} {\bibfnamefont {G.~C.}\ \bibnamefont
  {Lau}}, \bibinfo {author} {\bibfnamefont {B.~G.}\ \bibnamefont {Ueland}},
  \bibinfo {author} {\bibfnamefont {R.~S.}\ \bibnamefont {Freitas}}, \bibinfo
  {author} {\bibfnamefont {M.~L.}\ \bibnamefont {Dahlberg}}, \bibinfo {author}
  {\bibfnamefont {P.}~\bibnamefont {Schiffer}},\ and\ \bibinfo {author}
  {\bibfnamefont {R.~J.}\ \bibnamefont {Cava}},\ }\bibfield  {title} {\bibinfo
  {title} {Magnetic characterization of the sawtooth-lattice olivines
  {Z}n{L}$_{2}${S}$_{4}$ ({L}={E}r, {T}m, {Y}b)},\ }\href
  {https://doi.org/10.1103/PhysRevB.73.012413} {\bibfield  {journal} {\bibinfo
  {journal} {Phys. Rev. B}\ }\textbf {\bibinfo {volume} {73}},\ \bibinfo
  {pages} {012413} (\bibinfo {year} {2006})}\BibitemShut {NoStop}%
\bibitem [{\citenamefont {Kikuchi}\ \emph {et~al.}(2011)\citenamefont
  {Kikuchi}, \citenamefont {Fujii}, \citenamefont {Takahashi}, \citenamefont
  {Azuma}, \citenamefont {Shimakawa}, \citenamefont {Taniguchi}, \citenamefont
  {Matsuo},\ and\ \citenamefont {Kindo}}]{Kikuchi11}%
  \BibitemOpen
  \bibfield  {author} {\bibinfo {author} {\bibfnamefont {H.}~\bibnamefont
  {Kikuchi}}, \bibinfo {author} {\bibfnamefont {Y.}~\bibnamefont {Fujii}},
  \bibinfo {author} {\bibfnamefont {D.}~\bibnamefont {Takahashi}}, \bibinfo
  {author} {\bibfnamefont {M.}~\bibnamefont {Azuma}}, \bibinfo {author}
  {\bibfnamefont {Y.}~\bibnamefont {Shimakawa}}, \bibinfo {author}
  {\bibfnamefont {T.}~\bibnamefont {Taniguchi}}, \bibinfo {author}
  {\bibfnamefont {A.}~\bibnamefont {Matsuo}},\ and\ \bibinfo {author}
  {\bibfnamefont {K.}~\bibnamefont {Kindo}},\ }\bibfield  {title} {\bibinfo
  {title} {Spin gapped behavior of a frustrated delta chain compound
  euchroite},\ }\href {https://doi.org/10.1088/1742-6596/320/1/012045}
  {\bibfield  {journal} {\bibinfo  {journal} {J. Phys.: Conf. Ser.}\ }\textbf
  {\bibinfo {volume} {320}},\ \bibinfo {pages} {012045} (\bibinfo {year}
  {2011})}\BibitemShut {NoStop}%
\bibitem [{\citenamefont {White}\ \emph {et~al.}(2012)\citenamefont {White},
  \citenamefont {Honda}, \citenamefont {Kimura}, \citenamefont {Kimura},
  \citenamefont {Niedermayer}, \citenamefont {Zaharko}, \citenamefont {Poole},
  \citenamefont {Roessli},\ and\ \citenamefont {Kenzelmann}}]{White12}%
  \BibitemOpen
  \bibfield  {author} {\bibinfo {author} {\bibfnamefont {J.~S.}\ \bibnamefont
  {White}}, \bibinfo {author} {\bibfnamefont {T.}~\bibnamefont {Honda}},
  \bibinfo {author} {\bibfnamefont {K.}~\bibnamefont {Kimura}}, \bibinfo
  {author} {\bibfnamefont {T.}~\bibnamefont {Kimura}}, \bibinfo {author}
  {\bibfnamefont {C.}~\bibnamefont {Niedermayer}}, \bibinfo {author}
  {\bibfnamefont {O.}~\bibnamefont {Zaharko}}, \bibinfo {author} {\bibfnamefont
  {A.}~\bibnamefont {Poole}}, \bibinfo {author} {\bibfnamefont
  {B.}~\bibnamefont {Roessli}},\ and\ \bibinfo {author} {\bibfnamefont
  {M.}~\bibnamefont {Kenzelmann}},\ }\bibfield  {title} {\bibinfo {title}
  {Coupling of {{Magnetic}} and {{Ferroelectric Hysteresis}} by a
  {{Multicomponent Magnetic Structure}} in {M}n$_{2}${G}e{O}$_{4}$},\ }\href
  {https://doi.org/10.1103/PhysRevLett.108.077204} {\bibfield  {journal}
  {\bibinfo  {journal} {Phys. Rev. Lett.}\ }\textbf {\bibinfo {volume} {108}},\
  \bibinfo {pages} {077204} (\bibinfo {year} {2012})}\BibitemShut {NoStop}%
\bibitem [{\citenamefont {Garlea}\ \emph {et~al.}(2014)\citenamefont {Garlea},
  \citenamefont {Sanjeewa}, \citenamefont {McGuire}, \citenamefont {Kumar},
  \citenamefont {Sulejmanovic}, \citenamefont {He},\ and\ \citenamefont
  {Hwu}}]{Garlea14}%
  \BibitemOpen
  \bibfield  {author} {\bibinfo {author} {\bibfnamefont {V.~O.}\ \bibnamefont
  {Garlea}}, \bibinfo {author} {\bibfnamefont {L.~D.}\ \bibnamefont
  {Sanjeewa}}, \bibinfo {author} {\bibfnamefont {M.~A.}\ \bibnamefont
  {McGuire}}, \bibinfo {author} {\bibfnamefont {P.}~\bibnamefont {Kumar}},
  \bibinfo {author} {\bibfnamefont {D.}~\bibnamefont {Sulejmanovic}}, \bibinfo
  {author} {\bibfnamefont {J.}~\bibnamefont {He}},\ and\ \bibinfo {author}
  {\bibfnamefont {S.-J.}\ \bibnamefont {Hwu}},\ }\bibfield  {title} {\bibinfo
  {title} {Complex magnetic behavior of the sawtooth {{Fe}} chains in
  {R}b$_{2}${F}e$_{2}${O}({A}s{O}$_{4}$)$_{2}$},\ }\href
  {https://doi.org/10.1103/PhysRevB.89.014426} {\bibfield  {journal} {\bibinfo
  {journal} {Phys. Rev. B}\ }\textbf {\bibinfo {volume} {89}},\ \bibinfo
  {pages} {014426} (\bibinfo {year} {2014})}\BibitemShut {NoStop}%
\bibitem [{\citenamefont {May}\ \emph {et~al.}(2016)\citenamefont {May},
  \citenamefont {Calder}, \citenamefont {Parker}, \citenamefont {Sales},\ and\
  \citenamefont {McGuire}}]{May16}%
  \BibitemOpen
  \bibfield  {author} {\bibinfo {author} {\bibfnamefont {A.~F.}\ \bibnamefont
  {May}}, \bibinfo {author} {\bibfnamefont {S.}~\bibnamefont {Calder}},
  \bibinfo {author} {\bibfnamefont {D.~S.}\ \bibnamefont {Parker}}, \bibinfo
  {author} {\bibfnamefont {B.~C.}\ \bibnamefont {Sales}},\ and\ \bibinfo
  {author} {\bibfnamefont {M.~A.}\ \bibnamefont {McGuire}},\ }\bibfield
  {title} {\bibinfo {title} {Competing magnetic ground states and their
  coupling to the crystal lattice in {{CuFe$_2$Ge$_2$}}},\ }\href
  {https://doi.org/10.1038/srep35325} {\bibfield  {journal} {\bibinfo
  {journal} {Sci Rep}\ }\textbf {\bibinfo {volume} {6}},\ \bibinfo {pages}
  {35325} (\bibinfo {year} {2016})}\BibitemShut {NoStop}%
\bibitem [{\citenamefont {Baniodeh}\ \emph {et~al.}(2018)\citenamefont
  {Baniodeh}, \citenamefont {Magnani}, \citenamefont {Lan}, \citenamefont
  {Buth}, \citenamefont {Anson}, \citenamefont {Richter}, \citenamefont
  {Affronte}, \citenamefont {Schnack},\ and\ \citenamefont
  {Powell}}]{Baniodeh18}%
  \BibitemOpen
  \bibfield  {author} {\bibinfo {author} {\bibfnamefont {A.}~\bibnamefont
  {Baniodeh}}, \bibinfo {author} {\bibfnamefont {N.}~\bibnamefont {Magnani}},
  \bibinfo {author} {\bibfnamefont {Y.}~\bibnamefont {Lan}}, \bibinfo {author}
  {\bibfnamefont {G.}~\bibnamefont {Buth}}, \bibinfo {author} {\bibfnamefont
  {C.~E.}\ \bibnamefont {Anson}}, \bibinfo {author} {\bibfnamefont
  {J.}~\bibnamefont {Richter}}, \bibinfo {author} {\bibfnamefont
  {M.}~\bibnamefont {Affronte}}, \bibinfo {author} {\bibfnamefont
  {J.}~\bibnamefont {Schnack}},\ and\ \bibinfo {author} {\bibfnamefont {A.~K.}\
  \bibnamefont {Powell}},\ }\bibfield  {title} {\bibinfo {title} {High spin
  cycles: topping the spin record for a single molecule verging on quantum
  criticality},\ }\href@noop {} {\bibfield  {journal} {\bibinfo  {journal} {npj
  Quantum Materials}\ }\textbf {\bibinfo {volume} {3}},\ \bibinfo {pages} {10}
  (\bibinfo {year} {2018})}\BibitemShut {NoStop}%
\bibitem [{\citenamefont {Heinze}\ \emph
  {et~al.}(2019{\natexlab{b}})\citenamefont {Heinze}, \citenamefont {Jeschke},
  \citenamefont {Metavitsiadis}, \citenamefont {Reehuis}, \citenamefont
  {Feyerherm}, \citenamefont {Hoffmann}, \citenamefont {Wolter}, \citenamefont
  {Ding}, \citenamefont {Zapf}, \citenamefont {Moya}, \citenamefont {Weickert},
  \citenamefont {Jaime}, \citenamefont {Rule}, \citenamefont {Menzel},
  \citenamefont {Valentí}, \citenamefont {Brenig},\ and\ \citenamefont
  {S\"{u}llow}}]{Heinze19-2}%
  \BibitemOpen
  \bibfield  {author} {\bibinfo {author} {\bibfnamefont {L.}~\bibnamefont
  {Heinze}}, \bibinfo {author} {\bibfnamefont {H.}~\bibnamefont {Jeschke}},
  \bibinfo {author} {\bibfnamefont {A.}~\bibnamefont {Metavitsiadis}}, \bibinfo
  {author} {\bibfnamefont {M.}~\bibnamefont {Reehuis}}, \bibinfo {author}
  {\bibfnamefont {R.}~\bibnamefont {Feyerherm}}, \bibinfo {author}
  {\bibfnamefont {J.-U.}\ \bibnamefont {Hoffmann}}, \bibinfo {author}
  {\bibfnamefont {A.}~\bibnamefont {Wolter}}, \bibinfo {author} {\bibfnamefont
  {X.}~\bibnamefont {Ding}}, \bibinfo {author} {\bibfnamefont {V.}~\bibnamefont
  {Zapf}}, \bibinfo {author} {\bibfnamefont {C.}~\bibnamefont {Moya}}, \bibinfo
  {author} {\bibfnamefont {F.}~\bibnamefont {Weickert}}, \bibinfo {author}
  {\bibfnamefont {M.}~\bibnamefont {Jaime}}, \bibinfo {author} {\bibfnamefont
  {K.}~\bibnamefont {Rule}}, \bibinfo {author} {\bibfnamefont {D.}~\bibnamefont
  {Menzel}}, \bibinfo {author} {\bibfnamefont {R.}~\bibnamefont {Valentí}},
  \bibinfo {author} {\bibfnamefont {W.}~\bibnamefont {Brenig}},\ and\ \bibinfo
  {author} {\bibfnamefont {S.}~\bibnamefont {S\"{u}llow}},\ }\bibfield  {title}
  {\bibinfo {title} {Atacamite {{Cu$_2$Cl(OH)$_3$}}: A model compound for the
  $s=1/2$ sawtooth chain?},\ }\href@noop {} {\bibfield  {journal} {\bibinfo
  {journal} {arXiv preprint arXiv:1904.07820}\ } (\bibinfo {year}
  {2019}{\natexlab{b}})}\BibitemShut {NoStop}%
\bibitem [{\citenamefont {Gnezdilov}\ \emph {et~al.}(2019)\citenamefont
  {Gnezdilov}, \citenamefont {Pashkevich}, \citenamefont {Kurnosov},
  \citenamefont {Zhuravlev}, \citenamefont {Wulferding}, \citenamefont
  {Lemmens}, \citenamefont {Menzel}, \citenamefont {Kozlyakova}, \citenamefont
  {Akhrorov}, \citenamefont {Kuznetsova}, \citenamefont {Berdonosov},
  \citenamefont {Dolgikh}, \citenamefont {Volkova},\ and\ \citenamefont
  {Vasiliev}}]{Gnezdilov19}%
  \BibitemOpen
  \bibfield  {author} {\bibinfo {author} {\bibfnamefont {V.~P.}\ \bibnamefont
  {Gnezdilov}}, \bibinfo {author} {\bibfnamefont {Y.~G.}\ \bibnamefont
  {Pashkevich}}, \bibinfo {author} {\bibfnamefont {V.~S.}\ \bibnamefont
  {Kurnosov}}, \bibinfo {author} {\bibfnamefont {O.~V.}\ \bibnamefont
  {Zhuravlev}}, \bibinfo {author} {\bibfnamefont {D.}~\bibnamefont
  {Wulferding}}, \bibinfo {author} {\bibfnamefont {P.}~\bibnamefont {Lemmens}},
  \bibinfo {author} {\bibfnamefont {D.}~\bibnamefont {Menzel}}, \bibinfo
  {author} {\bibfnamefont {E.~S.}\ \bibnamefont {Kozlyakova}}, \bibinfo
  {author} {\bibfnamefont {A.~Y.}\ \bibnamefont {Akhrorov}}, \bibinfo {author}
  {\bibfnamefont {E.~S.}\ \bibnamefont {Kuznetsova}}, \bibinfo {author}
  {\bibfnamefont {P.~S.}\ \bibnamefont {Berdonosov}}, \bibinfo {author}
  {\bibfnamefont {V.~A.}\ \bibnamefont {Dolgikh}}, \bibinfo {author}
  {\bibfnamefont {O.~S.}\ \bibnamefont {Volkova}},\ and\ \bibinfo {author}
  {\bibfnamefont {A.~N.}\ \bibnamefont {Vasiliev}},\ }\bibfield  {title}
  {\bibinfo {title} {Flat-band spin dynamics and phonon anomalies of the
  saw-tooth spin-chain system {{Fe$_2$O(SeO$_3$)$_2$}}},\ }\href
  {https://doi.org/10.1103/PhysRevB.99.064413} {\bibfield  {journal} {\bibinfo
  {journal} {Phys. Rev. B}\ }\textbf {\bibinfo {volume} {99}},\ \bibinfo
  {pages} {064413} (\bibinfo {year} {2019})}\BibitemShut {NoStop}%
\bibitem [{\citenamefont {Toriyama}\ \emph {et~al.}(2014)\citenamefont
  {Toriyama}, \citenamefont {Nakayama}, \citenamefont {Konishi},\ and\
  \citenamefont {Ohta}}]{Toriyama14}%
  \BibitemOpen
  \bibfield  {author} {\bibinfo {author} {\bibfnamefont {T.}~\bibnamefont
  {Toriyama}}, \bibinfo {author} {\bibfnamefont {T.}~\bibnamefont {Nakayama}},
  \bibinfo {author} {\bibfnamefont {T.}~\bibnamefont {Konishi}},\ and\ \bibinfo
  {author} {\bibfnamefont {Y.}~\bibnamefont {Ohta}},\ }\bibfield  {title}
  {\bibinfo {title} {Charge and orbital orderings associated with
  metal-insulator transition in {{V$_6$O$_{13}$}}},\ }\href
  {https://doi.org/10.1103/PhysRevB.90.085131} {\bibfield  {journal} {\bibinfo
  {journal} {Phys. Rev. B}\ }\textbf {\bibinfo {volume} {90}},\ \bibinfo
  {pages} {085131} (\bibinfo {year} {2014})}\BibitemShut {NoStop}%
\bibitem [{\citenamefont {Tonegawa}\ and\ \citenamefont
  {Kaburagi}(2004)}]{tonegawa04}%
  \BibitemOpen
  \bibfield  {author} {\bibinfo {author} {\bibfnamefont {T.}~\bibnamefont
  {Tonegawa}}\ and\ \bibinfo {author} {\bibfnamefont {M.}~\bibnamefont
  {Kaburagi}},\ }\bibfield  {title} {\bibinfo {title} {Ground-state properties
  of an {S}=1/2 {$\Delta$}-chain with ferro- and antiferromagnetic
  interactions},\ }\href
  {https://doi.org/https://doi.org/10.1016/j.jmmm.2003.11.367} {\bibfield
  {journal} {\bibinfo  {journal} {Journal of Magnetism and Magnetic Materials}\
  }\textbf {\bibinfo {volume} {272-276}},\ \bibinfo {pages} {898 } (\bibinfo
  {year} {2004})}\BibitemShut {NoStop}%
\bibitem [{\citenamefont {Inagaki}\ \emph {et~al.}(2005)\citenamefont
  {Inagaki}, \citenamefont {Narumi}, \citenamefont {Kindo}, \citenamefont
  {Kikuchi}, \citenamefont {Kamikawa}, \citenamefont {Kunimoto}, \citenamefont
  {Okubo}, \citenamefont {Ohta}, \citenamefont {Saito}, \citenamefont {Azuma},
  \citenamefont {Takano}, \citenamefont {Nojiri}, \citenamefont {Kaburagi},\
  and\ \citenamefont {Tonegawa}}]{Inagaki05}%
  \BibitemOpen
  \bibfield  {author} {\bibinfo {author} {\bibfnamefont {Y.}~\bibnamefont
  {Inagaki}}, \bibinfo {author} {\bibfnamefont {Y.}~\bibnamefont {Narumi}},
  \bibinfo {author} {\bibfnamefont {K.}~\bibnamefont {Kindo}}, \bibinfo
  {author} {\bibfnamefont {H.}~\bibnamefont {Kikuchi}}, \bibinfo {author}
  {\bibfnamefont {T.}~\bibnamefont {Kamikawa}}, \bibinfo {author}
  {\bibfnamefont {T.}~\bibnamefont {Kunimoto}}, \bibinfo {author}
  {\bibfnamefont {S.}~\bibnamefont {Okubo}}, \bibinfo {author} {\bibfnamefont
  {H.}~\bibnamefont {Ohta}}, \bibinfo {author} {\bibfnamefont {T.}~\bibnamefont
  {Saito}}, \bibinfo {author} {\bibfnamefont {M.}~\bibnamefont {Azuma}},
  \bibinfo {author} {\bibfnamefont {M.}~\bibnamefont {Takano}}, \bibinfo
  {author} {\bibfnamefont {H.}~\bibnamefont {Nojiri}}, \bibinfo {author}
  {\bibfnamefont {M.}~\bibnamefont {Kaburagi}},\ and\ \bibinfo {author}
  {\bibfnamefont {T.}~\bibnamefont {Tonegawa}},\ }\bibfield  {title} {\bibinfo
  {title} {Ferro-antiferromagnetic delta-chain system studied by high field
  magnetization measurements},\ }\href {https://doi.org/10.1143/JPSJ.74.2831}
  {\bibfield  {journal} {\bibinfo  {journal} {J. Phys. Soc. Jpn.}\ }\textbf
  {\bibinfo {volume} {74}},\ \bibinfo {pages} {2831} (\bibinfo {year}
  {2005})}\BibitemShut {NoStop}%
\bibitem [{\citenamefont {Kaburagi}\ \emph {et~al.}(2005)\citenamefont
  {Kaburagi}, \citenamefont {Tonegawa},\ and\ \citenamefont
  {Kang}}]{Kaburagi05}%
  \BibitemOpen
  \bibfield  {author} {\bibinfo {author} {\bibfnamefont {M.}~\bibnamefont
  {Kaburagi}}, \bibinfo {author} {\bibfnamefont {T.}~\bibnamefont {Tonegawa}},\
  and\ \bibinfo {author} {\bibfnamefont {M.}~\bibnamefont {Kang}},\ }\bibfield
  {title} {\bibinfo {title} {Ground state phase diagrams of an anisotropic
  spin-1/2 {$\Delta$}-chain with ferro- and antiferromagnetic interactions},\
  }\href {https://doi.org/10.1063/1.1851893} {\bibfield  {journal} {\bibinfo
  {journal} {Journal of Applied Physics}\ }\textbf {\bibinfo {volume} {97}},\
  \bibinfo {pages} {10B306} (\bibinfo {year} {2005})}\BibitemShut {NoStop}%
\bibitem [{\citenamefont {Krivnov}\ \emph {et~al.}(2014)\citenamefont
  {Krivnov}, \citenamefont {Dmitriev}, \citenamefont {Nishimoto}, \citenamefont
  {Drechsler},\ and\ \citenamefont {Richter}}]{krivnov14}%
  \BibitemOpen
  \bibfield  {author} {\bibinfo {author} {\bibfnamefont {V.~Y.}\ \bibnamefont
  {Krivnov}}, \bibinfo {author} {\bibfnamefont {D.~V.}\ \bibnamefont
  {Dmitriev}}, \bibinfo {author} {\bibfnamefont {S.}~\bibnamefont {Nishimoto}},
  \bibinfo {author} {\bibfnamefont {S.-L.}\ \bibnamefont {Drechsler}},\ and\
  \bibinfo {author} {\bibfnamefont {J.}~\bibnamefont {Richter}},\ }\bibfield
  {title} {\bibinfo {title} {Delta chain with ferromagnetic and
  antiferromagnetic interactions at the critical point},\ }\href
  {https://doi.org/10.1103/PhysRevB.90.014441} {\bibfield  {journal} {\bibinfo
  {journal} {Phys. Rev. B}\ }\textbf {\bibinfo {volume} {90}},\ \bibinfo
  {pages} {014441} (\bibinfo {year} {2014})}\BibitemShut {NoStop}%
\bibitem [{\citenamefont {Sachdev}(2008)}]{Sachdev08}%
  \BibitemOpen
  \bibfield  {author} {\bibinfo {author} {\bibfnamefont {S.}~\bibnamefont
  {Sachdev}},\ }\bibfield  {title} {\bibinfo {title} {Quantum magnetism and
  criticality},\ }\href@noop {} {\bibfield  {journal} {\bibinfo  {journal}
  {Nature Physics}\ }\textbf {\bibinfo {volume} {4}},\ \bibinfo {pages} {173}
  (\bibinfo {year} {2008})}\BibitemShut {NoStop}%
\bibitem [{\citenamefont {Roch}\ \emph {et~al.}(2008)\citenamefont {Roch},
  \citenamefont {Florens}, \citenamefont {Bouchiat}, \citenamefont
  {Wernsdorfer},\ and\ \citenamefont {Balestro}}]{Roch08}%
  \BibitemOpen
  \bibfield  {author} {\bibinfo {author} {\bibfnamefont {N.}~\bibnamefont
  {Roch}}, \bibinfo {author} {\bibfnamefont {S.}~\bibnamefont {Florens}},
  \bibinfo {author} {\bibfnamefont {V.}~\bibnamefont {Bouchiat}}, \bibinfo
  {author} {\bibfnamefont {W.}~\bibnamefont {Wernsdorfer}},\ and\ \bibinfo
  {author} {\bibfnamefont {F.}~\bibnamefont {Balestro}},\ }\bibfield  {title}
  {\bibinfo {title} {Quantum phase transition in a single-molecule quantum
  dot},\ }\href@noop {} {\bibfield  {journal} {\bibinfo  {journal} {Nature}\
  }\textbf {\bibinfo {volume} {453}},\ \bibinfo {pages} {633} (\bibinfo {year}
  {2008})}\BibitemShut {NoStop}%
\bibitem [{\citenamefont {Monti}\ and\ \citenamefont
  {S{\"u}to}(1991)}]{Monti91}%
  \BibitemOpen
  \bibfield  {author} {\bibinfo {author} {\bibfnamefont {F.}~\bibnamefont
  {Monti}}\ and\ \bibinfo {author} {\bibfnamefont {A.}~\bibnamefont
  {S{\"u}to}},\ }\bibfield  {title} {\bibinfo {title} {Spin-$\frac{1}{2}$
  {H}eisenberg model on {$\Delta$} trees},\ }\href
  {https://doi.org/https://doi.org/10.1016/0375-9601(91)90937-4} {\bibfield
  {journal} {\bibinfo  {journal} {Physics Letters A}\ }\textbf {\bibinfo
  {volume} {156}},\ \bibinfo {pages} {197 } (\bibinfo {year}
  {1991})}\BibitemShut {NoStop}%
\bibitem [{\citenamefont {Nakamura}\ and\ \citenamefont
  {Kubo}(1996)}]{Nakamura96}%
  \BibitemOpen
  \bibfield  {author} {\bibinfo {author} {\bibfnamefont {T.}~\bibnamefont
  {Nakamura}}\ and\ \bibinfo {author} {\bibfnamefont {K.}~\bibnamefont
  {Kubo}},\ }\bibfield  {title} {\bibinfo {title} {Elementary excitations in
  the $\delta$ chain},\ }\href {https://doi.org/10.1103/PhysRevB.53.6393}
  {\bibfield  {journal} {\bibinfo  {journal} {Phys. Rev. B}\ }\textbf {\bibinfo
  {volume} {53}},\ \bibinfo {pages} {6393} (\bibinfo {year}
  {1996})}\BibitemShut {NoStop}%
\bibitem [{\citenamefont {{Blundell, S. A.}}\ and\ \citenamefont
  {{N\'u\~nez-Regueiro, M. D.}}(2003)}]{Blundell03}%
  \BibitemOpen
  \bibfield  {author} {\bibinfo {author} {\bibnamefont {{Blundell, S. A.}}}\
  and\ \bibinfo {author} {\bibnamefont {{N\'u\~nez-Regueiro, M. D.}}},\
  }\bibfield  {title} {\bibinfo {title} {Quantum topological excitations: from
  the sawtooth lattice to the {H}eisenberg chain},\ }\href
  {https://doi.org/10.1140/epjb/e2003-00054-2} {\bibfield  {journal} {\bibinfo
  {journal} {Eur. Phys. J. B}\ }\textbf {\bibinfo {volume} {31}},\ \bibinfo
  {pages} {453} (\bibinfo {year} {2003})}\BibitemShut {NoStop}%
\bibitem [{\citenamefont {Richter}\ \emph {et~al.}(2008)\citenamefont
  {Richter}, \citenamefont {Derzhko},\ and\ \citenamefont
  {Honecker}}]{Richter08}%
  \BibitemOpen
  \bibfield  {author} {\bibinfo {author} {\bibfnamefont {J.}~\bibnamefont
  {Richter}}, \bibinfo {author} {\bibfnamefont {O.}~\bibnamefont {Derzhko}},\
  and\ \bibinfo {author} {\bibfnamefont {A.}~\bibnamefont {Honecker}},\
  }\bibfield  {title} {\bibinfo {title} {The sawtooth chain: {{From
  Heisenberg}} spins to {{Hubbard}} electrons},\ }\href
  {https://doi.org/10.1142/S0217979208050176} {\bibfield  {journal} {\bibinfo
  {journal} {Int. J. Mod. Phys. B}\ }\textbf {\bibinfo {volume} {22}},\
  \bibinfo {pages} {4418} (\bibinfo {year} {2008})}\BibitemShut {NoStop}%
\bibitem [{\citenamefont {Chen}\ \emph {et~al.}(2001)\citenamefont {Chen},
  \citenamefont {B\"uttner},\ and\ \citenamefont {Voit}}]{Chen01}%
  \BibitemOpen
  \bibfield  {author} {\bibinfo {author} {\bibfnamefont {S.}~\bibnamefont
  {Chen}}, \bibinfo {author} {\bibfnamefont {H.}~\bibnamefont {B\"uttner}},\
  and\ \bibinfo {author} {\bibfnamefont {J.}~\bibnamefont {Voit}},\ }\bibfield
  {title} {\bibinfo {title} {Phase diagram of an asymmetric spin ladder},\
  }\href {https://doi.org/10.1103/PhysRevLett.87.087205} {\bibfield  {journal}
  {\bibinfo  {journal} {Phys. Rev. Lett.}\ }\textbf {\bibinfo {volume} {87}},\
  \bibinfo {pages} {087205} (\bibinfo {year} {2001})}\BibitemShut {NoStop}%
\bibitem [{\citenamefont {Chen}\ \emph {et~al.}(2003)\citenamefont {Chen},
  \citenamefont {B\"uttner},\ and\ \citenamefont {Voit}}]{Chen03}%
  \BibitemOpen
  \bibfield  {author} {\bibinfo {author} {\bibfnamefont {S.}~\bibnamefont
  {Chen}}, \bibinfo {author} {\bibfnamefont {H.}~\bibnamefont {B\"uttner}},\
  and\ \bibinfo {author} {\bibfnamefont {J.}~\bibnamefont {Voit}},\ }\bibfield
  {title} {\bibinfo {title} {Ground state and excitation of an asymmetric spin
  ladder model},\ }\href {https://doi.org/10.1103/PhysRevB.67.054412}
  {\bibfield  {journal} {\bibinfo  {journal} {Phys. Rev. B}\ }\textbf {\bibinfo
  {volume} {67}},\ \bibinfo {pages} {054412} (\bibinfo {year}
  {2003})}\BibitemShut {NoStop}%
\bibitem [{\citenamefont {White}(1993)}]{White92}%
  \BibitemOpen
  \bibfield  {author} {\bibinfo {author} {\bibfnamefont {S.~R.}\ \bibnamefont
  {White}},\ }\bibfield  {title} {\bibinfo {title} {Density-matrix algorithms
  for quantum renormalization groups},\ }\href
  {https://doi.org/10.1103/PhysRevB.48.10345} {\bibfield  {journal} {\bibinfo
  {journal} {Phys. Rev. B}\ }\textbf {\bibinfo {volume} {48}},\ \bibinfo
  {pages} {10345} (\bibinfo {year} {1993})}\BibitemShut {NoStop}%
\bibitem [{\citenamefont {Jeckelmann}(2002)}]{Jeckelmann02}%
  \BibitemOpen
  \bibfield  {author} {\bibinfo {author} {\bibfnamefont {E.}~\bibnamefont
  {Jeckelmann}},\ }\bibfield  {title} {\bibinfo {title} {Dynamical
  density-matrix renormalization-group method},\ }\href
  {https://doi.org/10.1103/PhysRevB.66.045114} {\bibfield  {journal} {\bibinfo
  {journal} {Phys. Rev. B}\ }\textbf {\bibinfo {volume} {66}},\ \bibinfo
  {pages} {045114} (\bibinfo {year} {2002})}\BibitemShut {NoStop}%
\bibitem [{\citenamefont {McCulloch}(2008)}]{mcculloch08}%
  \BibitemOpen
  \bibfield  {author} {\bibinfo {author} {\bibfnamefont {I.~P.}\ \bibnamefont
  {McCulloch}},\ }\href@noop {} {\bibinfo {title} {Infinite size density matrix
  renormalization group, revisited}} (\bibinfo {year} {2008}),\ \Eprint
  {https://arxiv.org/abs/0804.2509} {arXiv:0804.2509 [cond-mat.str-el]}
  \BibitemShut {NoStop}%
\bibitem [{\citenamefont {Schollw\"ock}(2011)}]{Schollwock11}%
  \BibitemOpen
  \bibfield  {author} {\bibinfo {author} {\bibfnamefont {U.}~\bibnamefont
  {Schollw\"ock}},\ }\bibfield  {title} {\bibinfo {title} {The density-matrix
  renormalization group in the age of matrix product states},\ }\href
  {https://doi.org/https://doi.org/10.1016/j.aop.2010.09.012} {\bibfield
  {journal} {\bibinfo  {journal} {Annals of Physics}\ }\textbf {\bibinfo
  {volume} {326}},\ \bibinfo {pages} {96} (\bibinfo {year} {2011})}\BibitemShut
  {NoStop}%
\bibitem [{\citenamefont {Bursill}\ \emph {et~al.}(1995)\citenamefont
  {Bursill}, \citenamefont {Gehringt}, \citenamefont {Farnellt}, \citenamefont
  {Parkinson}, \citenamefont {Xian},\ and\ \citenamefont {Zeng}}]{bursill95}%
  \BibitemOpen
  \bibfield  {author} {\bibinfo {author} {\bibfnamefont {R.}~\bibnamefont
  {Bursill}}, \bibinfo {author} {\bibfnamefont {G.}~\bibnamefont {Gehringt}},
  \bibinfo {author} {\bibfnamefont {D.}~\bibnamefont {Farnellt}}, \bibinfo
  {author} {\bibfnamefont {J.}~\bibnamefont {Parkinson}}, \bibinfo {author}
  {\bibfnamefont {T.}~\bibnamefont {Xian}},\ and\ \bibinfo {author}
  {\bibfnamefont {C.}~\bibnamefont {Zeng}},\ }\bibfield  {title} {\bibinfo
  {title} {Numerical and approximate analytical results for the frustrated
  spin-1/2 quantum spin chain},\ }\href
  {https://doi.org/10.1088/0953-8984/7/45/016} {\bibfield  {journal} {\bibinfo
  {journal} {J. Phys.: Condens. Matter}\ }\textbf {\bibinfo {volume} {7}},\
  \bibinfo {pages} {8605} (\bibinfo {year} {1995})}\BibitemShut {NoStop}%
\bibitem [{\citenamefont {Bader}\ and\ \citenamefont
  {Schilling}(1979)}]{bader79}%
  \BibitemOpen
  \bibfield  {author} {\bibinfo {author} {\bibfnamefont {H.~P.}\ \bibnamefont
  {Bader}}\ and\ \bibinfo {author} {\bibfnamefont {R.}~\bibnamefont
  {Schilling}},\ }\bibfield  {title} {\bibinfo {title} {Conditions for a
  ferromagnetic ground state of {H}eisenberg hamiltonians},\ }\href
  {https://doi.org/10.1103/PhysRevB.19.3556} {\bibfield  {journal} {\bibinfo
  {journal} {Phys. Rev. B}\ }\textbf {\bibinfo {volume} {19}},\ \bibinfo
  {pages} {3556} (\bibinfo {year} {1979})}\BibitemShut {NoStop}%
\bibitem [{\citenamefont {H\"artel}\ \emph {et~al.}(2008)\citenamefont
  {H\"artel}, \citenamefont {Richter}, \citenamefont {Ihle},\ and\
  \citenamefont {Drechsler}}]{hartel08}%
  \BibitemOpen
  \bibfield  {author} {\bibinfo {author} {\bibfnamefont {M.}~\bibnamefont
  {H\"artel}}, \bibinfo {author} {\bibfnamefont {J.}~\bibnamefont {Richter}},
  \bibinfo {author} {\bibfnamefont {D.}~\bibnamefont {Ihle}},\ and\ \bibinfo
  {author} {\bibfnamefont {S.-L.}\ \bibnamefont {Drechsler}},\ }\bibfield
  {title} {\bibinfo {title} {Thermodynamics of a one-dimensional frustrated
  spin-$\frac{1}{2}$ {H}eisenberg ferromagnet},\ }\href
  {https://doi.org/10.1103/PhysRevB.78.174412} {\bibfield  {journal} {\bibinfo
  {journal} {Phys. Rev. B}\ }\textbf {\bibinfo {volume} {78}},\ \bibinfo
  {pages} {174412} (\bibinfo {year} {2008})}\BibitemShut {NoStop}%
\bibitem [{\citenamefont {Nersesyan}\ \emph {et~al.}(1998)\citenamefont
  {Nersesyan}, \citenamefont {Gogolin},\ and\ \citenamefont
  {E{\ss}ler}}]{nersesyan98}%
  \BibitemOpen
  \bibfield  {author} {\bibinfo {author} {\bibfnamefont {A.~A.}\ \bibnamefont
  {Nersesyan}}, \bibinfo {author} {\bibfnamefont {A.~O.}\ \bibnamefont
  {Gogolin}},\ and\ \bibinfo {author} {\bibfnamefont {F.~H.~L.}\ \bibnamefont
  {E{\ss}ler}},\ }\bibfield  {title} {\bibinfo {title} {Incommensurate spin
  correlations in spin-1/2 frustrated two-leg {H}eisenberg ladders},\ }\href
  {https://doi.org/10.1103/PhysRevLett.81.910} {\bibfield  {journal} {\bibinfo
  {journal} {Phys. Rev. Lett.}\ }\textbf {\bibinfo {volume} {81}},\ \bibinfo
  {pages} {910} (\bibinfo {year} {1998})}\BibitemShut {NoStop}%
\bibitem [{\citenamefont {Sirker}\ \emph {et~al.}(2011)\citenamefont {Sirker},
  \citenamefont {Krivnov}, \citenamefont {Dmitriev}, \citenamefont {Herzog},
  \citenamefont {Janson}, \citenamefont {Nishimoto}, \citenamefont
  {Drechsler},\ and\ \citenamefont {Richter}}]{sirker11}%
  \BibitemOpen
  \bibfield  {author} {\bibinfo {author} {\bibfnamefont {J.}~\bibnamefont
  {Sirker}}, \bibinfo {author} {\bibfnamefont {V.~Y.}\ \bibnamefont {Krivnov}},
  \bibinfo {author} {\bibfnamefont {D.~V.}\ \bibnamefont {Dmitriev}}, \bibinfo
  {author} {\bibfnamefont {A.}~\bibnamefont {Herzog}}, \bibinfo {author}
  {\bibfnamefont {O.}~\bibnamefont {Janson}}, \bibinfo {author} {\bibfnamefont
  {S.}~\bibnamefont {Nishimoto}}, \bibinfo {author} {\bibfnamefont {S.-L.}\
  \bibnamefont {Drechsler}},\ and\ \bibinfo {author} {\bibfnamefont
  {J.}~\bibnamefont {Richter}},\ }\bibfield  {title} {\bibinfo {title}
  {${{J}}_1-{{J}}_{2}$ {H}eisenber model at and close to its $z$=4 quantum
  critical point},\ }\href {https://doi.org/10.1103/PhysRevB.84.144403}
  {\bibfield  {journal} {\bibinfo  {journal} {Phy. Rev. B}\ }\textbf {\bibinfo
  {volume} {84}},\ \bibinfo {pages} {144403} (\bibinfo {year}
  {2011})}\BibitemShut {NoStop}%
\bibitem [{\citenamefont {Oshikawa}(1992)}]{oshikawa92}%
  \BibitemOpen
  \bibfield  {author} {\bibinfo {author} {\bibfnamefont {M.}~\bibnamefont
  {Oshikawa}},\ }\bibfield  {title} {\bibinfo {title} {Hidden {Z2*Z2} symmetry
  in quantum spin chains with arbitrary integer spin},\ }\href
  {https://doi.org/10.1088/0953-8984/4/36/019} {\bibfield  {journal} {\bibinfo
  {journal} {J. Phys.: Condens. Matter}\ }\textbf {\bibinfo {volume} {4}},\
  \bibinfo {pages} {7469} (\bibinfo {year} {1992})}\BibitemShut {NoStop}%
\bibitem [{\citenamefont {Itoi}\ and\ \citenamefont {Qin}(2001)}]{itoi01}%
  \BibitemOpen
  \bibfield  {author} {\bibinfo {author} {\bibfnamefont {C.}~\bibnamefont
  {Itoi}}\ and\ \bibinfo {author} {\bibfnamefont {S.}~\bibnamefont {Qin}},\
  }\bibfield  {title} {\bibinfo {title} {Strongly reduced gap in the zigzag
  spin chain with a ferromagnetic interchain coupling},\ }\href
  {https://doi.org/10.1103/PhysRevB.63.224423} {\bibfield  {journal} {\bibinfo
  {journal} {Phys. Rev. B}\ }\textbf {\bibinfo {volume} {63}},\ \bibinfo
  {pages} {224423} (\bibinfo {year} {2001})}\BibitemShut {NoStop}%
\bibitem [{\citenamefont {Dmitriev}\ and\ \citenamefont
  {Krivnov}(2016)}]{Dmitriev16}%
  \BibitemOpen
  \bibfield  {author} {\bibinfo {author} {\bibfnamefont {D.~V.}\ \bibnamefont
  {Dmitriev}}\ and\ \bibinfo {author} {\bibfnamefont {V.~Y.}\ \bibnamefont
  {Krivnov}},\ }\bibfield  {title} {\bibinfo {title} {Ferrimagnetism in delta
  chain with anisotropic ferromagnetic and antiferromagnetic interactions},\
  }\href {https://doi.org/10.1088/0953-8984/28/50/506002} {\bibfield  {journal}
  {\bibinfo  {journal} {Journal of Physics: Condensed Matter}\ }\textbf
  {\bibinfo {volume} {28}},\ \bibinfo {pages} {506002} (\bibinfo {year}
  {2016})}\BibitemShut {NoStop}%
\bibitem [{\citenamefont {Karbach}\ \emph {et~al.}(1997)\citenamefont
  {Karbach}, \citenamefont {M\"uller}, \citenamefont {Bougourzi}, \citenamefont
  {Fledderjohann},\ and\ \citenamefont {M\"utter}}]{Karbach97}%
  \BibitemOpen
  \bibfield  {author} {\bibinfo {author} {\bibfnamefont {M.}~\bibnamefont
  {Karbach}}, \bibinfo {author} {\bibfnamefont {G.}~\bibnamefont {M\"uller}},
  \bibinfo {author} {\bibfnamefont {A.~H.}\ \bibnamefont {Bougourzi}}, \bibinfo
  {author} {\bibfnamefont {A.}~\bibnamefont {Fledderjohann}},\ and\ \bibinfo
  {author} {\bibfnamefont {K.-H.}\ \bibnamefont {M\"utter}},\ }\bibfield
  {title} {\bibinfo {title} {Two-spinon dynamic structure factor of the
  one-dimensional {S}=1/2 {H}eisenberg antiferromagnet},\ }\href
  {https://doi.org/10.1103/PhysRevB.55.12510} {\bibfield  {journal} {\bibinfo
  {journal} {Phys. Rev. B}\ }\textbf {\bibinfo {volume} {55}},\ \bibinfo
  {pages} {12510} (\bibinfo {year} {1997})}\BibitemShut {NoStop}%
\bibitem [{\citenamefont {Marshall}\ and\ \citenamefont
  {Peierls}(1955)}]{Marshall55}%
  \BibitemOpen
  \bibfield  {author} {\bibinfo {author} {\bibfnamefont {W.}~\bibnamefont
  {Marshall}}\ and\ \bibinfo {author} {\bibfnamefont {R.~E.}\ \bibnamefont
  {Peierls}},\ }\bibfield  {title} {\bibinfo {title} {Antiferromagnetism},\
  }\href {https://doi.org/10.1098/rspa.1955.0200} {\bibfield  {journal}
  {\bibinfo  {journal} {Proceedings of the Royal Society of London. Series A.
  Mathematical and Physical Sciences}\ }\textbf {\bibinfo {volume} {232}},\
  \bibinfo {pages} {48} (\bibinfo {year} {1955})}\BibitemShut {NoStop}%
\bibitem [{\citenamefont {Lieb}\ and\ \citenamefont {Mattis}(1962)}]{Lieb62}%
  \BibitemOpen
  \bibfield  {author} {\bibinfo {author} {\bibfnamefont {E.}~\bibnamefont
  {Lieb}}\ and\ \bibinfo {author} {\bibfnamefont {D.}~\bibnamefont {Mattis}},\
  }\bibfield  {title} {\bibinfo {title} {Ordering {{Energy Levels}} of
  {{Interacting Spin Systems}}},\ }\href {https://doi.org/10.1063/1.1724276}
  {\bibfield  {journal} {\bibinfo  {journal} {Journal of Mathematical Physics}\
  }\textbf {\bibinfo {volume} {3}},\ \bibinfo {pages} {749} (\bibinfo {year}
  {1962})}\BibitemShut {NoStop}%
\bibitem [{\citenamefont {Furuya}\ and\ \citenamefont
  {Giamarchi}(2014)}]{Furuya14}%
  \BibitemOpen
  \bibfield  {author} {\bibinfo {author} {\bibfnamefont {S.~C.}\ \bibnamefont
  {Furuya}}\ and\ \bibinfo {author} {\bibfnamefont {T.}~\bibnamefont
  {Giamarchi}},\ }\bibfield  {title} {\bibinfo {title} {Spontaneously
  magnetized {{Tomonaga}}-{{Luttinger}} liquid in frustrated quantum
  antiferromagnets},\ }\href {https://doi.org/10.1103/PhysRevB.89.205131}
  {\bibfield  {journal} {\bibinfo  {journal} {Phys. Rev. B}\ }\textbf {\bibinfo
  {volume} {89}},\ \bibinfo {pages} {205131} (\bibinfo {year}
  {2014})}\BibitemShut {NoStop}%
\bibitem [{\citenamefont {Nishimoto}\ \emph {et~al.}(2012)\citenamefont
  {Nishimoto}, \citenamefont {Drechsler}, \citenamefont {Kuzian}, \citenamefont
  {Richter}, \citenamefont {M{\'{a}}lek}, \citenamefont {Schmitt},
  \citenamefont {van~den Brink},\ and\ \citenamefont {Rosner}}]{linarite_2012}%
  \BibitemOpen
  \bibfield  {author} {\bibinfo {author} {\bibfnamefont {S.}~\bibnamefont
  {Nishimoto}}, \bibinfo {author} {\bibfnamefont {S.-L.}\ \bibnamefont
  {Drechsler}}, \bibinfo {author} {\bibfnamefont {R.}~\bibnamefont {Kuzian}},
  \bibinfo {author} {\bibfnamefont {J.}~\bibnamefont {Richter}}, \bibinfo
  {author} {\bibfnamefont {J.}~\bibnamefont {M{\'{a}}lek}}, \bibinfo {author}
  {\bibfnamefont {M.}~\bibnamefont {Schmitt}}, \bibinfo {author} {\bibfnamefont
  {J.}~\bibnamefont {van~den Brink}},\ and\ \bibinfo {author} {\bibfnamefont
  {H.}~\bibnamefont {Rosner}},\ }\bibfield  {title} {\bibinfo {title} {The
  strength of frustration and quantum fluctuations in {LiVCuO}$_4$},\ }\href
  {https://doi.org/10.1209/0295-5075/98/37007} {\bibfield  {journal} {\bibinfo
  {journal} {{EPL} (Europhysics Letters)}\ }\textbf {\bibinfo {volume} {98}},\
  \bibinfo {pages} {37007} (\bibinfo {year} {2012})}\BibitemShut {NoStop}%
\bibitem [{\citenamefont {Hida}(2007)}]{Hida07}%
  \BibitemOpen
  \bibfield  {author} {\bibinfo {author} {\bibfnamefont {K.}~\bibnamefont
  {Hida}},\ }\bibfield  {title} {\bibinfo {title} {Ferrimagnetic {{States}} in
  {{S}}=1/2 {{Frustrated Heisenberg Chains}} with {{Period}} 3 {{Exchange
  Modulation}}},\ }\href {https://doi.org/10.1088/0953-8984/19/14/145225}
  {\bibfield  {journal} {\bibinfo  {journal} {J. Phys.: Condens. Matter}\
  }\textbf {\bibinfo {volume} {19}},\ \bibinfo {pages} {145225} (\bibinfo
  {year} {2007})}\BibitemShut {NoStop}%
\bibitem [{\citenamefont {Hida}\ and\ \citenamefont {Takano}(2008)}]{Hida08}%
  \BibitemOpen
  \bibfield  {author} {\bibinfo {author} {\bibfnamefont {K.}~\bibnamefont
  {Hida}}\ and\ \bibinfo {author} {\bibfnamefont {K.}~\bibnamefont {Takano}},\
  }\bibfield  {title} {\bibinfo {title} {Frustration-induced quantum phases in
  mixed spin chain with frustrated side chains},\ }\href
  {https://doi.org/10.1103/PhysRevB.78.064407} {\bibfield  {journal} {\bibinfo
  {journal} {Phys. Rev. B}\ }\textbf {\bibinfo {volume} {78}},\ \bibinfo
  {pages} {064407} (\bibinfo {year} {2008})}\BibitemShut {NoStop}%
\bibitem [{\citenamefont {Li}\ and\ \citenamefont {Haldane}(2008)}]{li08}%
  \BibitemOpen
  \bibfield  {author} {\bibinfo {author} {\bibfnamefont {H.}~\bibnamefont
  {Li}}\ and\ \bibinfo {author} {\bibfnamefont {F.}~\bibnamefont {Haldane}},\
  }\bibfield  {title} {\bibinfo {title} {{Entanglement Spectrum as a
  Generalization of Entanglement Entropy: Identification of Topological Order
  in Non-Abelian Fractional Quantum Hall Effect States}},\ }\href
  {https://doi.org/10.1103/PhysRevLett.101.010504} {\bibfield  {journal}
  {\bibinfo  {journal} {Phys. Rev. Lett.}\ }\textbf {\bibinfo {volume} {101}},\
  \bibinfo {pages} {010504} (\bibinfo {year} {2008})}\BibitemShut {NoStop}%
\bibitem [{\citenamefont {den Nijs}\ and\ \citenamefont
  {Rommelse}(1989)}]{deNijs89}%
  \BibitemOpen
  \bibfield  {author} {\bibinfo {author} {\bibfnamefont {M.}~\bibnamefont {den
  Nijs}}\ and\ \bibinfo {author} {\bibfnamefont {K.}~\bibnamefont {Rommelse}},\
  }\bibfield  {title} {\bibinfo {title} {Preroughening transitions in crystal
  surfaces and valence-bond phases in quantum spin chains},\ }\href
  {https://doi.org/10.1103/PhysRevB.40.4709} {\bibfield  {journal} {\bibinfo
  {journal} {Phys. Rev. B}\ }\textbf {\bibinfo {volume} {40}},\ \bibinfo
  {pages} {4709} (\bibinfo {year} {1989})}\BibitemShut {NoStop}%
\bibitem [{\citenamefont {White}\ and\ \citenamefont {Huse}(1993)}]{White93}%
  \BibitemOpen
  \bibfield  {author} {\bibinfo {author} {\bibfnamefont {S.~R.}\ \bibnamefont
  {White}}\ and\ \bibinfo {author} {\bibfnamefont {D.~A.}\ \bibnamefont
  {Huse}},\ }\bibfield  {title} {\bibinfo {title} {Numerical
  renormalization-group study of low-lying eigenstates of the antiferromagnetic
  {S=1} {H}eisenberg chain},\ }\href {https://doi.org/10.1103/PhysRevB.48.3844}
  {\bibfield  {journal} {\bibinfo  {journal} {Phys. Rev. B}\ }\textbf {\bibinfo
  {volume} {48}},\ \bibinfo {pages} {3844} (\bibinfo {year}
  {1993})}\BibitemShut {NoStop}%
\bibitem [{ITe()}]{ITensor}%
  \BibitemOpen
  \href@noop {} {\bibinfo {title} {https://itensor.org/}}\BibitemShut {NoStop}%
\end{thebibliography}%

\end{document}